\documentclass[amsmath,amssymb,amsfonts,prd,twocolumn,superscriptaddress,nofootinbib,longbibliography]{revtex4-1}

\usepackage{graphicx}
\usepackage{multirow}
\usepackage{longtable}
\usepackage{color}

\newcommand{\beq}{\begin{equation}}
\newcommand{\eeq}{\end{equation}}
\newcommand{\beqa}{\begin{eqnarray}}
\newcommand{\eeqa}{\end{eqnarray}}

\newcommand{\lexp}{\mathop{\langle}}

\newcommand{\rexpc}{\mathop{\rangle_c}}

\def\dD{\delta_{\rm D}}

\def\tf{\tilde \phi}

\font\BF=cmmib10
\font\BFs=cmmib10 scaled 833
\def\k{{\hbox{\BF k}}}
\def\x{{\hbox{\BF x}}}

\def\ks{{\hbox{\BFs k}}}
\def\xs{{\hbox{\BFs x}}}
\def\p{{\hbox{\BF p}}}
\def\q{{\hbox{\BF q}}}

 \def\la{\mathrel{\mathpalette\fun <}}
 \def\ga{\mathrel{\mathpalette\fun >}}
 \def\fun#1#2{\lower3.6pt\vbox{\baselineskip0pt\lineskip.9pt
        \ialign{$\mathsurround=0pt#1\hfill##\hfil$\crcr#2\crcr\sim\crcr}}}

\def\Mpc{\, h^{-1} \, {\rm Mpc}}
\def\Gpc{\, h^{-1} \, {\rm Gpc}}

\def\kvecMpc{\, h \, {\rm Mpc}^{-1}}
\def\kMpc{\, h \, {\rm Mpc}^{-1}}
\def\fnl{f_{\rm NL}}
\def\gnl{g_{\rm NL}}
\def\tnl{\tau_{\rm NL}}

\def\mnras{MNRAS}

\def\apj{ApJ}
\def\apjs{ApJS}
\def\aap{A \& A}
\def\apjl{ApJ. Lett.}

\def\jcap{JCAP}

\def\prd{Phys. Rev. D}
\def\physrep{Phys. Rep.}

\begin{document}
%%%%%%%%%%%%%%%%%%%%%%%%%%%%%%%%%%%%%%%%%%%%%%%%%%%%%%%%%%%%%%%%%%%%%%%%%%%%%%%%%%%%%%

%
%
%==========================================================================
\title{Large-scale Bias and Efficient Generation of Initial Conditions \\ for Non-Local Primordial Non-Gaussianity}
%==========================================================================
%
%\vskip 1pc

% --- author list and order to decide ----
\author{Rom\'an Scoccimarro}
% \email{rs123@nyu.edu}
 \affiliation{Center for Cosmology and Particle Physics, Department of Physics,  New York University, NY 10003, New York, USA}
\author{Lam Hui}
\affiliation{ISCAP and Department of Physics, Columbia University, New York, NY 10027}
\author{Marc Manera}%
% \email{mm3812@nyu.edu}
 \affiliation{Center for Cosmology and Particle Physics, Department of Physics,  New York University, NY 10003, New York, USA}
 \affiliation{Institute of Cosmology and Gravitation, Dennis Sciama Building, University of Portsmouth, Burnaby Road, Portsmouth, PO1 3FX}
\author{Kwan Chuen Chan}
% \email{kcc274@nyu.edu}
\affiliation{Center for Cosmology and Particle Physics, Department of Physics,  New York University, NY 10003, New York, USA}

%:Abstract
%================
\begin{abstract}
%================
%
We study the scale-dependence of halo bias in generic (non-local) primordial non-Gaussian (PNG) initial conditions of the type motivated by inflation, parametrized by an arbitrary quadratic kernel. 
We first show how to generate  non-local PNG initial conditions with minimal overhead compared to local PNG models 
for a general class of primordial bispectra that can be written as linear combinations of separable templates.  We run cosmological simulations for the local, and non-local equilateral and orthogonal models and present results on the scale-dependence of halo bias. 
We also derive a general formula for the Fourier-space bias using the peak-background split (PBS) in the context of the excursion set approach to halos and discuss the difference and similarities with the known corresponding result  from local bias models.  
 Our PBS bias formula generalizes previous results in the literature to include non-Markovian effects and non-universality of the mass function and are in better agreement with measurements in numerical simulations than previous results for a variety of halo masses, redshifts and halo definitions. We also derive for the first time quadratic bias results for arbitrary non-local PNG, and show that non-linear bias loops give small corrections at large-scales. The resulting well-behaved perturbation theory paves the way to constrain non-local PNG from measurements of the power spectrum and bispectrum in galaxy redshift surveys.

\end{abstract}

\maketitle

\section{Introduction}

By far the leading scenario for explaining the initial seeds for the
formation of structure in the universe is inflation,
but we still don't know what kind of field(s) are responsible for it,
nor their Lagrangian.  For simplicity it is usually
assumed that inflation is driven by a single scalar field
in its vacuum state, that  slow rolls and has a canonical
kinetic term. Under these circumstances the fluctuations are very close to 
Gaussian, i.e. their reduced bispectrum is of the order of the tilt in
the scalar spectrum~\cite{2003JHEP...05..013M}. The tilt is known to
be rather small~\cite{2010arXiv1001.4538K}. In the language of
Eq.~(\ref{localfnl}) below, it results in a non-Gaussianity $\fnl \la
0.05$, which is unlikely to be ever detectable.

However, as discussed below, a large class of theories can produce a
Universe much like ours, but with small deviations
from Gaussianity that are potentially observable with future
experiments.  Producing interesting
constraints on the physics in the early Universe requires searches for
such deviations. The situation is analogous to particle physics, where two-point
functions describe freely propagating particles and higher-order
correlations (whose measurement is the goal of particle colliders)
describe the physical interactions. In cosmology, measuring the power
spectrum only gives information on free fields in the
inflationary background. To learn about  interactions and thus distinguish between models, it is essential that we measure
 higher-order correlation functions. In short, searching for
primordial non-Gaussianity (hereafter PNG) is akin to using accelerators to constrain
the interactions in the early universe. Therefore, detecting primordial
non-Gaussianity is our best chance to learn about the physics of
inflation. The prospects for the next decade are very promising~\cite{2009astro2010S.158K}.

Deviations from the assumptions enumerated above lead to observable
non-Gaussianities. The simplest type of non-Gaussianity arises by
going beyond single-field models.  A second scalar field, usually
called the ``curvaton'' could be light during inflation; such a field
would come to dominate the energy density of the universe after the
end of inflation, and then produce effectively a second
reheating~\cite{1997PhRvD..56..535L,*2002PhLB..524....5L,*EnqSlo0204,*2003PhRvD..67l1301B}.  In
addition, fluctuations could be generated during the reheating period
when the inflaton energy density is converted into standard model
particles with a fluctuating decay
rate~\cite{2003astro.ph..3614K,*2004PhRvD..69b3505D,*2004PhRvD..70h3004B}.  In all these
models, which are very closely related~\cite{2010JCAP...11..037A}, non-Gaussianities can be characterized by a primordial
gravitational potential at sub-horizon scales that obeys~\cite{SalBon9102,*SalBon9012}
\beq
\Phi(\x) = \phi(\x)+\fnl \Big([\phi(\x)]^2- \langle\phi^2\rangle \Big) 
 \label{localfnl}
\eeq
where $\phi$ is a random Gaussian field, leading to a bispectrum of the so-called {\em local} form,
\beq
B_\Phi^{\rm local} = 2 \fnl P_1 P_2 + {\rm cyc.}
\label{Blocal}
\eeq
where $P_i\equiv P_\phi(k_i)$ is the power spectrum, `cyc.' denotes
cyclic permutations of the $k_i$, and $\fnl$ could naturally be of
order  $\fnl \approx
5-30$~\cite{2004PhRvD..69h3505D,*2004PhRvD..69d3508Z}.  Primordial
non-Gaussianity at this level should be detectable through
measurements of the bispectrum of the CMB and in galaxy redshift
surveys. Current constraints on such local shape are $-10<\fnl^{\rm local}<74$ (95\% limit) from the CMB bispectrum in the WMAP7 dataset~\cite{2010arXiv1001.4538K}, and $-29 <\fnl^{\rm local}<70$ (95\% limit) from the scale-dependence of bias in the power spectrum of galaxies and quasars in the SDSS-II survey~\cite{2008JCAP...08..031S}. This constraint relies on the fact that local PNG generates a $1/k^2$ scale-dependent bias, as first  derived  in~\cite{2008PhRvD..77l3514D}, which justifiably has generated a lot of interest given the ever increasing volume of upcoming galaxy surveys\footnote{Of historical note, a quarter century ago \cite{1986ApJ...310...19G}  first  pointed out that PNG can lead to a scale-dependent bias growing at large scales, when they proposed a PNG model with a trispectrum that leads to a $1/k$ bias, but their result was somehow forgotten.}.

Another possible deviation from the simplest inflationary scenario, staying within single-field models, is to go beyond canonical kinetic terms; the primary examples in this case are DBI inflation~\cite{2004PhRvD..70l3505A,*2004PhRvD..70j3505S} and $k$-inflation~\cite{1999PhLB..458..209A,*2008JCAP...03..028L,*2010PhRvD..81d3502T}. In these models the bispectrum shape takes predominantly a form that peaks at equilateral configurations, the so-called {\em equilateral} shape~\cite{2007JCAP...01..002C}, and a small contribution of a different shape called the {\em orthogonal} shape~\cite{2010JCAP...01..028S}. For the convenience of numerical calculations, these shapes are approximated by {\em templates} which are a sum of factorizable contributions. The equilateral template bispectrum may be written as,
 \beqa
{1\over 6 \fnl}  B_\Phi^{\rm equil}  &=& - (P_1 P_2 + {\rm cyc.})  - 2 (P_1 P_2 P_3)^{2\over3} \nonumber \\ & & + ( P_1^{1\over3} P_2^{2\over3} P_3+ {\rm cyc.}) 
 \label{Bequil}
 \eeqa
where it should be noted that there are  a total of 3 cyclic permutations for the first term, and 6 for the last term.  Here $P_i\equiv P_\phi(k_i)=A k_i^{-3}$ is the power spectrum of the curvature perturbation (Bardeen potential), proportional to the Newtonian potential on subhorizon scales. The  orthogonal template bispectrum reads instead~\cite{2010JCAP...01..028S}

 \beqa
{1\over 6 \fnl}   B_\Phi^{\rm ortho} & = &-3 (P_1 P_2+ {\rm cyc.})    - 8 (P_1 P_2 P_3)^{2\over3} \nonumber \\ & & 
+ 3 (P_1^{1\over3} P_2^{2\over3} P_3+ {\rm cyc.}) 
 \label{Bortho}
 \eeqa
 More precisely, this template is only a good approximation to the orthogonal shape away from the squeezed limit, otherwise more complicated templates should be used~\cite{2010JCAP...01..028S,2011JCAP...02..006C}. This issue is relevant to the  calculation of the large-scale bias, as the more accurate template does not lead to a scale-dependent bias at low-$k$ whereas the simpler template in Eq.~(\ref{Bortho}) leads to a $1/k$ bias. On the other hand, such a behavior is interesting from a phenomenological point of view as it is in between scale independence and the $1/k^2$ of local PNG. Moreover, such a scale-dependent bias is realizable in some models of inflation~\cite{2010JCAP...04..027C,*2010PhRvD..81f3511C,*2011arXiv1106.1462C}. We will leave for future study more accurate templates and proceed with Eqs.~(\ref{Bequil}-\ref{Bortho}) in this paper. Current 95\% limits from the CMB bispectrum in WMAP7 are $-214<\fnl^{\rm equil}<266$ and  $-410<\fnl^{\rm ortho}<6$, respectively~\cite{2010arXiv1001.4538K}. Limits on these shapes from galaxy surveys are, $-419<\fnl^{\rm equil}<625$ and  $-179<\fnl^{\rm ortho}<6$, respectively~\cite{2011arXiv1104.5015X}.

While these are the most often discussed deviations from the simplest inflationary physics, two more of the assumptions enumerated above may be broken. Relaxing the assumption of initial vacuum states leads to the so-called {\em folded} shape~\cite{2007JCAP...01..002C,2008JCAP...05..001H,*2009JCAP...05..018M,*2010JCAP...02..001M,*AshShi1103}, which gives  
\beqa
{1\over 6 \fnl}  B_\Phi^{\rm folded} &=& (P_1 P_2 + {\rm cyc.})   + 3 (P_1 P_2 P_3)^{2\over3} \nonumber \\ & & 
- (P_1^{1\over3} P_2^{2\over3} P_3 + {\rm cyc.}) 
\label{Bfold}
 \eeqa
Note that this shape can be constructed as a linear combination of the equilateral and orthogonal shapes. Thus, from the point of view of constraining it in data, it is enough to consider the previous two non-local shapes. Finally, relaxing slow-roll can lead to more complicated shapes, as in resonant non-Gaussianity~\cite{2010arXiv1002.0833F}, which cannot be easily written in terms of the other shapes, due to its non-factorizable form. see also 

For a discussion of most of  these models from effective field theory  see~\cite{2008JHEP...03..014C}. Note that in addition, there could post-inflationary PNG contributions, such as preheating and phases transitions, which we ignore in this paper (see e.g.~\cite{EnqJokMaz0504,*BarCli0605,*BarCli0704,*ChaRaj0808,*ChaRaj0802,*BonFroHua0908,*FigCalKam1006,*RegShe1009}).

The current constraints on primordial non-Gaussianity from galaxy surveys come from the scale-dependence of the bias in the {\em power spectrum} of biased tracers such as galaxies and quasars, which are sensitive to the {\em local} form (Eq.~\ref{Blocal}) of the primordial bispectrum~\cite{2008PhRvD..77l3514D,2008ApJ...677L..77M,2008PhRvD..78l3534T,2008PhRvD..78l3507A,2008JCAP...08..031S}. Adding the galaxy bispectrum, not only enhances the constraining power of surveys, but also helps constrain other shapes (e.g. equilateral and orthogonal) where no strong scale-dependence of bias is induced. This in fact makes galaxy surveys in principle as powerful if not more than the CMB in the search for primordial non-Gaussianity~\cite{2007PhRvD..76h3004S,2009ApJ...703.1230J,2010JCAP...07..002N,2011JCAP...04..006B,EmiHaloBisp}. 

One of the main arguments  for being cautiously optimistic about large-scale structure (LSS) constraints is that while CMB gains more 2D modes by going to small scales, where other sources of non-Gaussianity become important,  LSS gets more {\em 3D modes} by going to larger scales where the physics becomes  simpler. But even if these forecasts were not achieved in practice, it is important to obtain constraints on PNG from large-scale structure (LSS) for several reasons: i) the kinds of surveys needed to accomplish this are already planned, mainly to study cosmic acceleration, ii) if there is a scale-dependence in $\fnl$ it is essential to have a wide coverage of scales and CMB+LSS does this very well, iii) something as fundamental as claiming that the initial conditions are non-Gaussian will require confirmation by an independent method.

In this paper we want to implement N-body simulations initial conditions for the different models discussed above and introduce an efficient algorithm for doing so. In addition, we provide more rigorous theoretical predictions for these models from the peak-background split and contrast these with measurements in numerical simulations, previous PBS predictions in the literature and those of local bias models.

A guide to the main results may be useful here. For readers interested in the algorithm for generating PNG initial conditions of non-local type (Section~\ref{GenBisp}), Eqs.~(\ref{KEQ}-\ref{KORT}) give our implentation for equilateral and orthogonal templates. Appendix~\ref{NSI} generalizes these results beyond scale-invariance and Appendix~\ref{ICsSDfnl} gives  a self-contained prescription for more general templates, including $\gnl$ PNG. For those interested in the peak-background split main result on scale-dependent bias in $\fnl$ models, it is  given by Eq.~(\ref{DeltaB}). Table~\ref{BiasParam} summarizes our notation for bias parameters.  Section~\ref{CompRes} discusses how this result differs from previous PBS calculations in the literature, and Section~\ref{NBcomp} discusses comparison against numerical simulations (see Figs.~\ref{BiasFOF0p2L}-\ref{BiasEq}).

\section{Method for Generating a Desired Bispectrum}
\label{GenBisp}
It is well known that the first term in Eqs.~(\ref{Bequil}-\ref{Bortho}) is easily generated by writing the Bardeen potential $\Phi$ as a local nonlinear function of an auxiliary Gaussian potential $\phi$, Eq.~(\ref{localfnl}), 
 and we would like to write similar transformations from $\phi$ to $\Phi$ for the full expressions in Eqs.~(\ref{Bequil}-\ref{Bortho}).
Since the bispectra in all cases scales as total power squared, we are interested in a quadratic non-local kernel $K$ such that,

\beq
\Phi = \phi +  f_{\rm NL}\ K[\phi,\phi]
\label{Kintro}
\eeq
which in Fourier space can be written as
\beqa
\Phi(\k)&=& \phi(\k) + \int  f_{\rm NL}\, [\delta_{\rm D}]\,K(\k_1,\k_2)\, \phi(\k_1) \phi(\k_2) \nonumber \\ & & 
\quad \quad \quad \quad d^3 k_1 d^3 k_2,
\label{KFourier}
\eeqa
where the Fourier-space kernel $K$ is dimensionless, and $[\delta_{\rm D}] =\delta_{\rm D}(\k-\k_{12}) $ ($\k_{12}\equiv \k_1+\k_2$).
For simplicity we assume that $\fnl$ is a constant independent of scale, although there are theoretical reasons to expect scale-dependence~\cite{2009JCAP...12..022S,*2010JCAP...02..034B,*2010JCAP...10..004B,*2011JCAP...01..006B,*Hua1012,2011JCAP...03..017S}. It is however straightforward to include scale-dependent PNG, see Appendix~\ref{ICsSDfnl} for the generalization.

The bispectrum of such a model reads simply,

\beq
B_\Phi = 2 f_{\rm NL} K_{12} P_1 P_2 + {\rm cyc.} 
\label{BfromK}
\eeq
where $K_{12}\equiv K(\k_1,\k_2)$, which from Eq.~(\ref{KFourier}) has the exchange symmetry $K_{12}=K_{21}$, i.e. $\k_1 \leftrightarrow \k_2$. The goal to generate initial conditions is to invert Eq.~(\ref{BfromK}) to find the kernel from a given bispectrum. Such inversion is not unique because the kernel describes pairwise couplings between two modes, while the bispectrum is the result of the ``average" (sum over cyclic permutations of two out of three modes in Eq.~\ref{BfromK}) over all such possible pairwise couplings. Therefore, there is no unique way of constructing the pairwise couplings themselves ($K$) using only the ``average" information provided by the bispectrum.

One  solution to this dilemma is to declare that the pairwise couplings are equal to the average regardless of the pair of momenta, that is, use the standard {\em reduced bispectrum} as the kernel,  i.e.  $K_{12}=B/(P_1P_2+P_2P_3+P_3P_1)$ (see~\cite{2010PhRvD..82j3002S}). However, while this inversion provides a unique kernel for a given bispectrum, it introduces other issues. This prescription has the much more restrictive symmetry under cyclic permutations of $k_1,k_2,k_3$. This means that two nearly antiparallel high-frequency modes $\k_1 \approx -\k_2$ couple with the same kernel than a low frequency $\k_3=-\k_1-\k_2$ to a high-frequency mode ($\k_1$ or $\k_2$), which seems a rather strong assumption particularly in the squeezed limit (where $k_1\approx k_2 \gg k_3$). See Appendix~\ref{sqz} for more discussion on this for the kernels we obtain in this paper.

Furthermore, using such ``average" kernel will in general give an incorrect configuration dependence to the ``snake" diagrams of the four-point function that depend on the square of the quadratic kernel  (usually characterized by the $\tau_{\rm NL}$ parameter, see~\cite{2006PhRvD..74l3519B}), since the averaging and squaring do not commute\footnote{
The ``snake''  trispectrum amplitude is, from Eq.~(\ref{KFourier}),  $T_{\rm snake} =(2\fnl)^2\ [K({\bf k}_{12},-{\bf k}_1)\, K({\bf k}_{12},{\bf k}_3)\, P_1\,P_{12}\,P_3 + {\rm 11~perm.}]$, where ${\bf k}_{12}\equiv {\bf k}_1+{\bf k}_2$ and $P_{ij}\equiv P(k_{ij})$.  Using the reduced bispectrum as a kernel gives a non-trivial constraint between this amplitude and the bispectrum and power spectrum. A well-known example of non-local (but non-primordial) NG  is that generated by gravitational evolution from Gaussian initial conditions~\cite{BerColGaz02}, in which  this constraint is violated. It would be remarkable if inflationary PNG satisfied this constraint. 
% On the other hand, it is not clear at present how important is this issue in practice, since the higher-order correlations arising from inflation cannot in general be considered as arising from the nonlinear evolution of a classical stochastic system, e.g.~\cite{2011JCAP...02..017B}. 
}. 
While one can of course fix this problem by adding a suitable cubic term ${\cal O}(\fnl^2\, L[\phi,\phi,\phi])$ which through the ``star" diagrams cancels the snake contribution (and adds the appropriate one), this is somewhat unnatural and would be costly numerically (as fixing the snake topology through star diagrams cannot be written in separable form). It seems  more desirable to have freedom at the level of the quadratic kernel that might be used to reproduce the correct bispectrum and the snake-topology contributions to the higher-point functions. Furthermore, one might be able to do so while keeping the kernel a sum over separable contributions, and thus considerably speeding up the generation of initial conditions. Here we provide a first attempt (ignoring four- and higher-point information) along these lines.

In this paper we start from the bispectrum templates and work to invert Eq.~(\ref{BfromK}). The main point of the bispectrum templates is to have expressions for the bispectrum which are factorizable functions of the $k_i$, to speed up numerical calculations. It is simple to construct an inversion procedure that keeps this basic property for the kernel, which  leads to fast implementation of initial conditions generation for N-body simulations. Let's focus for definiteness on the terms that have $\k_1 \leftrightarrow \k_2$ exchange symmetry. If one looks at the bispectrum contributions with this symmetry, there will be one solution for the kernel $K_{12}$ that generates each such terms; each distinct solution corresponds to matching $K_{12} P_1 P_2$ to the distinct bispectrum contributions of the form $(P_1^\alpha P_2^\beta +P_2^\alpha P_1^\beta)P_3^{2-\alpha-\beta}$ for distinct values of $\alpha,\beta$. This means that the number of solutions will equal the number of different exponents that the three power spectra are raised to, e.g. in the local model there are two exponents (one and zero) and thus there will be two separable solutions (see below). 

By performing linear combinations of the multiple solutions we construct a family of kernels that reproduce to tree-level the desired bispectrum. We then impose regularity constraints from loop corrections to restrict the free parameters so that the tree-level results are not spoiled at large scales. While our expression for the kernel in Fourier space is general, the expressions we present in the main text for the Bardeen potential $\Phi$ in terms of the Gaussian field $\phi$ are restricted to scale-invariant spectra. These expressions are generalized beyond scale-invariance in Appendix~\ref{NSI}. 

Since the templates we use (Eqs.~\ref{Bequil}-\ref{Bfold}) can be written as linear combinations of local form and two additional terms, we will first discuss how to generate each of these terms separately. Depending on the type of PNG one is interested in, these generators can then be summed up according to the desired linear combination.

Note that our methodology can be applied to any primordial bispectrum that can be written as sum of separable templates, i.e. $g_1(k_1)\, g_2(k_2)\, g_3(k_3)$ with $g_i$ arbitrary functions, not necessarily power laws, see Appendix~\ref{ICsSDfnl}. For other approaches to generating non-Gaussian fields see~\cite{2006astro.ph.12571S,2010PhRvD..82b3502F,2010JCAP...10..022W,2010arXiv1008.1730F,2011arXiv1102.3229W,2011JCAP...03..017S}.

\subsection{First Generator: Local non-Gaussianity}
\label{FirstGen}
 
Let's see how our method works in the simplest, well-known case of local NG. We require that
 
\beq
2 f_{\rm NL} K_{12} P_1 P_2 + {\rm cyc.} = 2 f_{\rm NL} P_1 P_2 + {\rm cyc.} 
\eeq
Note that there are {\em two} ways in which the RHS is, say, symmetric under $\k_1 \leftrightarrow \k_2$ exchange, one is the term $P_1P_2$ ($\alpha=\beta=1$), the other is the combination $P_3(P_1+P_2)$ (with $\alpha=1$ and $\beta=0$). Matching the contribution from $K_{12}$ to the first  leads to the trivial solution, $K_{12}=1$,  using the second one leads to the non-trivial one

\beq
K_{12}= {P_3\over 2} \Big( \frac{1}{P_1} + \frac{1}{P_2}\Big),
\label{Kloc2}
\eeq
where $P_3=P_{12}=P(|\k_1+\k_2|)$. This means that more generically one can generate the local bispectrum using the kernel

\beq
K_{12}^{\rm loc}= (1-u) + u\ {P_3\over 2} \Big( \frac{1}{P_1} + \frac{1}{P_2}\Big),
\label{Klocu}
\eeq
where $u$ is a free parameter, to be constrained later from one-loop power spectrum considerations, as the resulting (tree-level) bispectrum is independent of $u$. For a scale-invariant spectrum Eq.~(\ref{Klocu}) leads to a simple form for the Bardeen potential,

\beq
\Phi= \phi + f_{\rm NL}\Big[ (1-u)\, \phi^2 + u\, \nabla^{-2} \partial^{-1} (\phi \nabla^2 \partial \phi) \Big],
\label{Philoc}
\eeq

where the operators introduced can be most easily understood in their Fourier representation,

\beq
\partial \phi \equiv \sqrt{-\nabla^2}\phi (\x)\equiv  \int {\rm e}^{-i \k \cdot \x} \ k \, \phi(\k)\, d^3k
\label{sqrtbox}
\eeq
 and as usual,
 \beq
 \nabla^{-2} A(\x) \equiv -  \int {\rm e}^{-i \k \cdot \x} \Big(\frac{1}{k^2}\Big) \, A(\k)\, d^3k
 \label{lapinv}
 \eeq
and the inverse operation
 \beq
 \partial^{-1} A \equiv  \sqrt{-\nabla^{-2}}  A \equiv  \int {\rm e}^{-i \k \cdot \x} \Big(\frac{1}{k}\Big) \, A(\k)\, d^3k
 \label{nabalinv}
 \eeq
Note that all these auxiliary fields can be obtained by efficient Fast Fourier transforms (FFTs). For all our generators, not just local,  our algorithm only needs FFTs and basic arithmetic operations, avoiding cumbersome convolutions.

The choice $u=0$ in Eq.~(\ref{Philoc}) is often used to generate local PNG, and for a good reason, as we shall see soon.

\subsection{Second Generator}
\label{secGen}

For the second term in Eqs.~(\ref{Bequil}-\ref{Bfold}), we have (choosing the arbitrary normalization appropriately) 
\beq
2 f_{\rm NL} K_{12} P_1 P_2 + {\rm cyc.} = 6 f_{\rm NL} (P_1 P_2 P_3)^{2/3}
\label{secterm}
\eeq
and in this case we are dealing with a single totally symmetric bispectrum (no choice regarding exchange symmetry, i.e. we just have $\alpha=\beta=2/3$), thus there is a unique solution for the kernel. This can also be seen by rewriting Eq.~(\ref{secterm})  as

\beq
\hat{K}_{12} + {\rm cyc.} = 3,
\eeq

where $\hat{K}_{ij}\equiv K_{ij} (P_i P_j/P_{ij}^2)^{1/3}$ with $P_{ij}\equiv P(k_{ij})$, and thus since $k_i$ are arbitrary, $\hat{K}_{ij}=1$, which implies

\beq
K_{12}^{\rm 2nd} = \frac{P_3^{2/3}}{(P_1 P_2)^{1/3}}
\label{K2nd} 
\eeq

It is then easy to see that the second term in Eqs.~(\ref{Bequil}-\ref{Bortho}) is generated by the following nonlocal nonlinear function for scale-invariant spectra,
 
 \beq
 \Phi=\phi - f_{\rm NL} \nabla^{-2} (\partial \phi)^2,
 \label{2ndfnl}
 \eeq
which leads to a bispectrum for $\Phi$,
 
 \beq
 B_\Phi = 6 f_{\rm NL} (P_1 P_2 P_3)^{2/3}
 \label{B2nd}
 \eeq

\subsection{Third Generator}
\label{3rdterm}

For the third term in Eqs.~(\ref{Bequil}-\ref{Bfold}), we have 
\beq
2 f_{\rm NL} K_{12} P_1 P_2 + {\rm cyc.} = 2 f_{\rm NL} P_1^{1/3} P_2^{2/3} P_3 + {\rm cyc.}
\eeq
which contains three possible ways of arranging the $\k_1 \leftrightarrow \k_2$ symmetry, namely $\alpha=2/3,\beta=1$, $\alpha=1/3,\beta=1$, $\alpha=1/3,\beta=2/3$, 
and thus there are three solutions, respectively

\beq
K_{12}^{I}=  P_3^{1/3} ( P_1^{-1/3} + P_2^{-1/3}  ) 
\eeq
\beq
K_{12}^{II}= P_3^{2/3} ( P_1^{-2/3} + P_2^{-2/3}  ) 
\eeq
\beq
K_{12}^{III}= P_3\, ( P_1^{-2/3} P_2^{-1/3}+ P_1^{-1/3} P_2^{-2/3} )  
\eeq

Therefore, there is a two-parameter family of kernels that generate the desired bispectrum, i.e.

\beq
K_{12}^{\rm 3rd}= (1-t-s)\, K_{12}^{I}+ t\, K_{12}^{II} + s\, K_{12}^{III}
\label{K3rd}
\eeq
In terms of the Bardeen potential, we have that  the third term in Eqs.~(\ref{Bequil}-\ref{Bortho}) is generated by the following nonlocal nonlinear function (assuming scale-invariance) 
 
 \beqa
 \Phi&=&\phi + f_{\rm NL} \Big[ (1-t-s)\, \partial^{-1} (\phi\, \partial \phi)
 + t \, \nabla^{-2} (\phi \ \nabla^2 \phi) \nonumber \\ & &
 + s \,  \nabla^{-2}\partial^{-1} (\nabla^2\phi\ \partial \phi)
  \Big],
 \label{3rdfnl}
 \eeqa
which leads to a tree-level bispectrum for $\Phi$, 
 
 \beq
 B_\Phi = f_{\rm NL} P_1 P_2^{2/3} P_3^{1/3} + {\rm cyc.}
 \label{B3rd}
 \eeq
 independent of $s$ and $t$. The results for all the kernels so far are general, while those for $\Phi(\phi)$ assume scale invariance; the generalization of the latter is given in  Appendix~\ref{NSI}.

\subsection{Constraints from one-loop corrections}
\label{1Lconst}

While it is not surprising that a given bispectrum may be generated in different ways, leading to different higher-point functions at ${\cal O}(\fnl^2)$, one must require that loop corrections do not spoil the tree-level results at large scales (e.g. the primordial power spectrum and the bispectrum), and this may lead to constraints on the free parameters introduced above if the kernels are too singular in the infrared (IR). 
By one-loop correction, we mean specifically
the correction to the $\Phi$ power spectrum due to
the presence of PNG. And by tree-level, we mean
the contribution to the $\Phi$ power spectrum 
from the purely Gaussian part i.e. the $\phi$ 
power spectrum. In order to preserve the large-scale behavior that one wants (typically something
close to scale-invariant), one should make sure
the one-loop correction does not dominate over
the tree-level one in the IR.

It is easy to see that the requirement that the one-loop power spectrum does not alter the $k^{-3}$ behavior at large scales (i.e. we only accept corrections that diverge as $k^{-3}$ or slower as $k\rightarrow 0$) introduces non-trivial constraints. Consider for simplicity the local model. Using the kernel in Eq.~(\ref{Klocu}), we obtain for the one-loop corrections to the Bardeen potential power spectrum, which at tree-level reads $P_\Phi=P_\phi=A/k^3$,

\beqa
\delta P_\Phi &= & 2\, (A\,f_{\rm NL})^2 \times \int d^3q \, \Big[ 
\frac{(1-u)^2}{q^3|\k-\q|^3} + \frac{u(1-u)}{q^3\, k^3}  \nonumber \\
&\times &  {|\k-\q|^3 +q^3\over q^3}
+  \Big( \frac{u}{2k^3} \Big)^2\, \Big( {(|\k-\q|^3 +q^3)^2\over q^3\ |\k-\q|^3}\Big) \Big],  \nonumber\\ &&
\label{P1Llocu}
\eeqa
which gives a $k^{-6}$ correction to the spectrum with a divergent amplitude, unless $u=0$, that is, only the ``standard" local model kernel ($K=1$) is allowed. In principle such singular behaviors can be cancelled by adding appropriate terms of ${\cal O}(\fnl^2 \phi^3)$ in the Bardeen potential, but we ignore this possibility here as we have no ${\cal O}(\fnl^2)$ information, and unless required by symmetry this precise cancellation would introduce fine-tuning.

For more complicated bispectra with more than one type of term (e.g. as in equilateral and orthogonal models) the constraints follow similarly (with longer expressions). The generic linear combination of kernels that generate the desired bispectra given by Eqs.~(\ref{Bequil}) and~(\ref{Bortho}) to tree-level are (see Eqs.~\ref{Klocu},\ref{K2nd},\ref{K3rd})

\beq
K_{\rm EQ} = -3 K_{12}^{\rm loc} -2 K_{12}^{\rm 2nd} + 3 K_{12}^{\rm 3rd},
\label{KeqGen}
\eeq
\beq
K_{\rm ORT} = -9 K_{12}^{\rm loc} -8 K_{12}^{\rm 2nd} + 9 K_{12}^{\rm 3rd}.
\label{KortGen}
\eeq
The one-loop power spectrum leads to the following results for these kernels: for the most dangerous term at low-$k$ we have an amplitude

\beq
{1\over k^6}:\ \ \ \ \ \propto (u-2s)\ \ \ {\rm (EQ\ and\ ORT)}
\eeq
and choosing $s=u/2$ automatically makes the $k^{-5}$ amplitude vanish in both cases. For the $k^{-4}$ amplitude we have (after setting $s=u/2$),
\beq
{1\over k^4}:\ \ \ \ \ \propto 
\left\{ \begin{array}{rr}
 (1-3t)^2  &\mbox{{\rm (EQ)} } \\
   (4-9t)^2 &\mbox{ {\rm (ORT)}}
       \end{array} \right. 
\eeq
and again these choices, $t=1/3$ and $t=4/9$, make the respective $k^{-3}$ amplitudes vanish (another way to see this is to look at the squeezed limit of the kernels, see Appendix~\ref{sqz}). The only free parameter left after removing the $k^{-6}$ and $k^{-4}$ divergences is $u$. In principle this can be chosen to minimize (but not remove) the amplitude of the $k^{-2}$ one-loop correction, but there is no physical reason to do this as the tree-level result is more dominant in the infrared. The choice that would minimize such terms corresponds to
\beq
{1\over k^2}:\ \ \ \ \ u=
\left\{ \begin{array}{rr}
 {20\over 21}\approx 0.95  &\mbox{{\rm (EQ)} } \\ & \\
  {50\over 63} \approx 0.79 &\mbox{ {\rm (ORT)}}
       \end{array} \right. 
       \label{minimize}
\eeq

For simplicity, {\em instead of making this choice},  we rather use $u=0$ which removes the operators proportional to $\nabla^{-2} \partial^{-1}$ and simplifies our initial conditions generation. We will explore the impact of different choices of $u$ and other possible templates on the four-point function elsewhere. Appendix~\ref{sqz} provides other motivations for the choice of $u$. 

\subsection{Implementation}
\label{impl}

Therefore the kernels we use are (assuming scale-invariance, their straightforward generalization beyond this is given in  Appendix~\ref{NSI})

 \beqa
 \Phi_{\rm EQ}&=&\phi + f_{\rm NL} \Big[ -3 \phi^2 +4\, \partial^{-1} (\phi\, \partial \phi)  \nonumber \\ & &
 + 2 \, \nabla^{-2} (\phi \ \nabla^2 \phi)
 + 2 \,  \nabla^{-2}(\partial \phi)^2
  \Big],
 \label{KEQ}
 \eeqa
for the equilateral model and
 \beqa
 \Phi_{\rm ORT}&=&\phi + f_{\rm NL} \Big[ -9 \phi^2 +10\, \partial^{-1} (\phi\, \partial \phi)  \nonumber \\ & &
 + 8 \, \nabla^{-2} (\phi \ \nabla^2 \phi) + 8 \,  \nabla^{-2}(\partial \phi)^2
  \Big],
 \label{KORT}
 \eeqa
for the orthogonal case. As in Eq.~(\ref{localfnl}), one should subtract the expectation value of the quantity in square brackets to impose $\langle \Phi \rangle=0$. As discussed in Appendix~\ref{sqz} this subtracts an infinite  constant that is more IR divergent than in the local case (which is also infinite). In any case, the zero mode does not enter into the physics that just depends on derivatives of $\Phi$. However, requiring regularity for the kernel that enters into this expectation value may motivate further constraints on the free parameters, see Appendix~\ref{sqz} for further discussion.

 \begin{figure}[!t]
 \centering
 \includegraphics[ width=\linewidth]{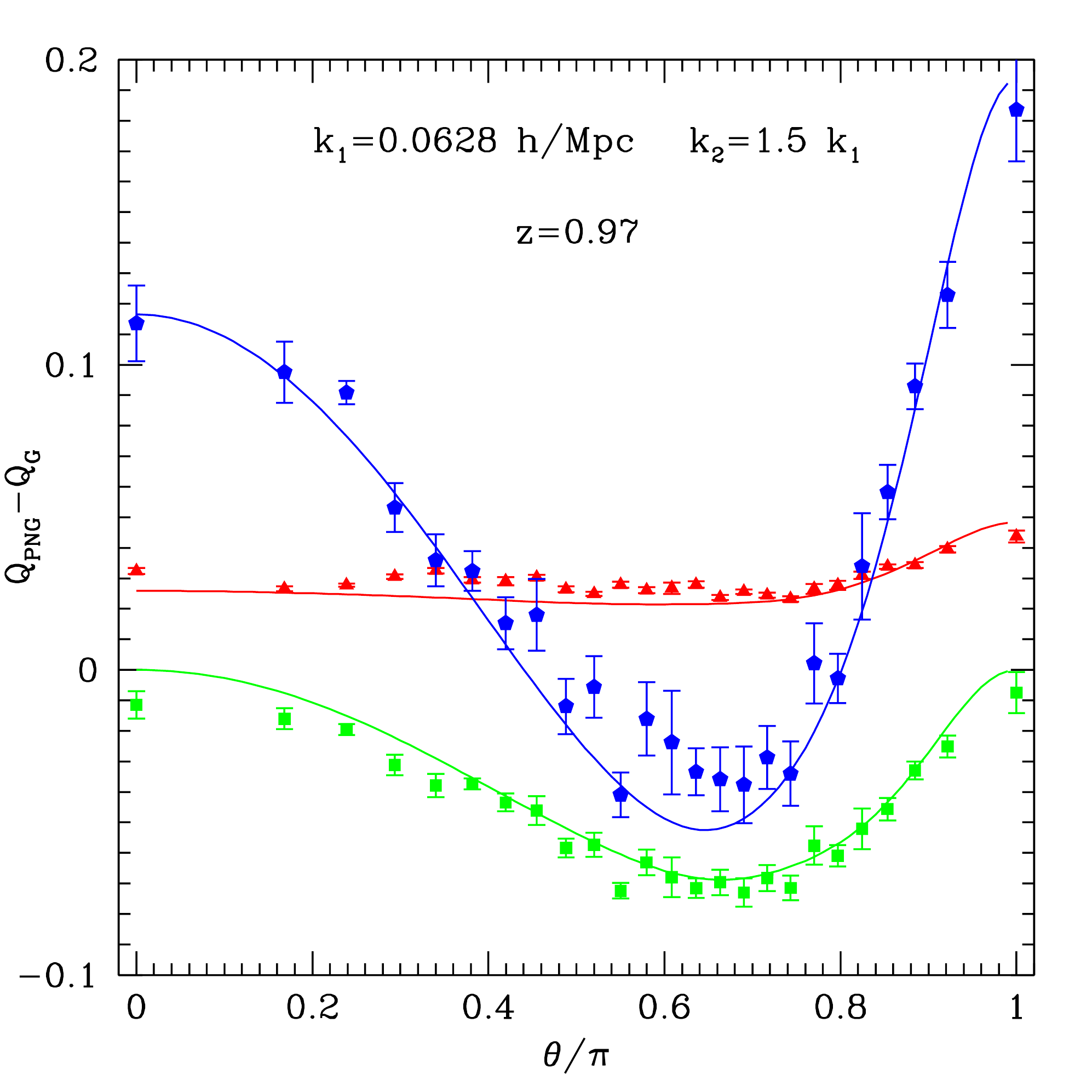}
 \caption{The difference in dark matter reduced bispectrum $Q$ from that in the Gaussian initial conditions case at redshift $z=0.97$ for triangles with sides $k_1=0.06 \kMpc$ and $k_2=1.5\, k_1$ as a function of angle $\theta$ between $\k_1$ and $\k_2$. The deviations seen in the N-body simulations agree with the expectations from linear perturbation theory evolution of the primordial bispectrum (solid) for all three models considered: orthogonal (blue pentagons, $\fnl=-400$), local (red triangles, $\fnl=100$) and equilateral (green squares, $\fnl=-400$), from top to bottom at $\theta=0,\pi$.}
 \label{dQics}
 \end{figure}

Given these kernels we generate initial conditions for numerical simulations. In Appendix~\ref{BispPhi} we present tests of our algorithm that show that we recover the correct bispectrum for the Bardeen potential $\Phi$ in the initial conditions for each of the operators appearing in Eqs.~(\ref{KEQ}-\ref{KORT}). Given the Bardeen potential, we generate initial conditions using second-order Lagrangian perturbation theory (2LPT), which provides a simple yet efficient way of minimizing transients from initial conditions~\cite{Sco98,*Sir0511,*2006MNRAS.373..369C,*2010MNRAS.403.1353C,*2010MNRAS.403.1859J,*2011arXiv1103.6031H} that arise from using linearized  Lagrangian perturbation theory (Zel'dovich approximation). This is important in particular when measuring higher-order statistics such as the bispectrum and the statistics of rare events, e.g. the mass function at high mass. For example, a 2LPT initial condition imposed at $z=49$ (our choice for the simulations we present here) is equivalent when measuring the mass function at $z=1,0$ to having run the same simulation with Zel'dovich initial conditions at an initial redshift of $z=2040, 4228$ respectively~\cite{LasDamasMF}. 

Our simulations are part of the LasDamas (Large Suite of Dark Matter Simulations) collaboration~\cite{LasDamasMF} runs, extended to PNG models. The cosmological parameters are $\Omega_m=0.25$, $\Omega_b=0.04$, $\Omega_\Lambda=0.75$, $h=0.7$, $n_s=1$ and $\sigma_8=0.8$. For this paper we present results based on 12 realizations of local (with $\fnl=100$), equilateral ($\fnl=-400$) and orthogonal ($\fnl=-400$) models run on a $2.4 \Gpc$  box with $1280^3$ particles, using the Gadget2 code~\cite{2005MNRAS.364.1105S}. For each PNG model we thus have a total volume of $166 (\Gpc)^3$, the largest to date, which will allow us to test theoretical predictions of large-scale bias to a greater accuracy than before. For such choices of $\fnl$ the skewness of the primordial density field is positive for the local and orthogonal case, while negative for the equilateral model. See~\cite{LasDamasS3S4} for a discussion of higher-order moments in these simulations and mock galaxy catalogs built from them.

In Figure~\ref{dQics} we show the difference in the matter density reduced bispectrum,
\beq
Q \equiv {B \over (P_1P_2+P_2P_3+P_3P_1)},
\label{Qdef}
\eeq
in each of the three models from the Gaussian initial conditions case at $z=0.97$ for triangles with sides $k_1=0.06 \kMpc$ and $k_2=1.5\, k_1$ as a function of angle $\theta$ between $\k_1$ and $\k_2$. The symbols (with error bars obtained from the scatter among 12 realizations) denote the measurements, while the solid lines correspond to the predictions of linear perturbation theory evolution of the primordial bispectrum for  each of the models (i.e. Eq.~\ref{Qdef} with $B$ and $P$ given by their primordial values scaled by the linear growth factor). We see a very good agreement, which is further evidence that the initial conditions in each case have been correctly generated. 

Our 2LPT-PNG initial conditions algorithm, being a sum of separable terms for the kernel, is very efficient. For non-local models the initial condition generation takes only 35\% longer than for local models, which for $N_{\rm par}=1280^3$ particles takes about 5 minutes in 320 cpus. This is several orders of magnitude faster than summation over modes methods that use non-separable kernels recently proposed in the literature~\cite{2011arXiv1102.3229W,2010JCAP...10..022W}, which scale as $N_{\rm par}^2$ (as opposed to $N_{\rm par} \ln N_{\rm par}$  coming from the use of FFTs in our case). Therefore, we can generate non-local initial conditions with minimal overhead over local models for {\em all} Fourier modes in the simulation box, without being forced to restrict the number of modes with PNG to a low-$k$ subset as in~\cite{2011arXiv1102.3229W,2010JCAP...10..022W}.

\section{The Peak-Background Split}

\subsection{Excursion-Set Basics}

We now turn to a derivation of the expected clustering in generic PNG models. For this purpose, it is useful to briefly review the peak-background split (PBS) argument that allows us to calculate the bias of collapsed objects~~\cite{1986ApJ...304...15B,1989MNRAS.237.1127C}. We will comment on how our approach differs from other accounts in the literature below, and also contrast the PBS predictions with local bias models. One of the main goals of this paper is to compute the relationship between halo and matter overdensities, i.e. the bias factors (see Eq.~\ref{deltahFS} below for a precise definition). 

In the excursion-set formalism~\cite{1991ApJ...379..440B}, halo formation can be described as a random walk of the smoothed linear density field $\delta$ as the smoothing radius goes from very large (infinitesimal variance $\sigma^2$, and thus tiny $\delta$) to crossing the linear threshold for collapse $\delta_c$ at some finite smoothing radius (which defines the mass-scale of the object). The random walk follows from the fact that changing the smoothing radius or variance (which can be thought of as the time variable in the random walk) one allows into $\delta$ smaller-scale Fourier modes which change $\delta$ stochastically. The first (smallest $\sigma^2$) crossing of the barrier described by $\delta_c$ is equated with halo formation.

Despite using a number of simplifying assumptions (random walks about generic points rather than peaks in the initial density field, simplified dynamics, etc; see e.g.~\cite{2011arXiv1105.1990P} for a recent discussion on this) the excursion set framework is very useful  as it provides a non-perturbative model (that includes e.g. exclusion effects at small scales~\cite{1999MNRAS.304..767S}) and at large scales is known to reproduce, in the Gaussian initial conditions case, the local bias perturbative expansion with linear and nonlinear bias parameters~\cite{MoJinWhi97,ScoSheHui01} which are in reasonably good agreement with numerical simulations~\cite{2007PhRvD..75f3512S,2010MNRAS.402..589M,2009arXiv0912.0446M}.

Within the excursion-set formulation of collapsed objects (dark matter halos) of mass $m$, their  number density per unit mass (obeying the standard normalization condition $\int (dn/dm)\, m \, dm=\bar{\rho}$) is given by

\beqa
\Big({dn\over dm}\Big) &=& \frac{\bar{\rho}}{m}\ \partial_m \int^{\delta_c}_{-\infty} \Pi_0(\delta_s,\sigma_m^2,\delta_c) d\delta_s,
\nonumber \\ &=& 
\frac{\bar{\rho}}{m} \Big| \frac{d \sigma_m^2}{dm} \Big| \times {\cal F}_0(\delta_c,\sigma_m^2)
\label{nLag}
\eeqa
where $\sigma_m^2$ is the variance of the small-scale density field smoothed with some filter at spatial scale $R$ (identified at first-crossing with the Lagrangian radius of the halo) with $m \equiv 4\pi \bar{\rho} a^3 R^3/3$. In Equation~(\ref{nLag})
\beq
{\cal F}_0(\delta_c,\sigma_m^2) \equiv -\frac{\partial}{\partial \sigma_m^2} \int^{\delta_c}_{-\infty} \Pi_0(\delta_s,\sigma_m^2,\delta_c) d\delta_s
\label{Fdef}
\eeq
is the probability of first crossing the linear-theory threshold for collapse $\delta_c$ between ``time" $\sigma_m^2$ and $\sigma_m^2 + d\sigma_m^2$ and $\Pi_0(\delta_s,\sigma_m^2,\delta_c)$ is the probability distribution for the small-scale density perturbation reaching $\delta_s$ by ``time" $\sigma_m^2$ given the initial condition that $\delta_s=0$ when $\sigma_m^2=0$, and it satisfies the absorbing barrier boundary condition $\Pi_0(\delta_c,\sigma_m^2,\delta_c)=0$. For Gaussian initial conditions and a top-hat filter in $k$-space $\Pi_0(\delta_s,\sigma_m^2,\delta_c)= P_G(\delta_s,\sigma_m^2)-P_G(2\delta_c-\delta_s,\sigma_m^2)$, where $P_G$ denotes the Gaussian PDF. The subscript in ${\Pi}_0$ denotes that $\Pi$ is computed for the initial conÁdition that $\delta_s=0$ when $\sigma_m^2=0$. 

Note that except to make contact with the literature, {\em we don't assume universality of the mass function}, which requires
\beq
\Pi_0(\delta_s,\sigma_m^2,\delta_c) = F({\delta_s\over\sigma_m},{\delta_c\over\sigma_m}), 
\label{Univ}
\eeq
in which case  $\sigma^2_m {\cal F}_0$ is a function of $\nu=\delta_c/\sigma_m$ alone, in fact 
\beq
\sigma^2_m \, {\cal F}_0(\delta_c,\sigma_m^2)= {\nu f(\nu) \over 2},
\label{nufnu}
\eeq
where $f(\nu)$ is the usual Gaussian factor in Press-Schechter theory for Gaussian initial conditions.

\subsection{Halo Bias and non-Markovian effects}

We are interested in calculating the clustering of objects (halos), which can be derived from their {\em conditional mass function} that describes how the abundance of halos responds to a large-scale perturbation $\delta_\ell$, i.e. conditional on the initial condition that at ``time" $\sigma_\ell^2$ the density field was $\delta_\ell$. The Lagrangian number density perturbation of halos is given by

\beq
(1+\delta_h^L)
= \frac{\partial_m \int_{-\infty}^{\delta_c} \Pi(\delta_s,\sigma_m^2,\delta_c;\delta_\ell,\sigma^2_\ell) d\delta_s}{ \partial_m \int_{-\infty}^{\delta_c} \Pi_0(\delta_s,\sigma_m^2,\delta_c) d\delta_s} 
\label{deltahL}
\eeq

where $\Pi(\delta_s,\sigma_m^2,\delta_c;\delta_\ell,\sigma^2_\ell)$ is the conditional probability for $\delta_s$ with absorbing boundary conditions at $\delta_c$ at time $\sigma_m^2$ {\em given} that  $\delta_s=\delta_\ell$ at time $\sigma_\ell^2$ when the smoothing scale was much larger (note that  $\sigma^2_\ell < \sigma_m^2$). As we mentioned above, it is important to stress that we use a zero subscript for {\em unconditional} quantities and we write 
\beq
\Pi_0(\delta_s,\delta_c,\sigma_m^2) \equiv \Pi(\delta_s,\delta_c,\sigma_m^2;0,0). 
\eeq
Since we are interested in the large-scale limit of the bias, we work in the regime where $\sigma^2_\ell \ll \sigma_m^2$ (scales much larger than the Lagrangian size of halos) and look for the relation between $\delta_h$ and $\delta_\ell$. For reference, the Lagrangian size of halos of mass $m=10^{14}, 10^{15} M_\odot/h$ is $R_L=7,15 \Mpc$. In this case the calculation simplifies significantly because we can approximate,

\beqa
\Pi(\delta_s,\sigma_m^2,\delta_c;\delta_\ell,\sigma^2_\ell) &\approx &\Pi(\delta_s,\sigma_m^2,\delta_c;\delta_\ell,0).
\label{Piapprox1}
\eeqa
Note that it is often used, or implicitly assumed, that one can write this as 
\beqa
\Pi(\delta_s,\sigma_m^2,\delta_c;\delta_\ell,0) &\approx &\Pi_0(\delta_s-\delta_\ell,\sigma_m^2,\delta_c-\delta_\ell),
\nonumber \\
\label{Piapprox2}
\eeqa
where the last approximation {\em only holds if non-Markovian effects can be neglected}, i.e. the statistics of the random walk depend only on the total amount walked ($\delta_s-\delta_\ell$) and to be walked ($\delta_c-\delta_s$) at a given time so one can use a shifted barrier ($\delta_c \rightarrow \delta_c-\delta_\ell$) from the unperturbed initial condition ($\delta_\ell=0$). This is an important simplification, because it means the statistics of collapsed objects can be obtained for small large-scale $\delta_\ell$ by Taylor expanding the {\em unconditional} mass function (which is determined by $\Pi_0$), and therefore the halo bias can be obtained from derivatives of the (unconditional) mass function. 

While Eq.~(\ref{Piapprox2}) is routinely assumed and the bias computed from derivatives of the unconditional mass function, there may be  corrections to the large-scale bias that will result from this. For peaks in a random Gaussian field, computing the bias from derivatives of the unconditional mass function does not hold in general~\cite{DesCroSco1011}, although it holds asymptotically at large-scales. In~\cite{2010arXiv1007.4201M} it is found that violations of Eq.~(\ref{Piapprox2}) for halos can alter the large-scale bias, but~\cite{2011arXiv1105.2261P} find that the shift $\delta_c-\delta_\ell$ in Eq.~(\ref{Piapprox2}) should be replaced in the non-Markovian case by $\delta_c-\alpha\, \delta_\ell$ where $\alpha$ depends on the filter and the power spectrum ($\alpha=1$ for a sharp-$k$ filter). This change, however, does not affect the {\em Fourier space} large-scale bias  (as opposed to the counts-in-cell bias), which is predicted rather accurately by the unmodified Eq.~(\ref{Piapprox2}) when compared to numerical random walks. In practice, though, there should be  non-Markovian corrections other than filter effects (e.g. PNG itself, or physics that is not included).  

In simulations with Gaussian initial conditions,~\cite{2010MNRAS.402..589M} tested how well the bias computed from the unconditional mass function agrees with the large-scale bias measured from simulations in real and Fourier space. In this case, however, the assumption of Markovianity (which converts the $\delta_\ell$ to a $\delta_c$ derivative) is not enough to do the numerical evaluation of the derivative and one needs to further assume universality of the mass function (to convert the $\delta_c$ to a $\nu=\delta_c/\sigma_m$ derivative). Proceeding in this way~\cite{2010MNRAS.402..589M}  found deviations at the $5-10\%$, particularly at high-mass. 

 As we shall see, fortunately, we will be able to proceed quite far for $\fnl$ PNG without the need of a full description of non-Markovian effects, although the same is not true for $\gnl$ PNG. For further discussion of non-Markovian effects see~\cite{1991ApJ...379..440B,1990MNRAS.243..133P,2010ApJ...711..907M,2010ApJ...717..515M,2010arXiv1007.4201M,2011arXiv1105.2261P,2011arXiv1105.1990P}.

\subsection{Modulation Bias from PNG in the PBS}
\label{modbias}

In the presence of primordial non-Gaussianity (PNG), different Fourier modes are no longer independent, and the probabilities $\Pi$ for the small scale density field fluctuations get modified by the presence of a long-wavelength mode of the Gaussian field $\phi$ as follows
 
\beq
 \Pi(\delta_s,\sigma_m^2,\delta_c;\delta_\ell,0) \rightarrow \Pi[\delta_s,\sigma^2(\phi),c_p(\phi),\delta_c;\delta_\ell(\phi),0],
\label{PiNG}
\eeq
where $c_p\equiv \langle \delta_s^p \rangle_c$ are the cumulants of the small-scale modes\footnote{Here we assume that $\Pi$ depends on the cumulants of $\delta_s$, but~\cite{2010ApJ...717..526M} find sub-leading corrections that depend on derivatives of correlators, which we ignore for simplicity.}. That is, unlike the case of Gaussian initial conditions where only the  mean becomes $\phi$-dependent, all higher-order cumulants now depend on the long-wavelength perturbation because small-scale modes are coupled to them through PNG. 

It is precisely this $\phi$-dependence that we derive in this subsection. For readers interested in skipping the details and going directly to section~\ref{linb} for the results on scale-dependent linear bias to leading order in $\fnl$, the main results needed are Eq.~(\ref{RelBisp}), and  Eqs.~(\ref{deltaHL2}-\ref{PiFirst}).

To proceed, although we are primarily interested in $\fnl$ PNG, we will  generalize the model to account for a cubic $\gnl$ kernel,
\beq
\Phi = \phi + \fnl \, K[\phi,\phi] + \gnl \, L[\phi,\phi,\phi]
\label{gnlPhi}
\eeq
To avoid very long expressions, we will suppress the contributions of the $\gnl$ kernel in most equations, except when the extension is a non-trivial generalization of the $\fnl$ kernel (see Eqs.~\ref{sigma-phi} and further). We will also  take into account all orders in PNG parameters ($\fnl^a \gnl^b$) to arrive at very general expressions (see Eq.~\ref{AllDers} and below).

If one is interested in first-order in $\fnl$ calculations,  computing the modulation of the small-scale variance $\sigma(\phi)$ by $\phi$ due to non-Gaussianity (linear in $\fnl$) is all that is needed, while we can evaluate the third-order cumulant $c_3$ (related to the skewness $s_3$ by $c_3=s_3\, \sigma^3$) ignoring the modulation by the large-scale mode $\phi$, since modulation of $c_3$ by the large-scale $\phi$ starts at the four-point function level, and thus is at least ${\cal O}(\fnl^2)$ or ${\cal O}(\gnl)$ for PNG models with cubic kernels. 
Similarly, we can put $c_4$ and higher-order cumulants to zero. However, for now we will keep effects up to ${\cal O}(\fnl^2)$ or ${\cal O}(\gnl)$ to see how our formulas apply to more general combination of quadratic and cubic PNG.

To calculate  the modulation effects, we perform the peak-background split (PBS) between small-scale (``peak", denoted by $s$ subscripts) and long-wavelength (``background", denoted by $\ell$ subscripts) perturbations. This split takes into account that the long-wavelength modes have been conditioned on to take values $\phi_\ell(\k)$ (that is, we only look at random walks that satisfy this large-scale constraint). The split is often done in real space, by choosing a top-hat filter at some  scale $R$ (larger than the Lagrangian radius of the halo $R_L$). This is the appropriate constraint when dealing with counts-in-cells statistics (at radius $R$), but here we are  interested in the power spectrum where all modes are weighted equally. Therefore, we need to impose a more natural constraint for Fourier-space statistics, i.e. a sharp-$k$ filter.   

Another choice we must make is on what variable to do the split. In the excursion set framework, the random walk is performed by the full, {\em non-Gaussian}, density field, which is equivalent to constraining the non-Gaussian $\Phi_\ell(\k)$ rather than $\phi_\ell(\k)$.  However, technically it is easier  to do the split in $\phi$ since for a Gaussian field long and short modes are independent, and this is presumably the reason why all past work on the subject has done so.   However, arguably the more physical variable is $\Phi$, after all this is what is supplied to the equations of motion of gravitational evolution that describe the physics of halo clustering. We shall see that there are  two instances where constraining $\phi_\ell$ differs from $\Phi_\ell$, but for simplicity we will do the split on $\phi$ where general results are easier to derive to all orders in PNG and discuss where appropriate how the results change when constraints on $\Phi$ are used instead.

We thus proceed with a split  on $\phi$ in Fourier space at scale $k_{\rm split}$, 
\beqa
\phi^{(s)}(\q) &=& \phi(\q)\, \Theta(q-k_{\rm split}),  \nonumber \\ 
\phi^{(\ell)}(\q) &=& \phi(\q)\, \Theta(k_{\rm split}-q),
\label{SharpPhi}
\eeqa
where $\Theta$ is a  step function, i.e. a sharp-$k$ filter. The splitting scale obviously must satisfy $k_{\rm split}R_L\la1$, where $R_L$ is the Lagrangian radius of the halo. {\em When computing the power spectrum at scale $k$, we will choose} $k_{\rm split}=k$, analogous to choosing the split scale to be $R$ when computing counts-in-cells statistics at scale $R$. Note that the precise choice of the splitting scale does not appear in the calculation until one computes one-loop corrections, e.g. one could have chosen e.g. $k_{\rm split}=2k$ and get the same tree-level results. Our choice means that when computing the power spectrum at scale $k$ we have integrated out all smaller-scale fluctuations, and loops contribute a small renormalization of the tree-level PBS bias parameters that results in their running up to scale $k$ (see Section~\ref{PBSvsLOC}). 
The choice of a sharp-$k$ filter rather than a top-hat split has the added benefit of avoiding $W_{\rm TH}(kR)$ factors that would otherwise appear in the power spectrum (with $R$ poorly determined) when computing loops. Therefore, in our treatment the variance of the large-scale modes is 
\beq
\sigma_\ell^2(k) \equiv \int^k_0 P(q) \, 4\pi q^2 dq.
\label{sigmaell}
\eeq

The small-scale field will be smoothed with a top-hat filter at the Lagrangian scale of the halo, as this is the usual way of identifying collapsed objects, and will be introduced shortly. We are thus imagining random walks where the filter changes from sharp-$k$ to top-hat as the Fourier modes added go from large to small-scale (the transition from one to the other filter should be done smoothly at $\simeq k$, but the details won't matter as long as $kR_L \ll 1$, which as we shall discuss in Section~\ref{PBSvsLOC} is in any case the regime of validity of the perturbation theory for bias. 
We write for the small-scale density perturbations in Fourier space,

\beq
\delta_s(\k)\equiv M(k) \Phi_s(\k) = M(k) \Big[ \phi_s(\k) +  f_{\rm NL}\ K^{(s)}_{\bf k}[\phi,\phi] \Big]
\label{ds}
\eeq
where the function $M(k)$ relates the density fluctuations to the Bardeen potential $\Phi$
\beq
M(k) \equiv {2\over 3} {D(z) T(k) \over \Omega_m^0 H_0^2}\, k^2 ,
\label{Mdef}
\eeq
where $\Omega_m^0$ and $H_0$ denote the $z=0$ values of matter density in units of critical and the Hubble constant, $D(z)$ is the growth factor in the matter era, with normalization $D=a$ at early times, where $a$ is the scale factor. $T(k)$ denotes the transfer function, and the gravitational potential obeying the Poisson equation at sub-horizon scales at redshift $z$ reads $\Phi_{\rm grav} = -(D/a)\, T(k)\, \Phi$. 

In order to properly perform the PBS at quadratic (and higher-order) we distinguish between the long and short wavelength limits of the kernel (which in general can be very different, see discussion in section~\ref{GenBisp} above, and Appendix~\ref{sqz} for examples). That is, in Eq.~(\ref{ds}) the upper index in $K$ denotes that the kernel is non-zero only when evaluated for large external momentum $k$, i.e. specifically 
\beqa
K^{(s)}_{\bf k}[\phi,\phi] &=& \int \delta_D(\k-\k_1-\k_2)\ K^{(s)}(\k_1,\k_2)\nonumber \\ &  \times & [\phi_s(\k_1) \phi_s(\k_2)+ 2 \phi_s(\k_1) \phi_\ell(\k_2)],
\nonumber \\ & & d^3k_1\, d^3 k_2
\label{Ks}
\eeqa
where we have used symmetry in $\k_1$ and $\k_2$ to write down two equivalent contributions from terms linear in $\phi_\ell$, and  $K^{(s)}$ is a high-pass filtered version of $K$ to ensure that the only small-scale $\phi_s(\k_1)$ and $\phi_s(\k_2)$ contributions included in the integral couple to a high-$k$ mode ($k>k_{\rm split}$), the missing contributions below the splitting scale are included in the  {\em large-scale} density perturbation for which we have,
\beq
\delta_\ell(\k) = M(k) \Big[ \phi_\ell(\k) +  f_{\rm NL}\ K^{(\ell)}_{\bf k}[\phi,\phi] \Big]
\label{dl}
\eeq
where the upper index in $K$ denotes that the kernel is non-zero only when evaluated for small external momentum $k$, 
\beqa
K^{(\ell)}_{\bf k}[\phi,\phi] &=& \int \delta_D(\k-\k_1-\k_2)\ K^{(\ell)}(\k_1,\k_2)\nonumber \\ &  \times & [\phi_s(\k_1) \phi_s(\k_2)+ \phi_\ell(\k_1) \phi_\ell(\k_2)],
\nonumber \\ & & d^3k_1\, d^3 k_2
\label{Kell}
\eeqa
where $K^{(\ell)}$ is a low-pass filtered version of $K$ to ensure that the only small-scale modes $\phi_s(\k_1)$ and $\phi_s(\k_2)$ that contribute in Eq.~(\ref{Kell}) are nearly opposite (and thus couple to a large scale mode $k$ below the splitting scale). By definition we have $K=K^{(s)}+K^{(\ell)}$. We can make all this explicit by writing (see Eq.~\ref{SharpPhi}),
\beqa
K^{(s)}(\k_1,\k_2) &=& K(\k_1,\k_2)\, \Theta(k_{12}-k_{\rm split}),  \nonumber \\ 
K^{(\ell)}(\k_1,\k_2) &=& K(\k_1,\k_2)\, \Theta(k_{\rm split}-k_{12}).
\label{Sharp}
\eeqa
Note our split at quadratic order automatically incorporates all contributions to large and small scale perturbations, ie. we {\em do not assume} e.g. for  local PNG that 
\beqa
\nabla^2 \Phi_s &\approx&  \nabla^2 \phi_s +2\fnl (\phi_\ell  \nabla^2 \phi_s + \nabla\phi_s \cdot \nabla\phi_s)\nonumber \\
&  \approx  & \nabla^2 \phi_s + 2\fnl \, \phi_\ell  \nabla^2 \phi_s
\label{bogus}
\eeqa
the latter often ``justified" by saying that $ \nabla\phi_s\approx 0$ at peaks. However, since we are dealing with peaks in the density then $ \nabla\phi_s$  is proportional to the velocity field, which does not vanish. Eq.~(\ref{bogus}) is an approximation at the level of $M$ (which involves the Laplacian), and there are similar approximations often made about the filter function that defines the mass scale of objects. Our expressions automatically keep all such terms, we shall see that they  contribute to scale-independent bias for local PNG~(see also Appendix~\ref{IBlocApp} for more details).

 Let us now comment on using $\phi$ rather than $\Phi$ (or $\delta$) as the split variable. One way in which differences can arise is when the kernel $K^{(\ell)}$ is singular enough in the squeezed limit that there are non-negligible contributions  to the large-scale modes $\Phi_\ell$ (or $\delta_\ell$) in Eq.~(\ref{dl}) coming from $\phi_s$  through the first term in Eq.~(\ref{Kell}). When this happens, constraining on $\phi_\ell$ and $\Phi_\ell$ can differ, as $\Phi_\ell$ cannot be well approximated as a function of just $\phi_\ell$, that is, fixing $\phi_\ell$ is not fixing the large-scale density field because the small-scale modes $\phi_s$ make a non-negligible contribution to it. We will find an example for the equilateral template where this can lead to different conclusions about the scale-dependent bias.

From Eq.~(\ref{ds}) we can now calculate all the needed modulations of the cumulants of the small-scale fluctuations, which will now be space-dependent due to the background of large-scale modes that breaks statistical translation invariance. For the variance after introducing smoothing on mass-scale $m$, keeping up to ${\cal O}(\fnl^2)$ and ${\cal O}(\gnl)$, 

\begin{widetext}

\beqa
\sigma^2(\phi_\ell) &=& \sigma_m^2 +  \int P_\phi(p)\, d^3p\ \Big[ 4 \fnl M_m(p)\, \varphi_p(\x) %\nonumber \\ & & 
+ 4 \,\fnl^2\ |\varphi_p(\x)|^2 + 6\,\gnl\, M_m(p)\, \varphi_p^{(2)}(\x)
\Big]  
 \equiv  \sigma_m^2 + \delta\sigma_m^2(\phi_\ell), \nonumber \\ & &
\label{sigma-phi}
\eeqa
where we neglected a $-\sigma^2_\ell$ contribution since we assume $\sigma^2_m\gg \sigma_\ell^2$, but this can be taken into account if the observation scale $k$ is not too far from the Lagrangian radius of the halo $R_L$. Although it is often argued that in doing the PBS the split scale must be much larger than $R_L$, this is not the case. If $kR_L$ is not much less than unity what happens is that the modes that have been conditioned on contribute to the halo as well, but that's not a problem as long as these conditioned ``long" modes are later averaged properly when computing halo clustering at $k$. What is difficult when $kR_L\rightarrow 1$ is to do this averaging {\em perturbatively}, see Section~\ref{PBSvsLOC}. 

In Eq.~(\ref{sigma-phi}) the first and second-order $\varphi_p$ fields are given by
\beq
\varphi_p(\x) \equiv \int M_m(\p-\q) K^{(s)}(-\p,\q) \phi_\ell(\q) {\rm e}^{-i{\bf q}\cdot{\bf x}} d^3q,
\label{varphi}
\eeq
%and
\beqa
\varphi_p^{(2)}(\x) & \equiv& \int M_m(\p-\q_{12})\, L^{(s)}(-\p,\q_1,\q_2) % \nonumber \\ & &
\times  \phi_\ell(\q_1) \phi_\ell(\q_2) {\rm e}^{-i{\bf q}_{12}\cdot{\bf x}} d^3q_1d^3q_2,
\label{chisq}
\eeqa
with $L$ being the cubic kernel describing $\gnl$ non-Gaussianity (see Eq.~\ref{gnlPhi}). In Eq.~(\ref{sigma-phi}),
\beqa
\sigma_m^2&=&\int M_m(k)^2 P_\phi(k) d^3k +2 \fnl^2  \int  M_m(k)^2 % \nonumber \\ & & 
[K^{(s)}(\k-\q,\q)]^2  P_\phi(|\k-\q|)  P_\phi(q) d^3k d^3q % \nonumber \\ & & 
\label{sigmam}
\eeqa
is the variance of the small-scale density fluctuations including non-Gaussian corrections and we have included filtering into a redefinition of $M$,
\beq
M_m(k)\equiv M(k) \, W_{\rm TH}(kR_L),
\eeq
where $R_L$ is the Lagrangian radius of halos of mass $m$ and $W_{\rm TH}$ is the Fourier transform of a top-hat window. Similarly, we can write for the third cumulant,

\beqa
c^{(3)}(\phi_\ell)= c^{(3)}_m &+& \int  P_\phi(k_1) d^3k_1\,  P_\phi(k_2) d^3k_2 \, \Big[ 
 24\fnl^2 \, M_m(k_1) \, M_m(|\k_1+\k_2|)\, K^{(s)}(\k_1,\k_2) \, \varphi_{k_2}(\x) \nonumber \\ & + &
  18\gnl \, M_m(k_1)\,  M_m(k_2) \, \varphi_{k_1k_2}(\x)
\Big] 
\equiv c_m^{(3)} + \delta c_m^{(3)}(\phi_\ell)
\label{c3-phi}
\eeqa
where $c^{(3)}_m$ is the third-order cumulant calculated to the desired order in PNG (as in Eq.~\ref{sigmam} for $\sigma^2_m$), and
\beq
\varphi_{k_1k_2}(\x) \equiv \int L^{(s)}(\q,\k_1,\k_2) M_m(\k_{12}+\q) \phi_\ell(\q) {\rm e}^{-i{\bf q}\cdot{\bf x}} d^3q,
\label{vartheta}
\eeq
is  analogous to $\varphi_k$ but for the $\gnl$ kernel. All these $\varphi$-fields  can be written in terms of functional derivatives of small-scale density modes with respect to $\phi_s$ and admit a simple diagrammatic representations which help derive these expressions.
%\end{widetext}

What we need to compute the bias factors are the functional derivatives of these modulated cumulants with respect to large-scale fields in Fourier space, where we want the bias (as opposed to real space). For the variance we have,
\beq
\Big( {{\cal D} \sigma^2 \over {\cal D} \phi_\ell(\k)} \Big)_0 = {\rm e}^{-i\k \cdot \x} 4\fnl 
\int M_m(p) M_m(|\p-\k|) K^{(s)}(-\p,\k)P_\phi(p) d^3p =  \frac{{\rm e}^{-i\k \cdot \x} }{P_\phi(k)} \int 
B_{\widehat{\delta}\widehat{\delta}\phi}(\q,\k-\q,-\k) \, d^3q  
\label{LinRes}
\eeq
where $(\ldots)_0$ means evaluating at $\phi_\ell=0$ and  we introduced the cross-bispectrum between small-scale {\em smoothed} density perturbations $\widehat\delta(\k)\equiv M_m(k)\Phi(\k)$ and $\phi$. This result should not be surprising, as it is  expected from linear response, i.e. 
\beq
\Big( {{\cal D} \sigma^2 \over {\cal D} \phi_\ell(\k)} \Big)_0 = { \langle \sigma^2(\phi_\ell) \phi_\ell^*(\k) \rangle \over P_\phi(k)},  \ \ \ \ \  \ \ \ \ \ 
 \langle \sigma^2(\phi_\ell) \phi_\ell^*(\k) \rangle =   e^{-i\ks \cdot \xs}  \int 
B_{\widehat{\delta}\widehat{\delta}\phi}(\q,\k-\q,-\k) \, d^3q 
\label{RelBisp}
\eeq
Note that because $\phi(\k)$ is Gaussian, there is no contribution to the cross bispectrum that depends on $K^{(\ell)}$, only $K^{(s)}$ appears, and the two contributions can be written identically after a change of variables, as in Eq.~(\ref{LinRes}). This can be traced back to the fact that we conditioned on the $\phi_\ell$ modes rather than the large-scale non-Gaussian $\Phi_\ell$ (or $\delta_\ell$). 
This in contrast  with  local bias models, which predicts primordial non-Gaussianity enters only through the small-scale density bispectrum. We'll come back to this later. 

To make our expressions more compact in the generic PNG case (i.e. beyond quadratic $\fnl$ kernels) we will also use the second cumulant notation for the variance when convenient and slightly abuse notation for $p=1$ to denote $\delta_\ell$, i.e.
\beq
c^{(1)}\equiv \delta_\ell,  \ \ \ \ \ \ \ \ \ \  c^{(2)}\equiv \sigma^2(\phi_\ell), \ \ \ \ \ \ \ \ \ \  c^{(p)}\equiv \lexp\, [\delta_s(\phi_\ell)]^p \rexpc,
 \ \ \ \ \ \ \ \ \ \  c^{(2)}_m\equiv \sigma^2_m, \ \ \ \ \ \ \ \ \ \  c^{(p)}_m\equiv \lexp \delta_s^p \rexpc,
\label{redefCp} 
\eeq
thus one should keep in mind that $p=1$ is special (in the sense that derivatives with respect to it will be related to the conditional mass function in the expressions below, as opposed to unconditional).

To compute linear bias to leading order in $\fnl$, Eq.~(\ref{LinRes}) is all that is needed. To go beyond this, we must compute higher-order derivatives of the variance or higher-order cumulants. Rather than do case by case, we quote the most general result ($p,q\geq 1$),
\beqa
\Big( {{\cal D}^q c^{(p)} \over {\cal D} \phi_\ell(\k_1) \ldots {\cal D} \phi_\ell(\k_q)}  \Big)_0 & = &  
{ \langle c^{(p)}(\phi_\ell)\  \phi_\ell^*(\k_1) \ldots \phi_\ell^*(\k_q) \rangle_c \over P_\phi(k_1) \ldots P_\phi(k_q)} \nonumber \\ &=& 
 \frac{{\rm e}^{-i\, \k_{1; q} \cdot \x}  \int 
T^{(p+q)}_{\widehat{\delta}\ldots \widehat{\delta}\, \phi \ldots \phi}(\q_1,\ldots,\q_{p-1},\k_{1; q}-\q_{1; p-1},-\k_1,\ldots,-\k_q) \, d^3q_1 \ldots d^3q_{p-1} }{P_\phi(k_1) \ldots P_\phi(k_q)},
\nonumber \\ & \equiv & {\rm e}^{-i\, \k_{1; q} \cdot \x}\ I_{pq}(\k_1,\ldots,\k_q,m)
\label{AllDers}
\eeqa
where $\k_{1;q} \equiv \sum_{i=1}^{q} \k_i$, and 
 $T^{(n)}$ is the {\em primordial} $n^{\rm th}$ polyspectra, i.e. $T^{(3)}$ is the bispectrum, $T^{(4)}$ is the trispectrum. Note the subindices of $T^{(p+q)}$ denote that this expression involves the polyspectrum between $p$ $\widehat\delta$-fields (at large momenta) and $q$ $\phi_\ell$-fields (at low-$k$).  
For $p=1$ there is no smoothing, thus $\widehat\delta$ is replaced by $\delta$, according to Eq.~(\ref{redefCp}). Note that in deriving this result, it is crucial that $\phi_\ell$ is a Gaussian field, we will come back to this in Section~\ref{ConstPhi}.

For example, to calculate linear bias at ${\cal O}(\gnl)$ one needs the modulation of the third-cumulant by one large-scale field, i.e. $p=3$ and $q=1$ in the above formula, which involves the primordial trispectrum. One can compute directly the derivative from Eq.~(\ref{c3-phi}) and check that the appropriate (with all combinatoric factors) trispectrum given by Eq.~(\ref{AllDers}) is obtained. Similarly, to compute quadratic bias at ${\cal O}(\fnl^2)$ or ${\cal O}(\gnl)$, one needs the modulation of the variance by two large-scale fields ($p=q=2$ in the above formula), and this involves a different trispectrum, which can be checked directly by taking derivatives from Eq.~(\ref{sigma-phi}). Lastly, we note that for $p=1$, Eq.~(\ref{AllDers}) is proportional to  the  PNG kernels at order $q$, 
\beq
I_{11}(k,m)=M(k), \ \ \ \ \ \ \ \ \ \ I_{12}=2 \,M(k_{12})\, K^{(\ell)}(\k_1,\k_2), \ \ \ \ \ \ \ \ \ \  I_{1q}= q!\ M(k_{1;q})\ K^{(\ell)}(\k_1,\ldots ,\k_q)
\label{KfromI}
\eeq

We are now ready to calculate the bias for arbitrary PNG. We introduce shorthand notation to distinguish between real-space (where the excursion-set expressions were written) and Fourier space (where  we want to calculate the bias), otherwise confusion can arise. Let
\beq
A \equiv A(\x), \ \ \ \ \ \ \ \ \ \  \tilde{A} \equiv A(\k),
\label{notation}
\eeq
that is, unless the arguments make it clear, no argument means real space, no argument with a tilde denotes Fourier space. 

We can write the Lagrangian halo overdensity as an expansion over the large-scale $\phi$ modes,
\beqa
\delta_h^L & = & \int d^3k \frac{ \partial_m \int_{-\infty}^{\delta_c} d\delta_s\, ({\cal D} \Pi / {\cal D} \tf_\ell)_0} {
\partial_m \int_{-\infty}^{\delta_c}d\delta_s \, \Pi_0(\delta_s,\sigma^2,\delta_c) } \, \phi_\ell(\k) %\nonumber \\ &+ & 
+ {1\over 2} \int d^3k_1 d^3k_2 \frac{ \partial_m \int_{-\infty}^{\delta_c} d\delta_s\, ({\cal D}^2 \Pi / {\cal D} \tf_\ell {\cal D} \tf_\ell)_0} {
\partial_m \int_{-\infty}^{\delta_c}d\delta_s \, \Pi_0(\delta_s,\sigma^2,\delta_c) } \, \phi_\ell(\k_1) \,  \phi_\ell(\k_2) %\nonumber \\ & + & 
+ \ldots \nonumber \\ & & 
\label{deltaHL2}
\eeqa
where $(\ldots)_0$ means evaluating at $\phi_\ell=0$. Note that we expand only in the large-scale Gaussian field $\phi_\ell(\k)$. We shall see soon that the perturbation expansion reorganizes itself into local terms in $\delta_\ell(\x)$ (which {\em contains} PNG) and non-local terms in $\phi_\ell(\x)$. The former are the same terms present for Gaussian initial conditions that lend itself to a local (Eulerian or Lagrangian) bias description, while the latter become local in $\phi_\ell(\x)$ only for local PNG. Therefore,  {\em we do not assume} the nature of bias (local or nonlocal, and in which fields), it all follows from applying the PBS to the particular PNG model under consideration, which in our case they  are all described in terms of a  Gaussian field $\phi$ and its interactions (PNG kernels).

The first derivative required in Eq.~(\ref{deltaHL2}) is, to all orders in PNG
\beqa
\Big( {{\cal D} \Pi \over {\cal D} \phi_\ell(\k) } \Big)_0 &= &  \sum_{p=1}^{\infty}   \Big({\partial \Pi \over \partial c^{(p)}}\Big)_0  \Big( {{\cal D} c^{(p)} \over {\cal D} \phi_\ell(\k)} \Big)_0 =
\Big( {\partial \Pi \over \partial \delta_\ell} \Big)_0  \Big( {{\cal D} \delta_\ell \over {\cal D} \phi_\ell(\k)} \Big)_0+
% {\partial \Pi_0 \over \partial \sigma^2_m}\,   \Big( {{\cal D} \sigma^2_s \over {\cal D} \phi_\ell(\k)} \Big)_0 %\nonumber \\ & + &
 \sum_{p=2}^{\infty}  {\partial \Pi_0 \over \partial c_m^{(p)}}\,  \Big( {{\cal D} c^{(p)} \over {\cal D} \phi_\ell(\k)} \Big)_0  
\label{PiFirst}
\eeqa
where to first order in $\fnl$, as discussed above, only the first two terms contribute ($p=1,2$) while $p=3$ contributes to ${\cal O}(\fnl^2)$ and  ${\cal O}(\gnl)$, see Eqs.~(\ref{sigma-phi}) and~(\ref{c3-phi}). The $p=1$ contribution is the usual PBS bias present in the Gaussian initial conditions case, while the rest are new contributions due to PNG. Similarly, for the second derivative we have, to all orders in PNG
\beqa
\Big( {{\cal D}^2 \Pi \over {\cal D} \phi_\ell(\k_1)  {\cal D}\phi_\ell(\k_2) } \Big)_0 &=& 
 \sum_{p=1}^{\infty}   \Big({\partial \Pi \over \partial c^{(p)}}\Big)_0  \Big( {{\cal D}^2 c^{(p)} \over {\cal D} \phi_\ell(\k_1) {\cal D} \phi_\ell(\k_2)} \Big)_0 +
  \sum_{p,q}   \Big({\partial^2 \Pi \over \partial c^{(p)} \partial c^{(q)} }\Big)_0  
    \Big( {{\cal D} c^{(p)} \over {\cal D} \phi_\ell(\k_1)} \Big)_0  \Big( {{\cal D} c^{(q)} \over {\cal D} \phi_\ell(\k_2)} \Big)_0
\nonumber 
\\ && \\ &=&
\Big( {\partial^2 \Pi \over \partial \delta_\ell^2} \Big)_0 \  \Big( {{\cal D} \delta_\ell \over {\cal D} \phi_\ell(\k_1)} \Big)_0
  \Big( {{\cal D} \delta_\ell \over {\cal D} \phi_\ell(\k_2)}  \Big)_0 
  +\Big( {\partial \Pi \over \partial \delta_\ell} \Big)_0 \  \Big( {{\cal D}^2 \delta_\ell \over {\cal D} \phi_\ell(\k_1){\cal D} \phi_\ell(\k_2)} \Big)_0  \nonumber \\ &&
+ \ \Big( {\partial^2 \Pi \over \partial \delta_\ell \partial \sigma^2_m} \Big)_0 \ \Big[ 
\Big( {{\cal D} \delta_\ell \over {\cal D} \phi_\ell(\k_1)} \Big)_0
  \Big( {{\cal D} \sigma^2 \over {\cal D} \phi_\ell(\k_2)}  \Big)_0 + \Big( {{\cal D} \delta_\ell \over {\cal D} \phi_\ell(\k_2)} \Big)_0
  \Big( {{\cal D} \sigma^2 \over {\cal D} \phi_\ell(\k_1)}  \Big)_0  \Big] \nonumber \\ & & 
  + \
 {\partial^2 \Pi_0 \over \partial (\sigma_m^2)^2}  \  \Big( {{\cal D} \sigma^2 \over {\cal D} \phi_\ell(\k_1)} \Big)_0
  \Big( {{\cal D} \sigma^2 \over {\cal D} \phi_\ell(\k_2)}  \Big)_0
  +
 {\partial \Pi_0 \over \partial \sigma_m^2}  \  \Big( {{\cal D}^2 \sigma^2 \over {\cal D} \phi_\ell(\k_1) {\cal D} \phi_\ell(\k_2)}  \Big)_0 + \ldots
\label{PiSecond}
\eeqa
\end{widetext}
where in the second equality we singled out the most dominant contributions: again the first term here is present even for Gaussian initial conditions, the second and third are  ${\cal O}(\fnl)$, and the last line is ${\cal O}(\fnl^2)$ and  ${\cal O}(\gnl)$. 

We now have all the ingredients to write down general expressions for the PBS predicted linear and quadratic bias to all orders in PNG.

\subsection{Linear Bias}
\label{linb}

Let us first look at the linear bias terms, and include for now only the first derivative contributions in Eq.~(\ref{deltaHL2}). We are interested in the bias in Fourier space, which is easy to obtain from Eq.~(\ref{deltaHL2}) since the only $\x$-dependence comes from the first derivatives of $\Pi$ which from Eqs.~(\ref{PiFirst}) and~(\ref{AllDers}) means that they are just proportional to a plane wave. Thus we obtain for the {\em Fourier space} Lagrangian halo perturbation,
\beq
\widetilde\delta_h^L = \sum_{p=1}^\infty {\partial_m \Big[  I_{p1} \int (\partial \Pi/\partial c^{(p)})_0 \Big] \over \partial_m \int \Pi_0} \ \widetilde\phi_\ell,
\label{delF}
\eeq
where to simplify notation we omit arguments which are displayed in Eq.~(\ref{deltaHL2}). The sum over $p$ here includes all orders in PNG through  $I_{p1}$, the first term is the usual scale-independent linear Lagrangian bias obtained from derivatives of the conditional mass function
\beqa
p=1:\ \ \ \ \ b_{1L}^{(1)} &=&{\partial_m \int (\partial \Pi/\partial \delta_\ell)_0 \over \partial_m\int \Pi_0} \nonumber \\ &
= & \Big[{\partial \over \partial \delta_\ell}\ \ln  \Big({d n(\delta_\ell) \over d\ln m}\Big)\Big]_0
\label{Bpeq1}
\eeqa
whereas the next gives the possibly scale-dependent bias correction to ${\cal O}(\fnl)$,
\beqa
p=2: \ \ \ \ \ \ b_{1L}^{(2)}  &=&{\partial_m[ I_{21} \int \partial \Pi_0/\partial \sigma^2_m] \over M(k)\, \partial_m\int \Pi_0}  \nonumber \\ &= & 
 {\partial_{\sigma^2_m} [ I_{21} \, {\cal F}_0 ] \over M(k)\, {\cal F}_0 }
\label{Bpeq2}
\eeqa 
where in this last equality it is understood that any $m$-dependence inside the square brackets is rewritten in terms of $\sigma^2_m$. Going from the first-upcrossing rate ${\cal F}_0$ to the mass function we can rewrite this as,
\beq
 b_{1L}^{(2)}  = 
{\partial_m \Big[ I_{21}(k,m)\, \Big({d \, n\over d\ln m}\Big)\,  \Big({d\sigma^2_m \over dm}\Big)^{-1} \Big]   \over M(k)\ \Big({d\, n\over d\ln m}\Big) }
\label{DeltaB}
\eeq
This is our final prediction for the possibly scale-dependent bias correction due to PNG at leading order in $\fnl$ in terms of the mass function $(dn/d\ln m)$ and the variance smoothed at the mass-scale of the halo $\sigma^2_m$ (see Eq.~\ref{Mdef} for definition of $M$). Note that to leading order, since $I_{21}$ depends on the primordial bispectrum (Eq.~\ref{AllDers})
\beq
I_{21}(k,m) \equiv  {1\over P_\phi(k)} \int 
B_{\widehat{\delta}\widehat{\delta}\phi}(\q,\k-\q,-\k) \, d^3q ,
\label{IBdef}
\eeq
the mass function can be computed from {\em Gaussian} initial conditions, therefore Eq.~(\ref{DeltaB}) can be computed given a PNG model from Gaussian simulations alone {\em for any type of }PNG, without making any assumptions about universality of the mass function, Markovian behavior or having a specific implementation of the excursion-set approach. Note that our result, Eq.~(\ref{DeltaB}), is different from that in~\cite{2008JCAP...08..031S} where the scale-dependent bias for local PNG without assuming universality is given in terms of derivatives of the mass function with respect to $\sigma_8$, as that would require running Gaussian simulations for different normalizations. Our expression is for valid for general PNG (characterized by Eq.~\ref{IBdef}) rather than local and one only needs Gaussian simulations with the same $\sigma_8$ to make predictions.

\begin{table*}
\caption{Different bias parameters used in the text }
\begin{ruledtabular}
\begin{tabular}{c c c }
Symbol &   Meaning &  First appearance  \\
\hline
$b_{1}$ & linear Eulerian bias (includes scale-independent and dependent contributions) &  Eq.~(\ref{blin})  \\
$b_{1L}$ &  linear Lagrangian bias   (includes scale-independent and dependent contributions) &  Eq.~(\ref{blin})  \\
$b_{1L}^{(1)}$ & {\em scale-independent} contribution to linear Lagrangian bias  &  Eq.~(\ref{Bpeq1})  \\
$ b_{1L}^{(2)}$ & scale-dependent correction to $b_{1L}$ from variance modulation ${\cal O}(\fnl)$&  Eq.~(\ref{Bpeq2})  \\
$ b_{1L}^{(3)}$ & scale-dependent correction to $b_{1L}$  from third-moment modulation ${\cal O}(\gnl)$&  Eq.~(\ref{Bpeq3})  \\
$ \Delta b_{1}$ & sum of all contributions to scale-dependent  $b_{1}$ ($\Delta b_1 \simeq b_{1L}^{(2)}$ for $\fnl$ PNG) &  Eq.~(\ref{Db1Lapprox})  \\
$ \Delta b_{1L}$ & sum of all contributions to scale-dependent  $b_{1L}$, $\Delta b_{1L}=\Delta b_1$ &  Eq.~(\ref{Db1Lapprox})  \\
$b_{G}$ & bias measured in Gaussian simulations from halo-matter spectrum&  Fig.~\ref{BiasFOF0p2L}  \\
$b_{\rm res}$ & residual halo bias in simulations after {\em theoretical} scale-dependent bias is substracted &  Eq.~(\ref{bres}) \\
$\delta b_{1L}$ & PNG correction to the {\em scale-independent} linear Lagrangian bias  &  Eq.~(\ref{dbPNG}) \\
\hline
$b_{2}$ & quadratic Eulerian bias (includes scale-independent and dependent contributions) &  Eq.~(\ref{FGexp})  \\
$b_{2L}$ &  quadratic Lagrangian bias   (includes scale-independent and dependent contributions) &  Eq.~(\ref{b2Lkernel})  \\
$b_{2L}^{(1)}$ & {\em scale-independent} contribution to quadratic Lagrangian bias  &  Eq.~(\ref{b2Lsi})  \\
$ b_{2L}^{(1,2)}$ & scale-dependent correction to $b_{2L}$ from mean and variance modulation ${\cal O}(\fnl)$&  Eq.~(\ref{b2L12f0})  \\
$ b_{2L}^{(2,2)}$ & scale-dependent correction to $b_{2L}$ from variance-variance modulation ${\cal O}(\fnl^2)$&  Eq.~(\ref{b2L22f0})  \\
$ b_{2L}^{(2)}$ & scale-dependent correction to $b_{2L}$ from  variance modulation ${\cal O}(\gnl)$&  Eq.~(\ref{b2L2f0})  
\label{BiasParam}
\end{tabular} 
\end{ruledtabular} 
\end{table*}

That Eq.~(\ref{Bpeq2}) can be written in terms of the mass function without any assumptions about Markovian evolution and universality is  only because this correction comes from a modulation of the variance and thus involves $\partial \Pi_0/ \partial \sigma_m^2$ which gives the first up-crossing rate, and thus can be written back in terms of the mass function. This fact is a special property of $\fnl$-PNG, on the other hand, at cubic order
\beqa
p=3: \ \ \ \ \  b_{1L}^{(3)}   &=&{\partial_m[ I_{31} \int \partial \Pi_0/\partial c^{(3)}_m] \over M(k)\, \partial_m\int \Pi_0} 
\label{Bpeq3}
\eeqa 
and there is no simple way of proceeding further without knowing how $\Pi_0$ depends on the third moment of small-scale primordial fluctuations. Because $I_{31}$ depends on $m$, Eq.~(\ref{Bpeq3}) cannot be written solely in terms of derivatives of the mass function with respect to the third cumulant.  This is important for $\gnl$-PNG, for which this term makes the leading contribution to possibly scale-dependent bias. And note that Gaussian simulations are not useful in this case, one must resort to some kind of approximation, e.g. assuming that $\Pi_0$ can be replaced by the PDF of the density field (violating the boundary condition that $\Pi_0(\delta_c)=0$) and then expanding the PDF in Edgeworth series to compute its derivative with respect to $c^{(3)}$ (see~\cite{2011arXiv1105.3628D,2011arXiv1105.3476D,2011arXiv1106.0503S} for work along these lines) or use results from the excursion-set approach in~\cite{2010arXiv1007.1903D} to compute how $\Pi_0$ depends on $c^{(3)}$. We won't comment on this further, except to note that clearly in this case computing the scale-dependent bias contribution relies on a lot more assumptions than for $\fnl$-PNG where a clean result such as Eq.~(\ref{DeltaB}) is possible for arbitrary primordial bispectra. For a calculation of scale-dependent bias up to trispectrum contributions from local bias models see~\cite{2011arXiv1106.4404G}.

So far to calculate linear bias we have relied on Eq.~(\ref{delF}) which only includes first derivative contributions from Eq.~(\ref{deltaHL2}). In going from Eq.~(\ref{delF}) to Eq.~(\ref{Bpeq1}) we have implicitly assumed that $I_{11} \widetilde\phi_\ell = M \widetilde\phi_\ell = \widetilde\delta_\ell$ which is only true in the Gaussian case. However, it's easy to see that the second derivative contributions with $p=1$ in Eq.~(\ref{PiSecond}) contribute the non-Gaussian part of $\delta_\ell$ with the same bias coefficient, and thus Eq.~(\ref{Bpeq1}) holds when comparing the halo to the fully non-Gaussian matter perturbations (higher-order PNG terms inside $\delta_\ell$ get generated by higher-derivatives). This goes to show that, as stated in  Section~\ref{modbias}, the perturbation expansion can be reorganized in terms of $\phi_\ell$ and $\delta_\ell$. The remaining contributions coming from second derivatives  contribute to quadratic bias, and will be considered below in Section~\ref{QuadB}. 

Summarizing, we can write the {\em Eulerian}  linear bias in Fourier space as 
\beqa
b_1 &=& 1+b_{1L} \nonumber \\ & =&  1+ b_{1L}^{(1)}  +  b_{1L}^{(2)} +
 \sum_{p=3}^\infty {\partial_m[ I_{p1} \int \partial \Pi_0/\partial c^{(p)}_m] \over M(k)\, \partial_m\int \Pi_0} ,
 \nonumber \\ &=& b_1^{(1)} + b_1^{(2)} + \ldots 
\label{blin}
\eeqa
where $b_1^{(1)} =1+b_{1L}^{(1)}$, $b_1^{(p)} =b_{1L}^{(p)}$ ($p\geq 2$), and  
the first two terms $b_{1L}^{(1)} $ and $ b_{1L}^{(2)}$ can be computed from a given conditional and unconditional mass function (Eqs.~\ref{Bpeq1} and~\ref{DeltaB}), but the additional contributions from cubic and higher-order PNG cannot be simply related to mass functions without further assumptions.

From now on, we will concentrate on $\fnl$ PNG to leading order in $\fnl$, and thus the contribution to possibly scale-dependent bias (in either Lagrangian or Eulerian space) is 
\beq
\Delta b_{1} =\Delta b_{1L} \equiv  b_{1L}^{(2)} +  b_{1L}^{(3)}  + \ldots \simeq  b_{1L}^{(2)} ,
\label{Db1Lapprox}
\eeq
thus henceforth  we will use $\Delta b_{1}$ to denote the scale-dependent bias from $\fnl$ PNG. Table~\ref{BiasParam} summarizes our notation for bias parameters.

\subsection{Comparison with Known Results}
\label{CompRes}

Here we compare our results to known PBS results in the literature, for comparison with other than PBS approaches see Section~\ref{PBSvsLOC}.

Let us first evaluate these results for local PNG of $\fnl$ type, i.e. $K=1$ and no cubic or higher-order PNG.  To make contact with the literature, we work in  {\em the low-$k$ limit} where (see Eqs.~\ref{LinRes} and~\ref{IBdef})
\beq
I_{21}^{\rm loc}(k\rightarrow0,m)  = 4 \fnl \ \sigma_m^2 + {\cal O}(k^2)
\label{IBls}
\eeq
and then Eq.~(\ref{Bpeq2}) reads
\beq
\Delta b_{1} (k\rightarrow 0) = {4\fnl \over M(k)} \ \partial_{\ln \sigma_m^2} \ln ( \sigma_m^2 {\cal F}_0) %= 2\fnl\, (\nu^2-1)
\label{Db1Llowk}
\eeq
where again as in Eq.~(\ref{Bpeq2}) any dependence of ${\cal F}_0$ on $m$ is rewritten in terms of  $\sigma_m^2$. This result is still more general than those in the literature, which assume Markovian behavior and universality of the mass function to relate the scale-dependent bias amplitude to the scale-independent Lagrangian bias. To obtain this limit, let's assume Markovianity, in which case the PBS linear Lagrangian bias can be written as a derivative of the unconditional mass function
\beq
b_{1L}^{(1)} = - {\partial \over \partial \delta_c} \ln \Big( {dn \over d\ln m} \Big),
\label{b1Lmarkov}
\eeq
and in addition assuming universality (see Eq.~\ref{nufnu}) we can rewrite this as
\beq
b_{1L}^{(1)} = {2 \over  \delta_c} \partial_{\ln \sigma^2_m}  \ln ( \sigma_m^2 {\cal F}_0) ,
\label{b1LmarkovUniv}
\eeq
which reduces Eq.~(\ref{Db1Llowk}) to the well-known result~\cite{2008PhRvD..77l3514D,2008JCAP...08..031S}
\beq
\Delta b_{1}(k\rightarrow 0) = {2\fnl \over M(k)} \ \delta_c \, b_{1L}^{(1)} 
\label{Db1Lstd}
\eeq
In section~\ref{NBcomp} we will compare N-Body simulations to this standard result and our more general result, Eq.~(\ref{DeltaB}). The latter incorporates corrections to this formula due to three effects: i) non-universality of the mass function, ii) non-Markovian behavior, iii) beyond-leading order corrections in $k$ to Eq.~(\ref{IBls}) coming from $M$ (see Eq.~\ref{bogus}), the transfer function and the smoothing kernel. These lead to corrections to the {\em scale-independent} bias to ${\cal O}(\fnl)$, see Appendix~\ref{IBlocApp} for more details. 

\begin{figure}[!t]
 \centering
 \includegraphics[ width=0.95\linewidth]{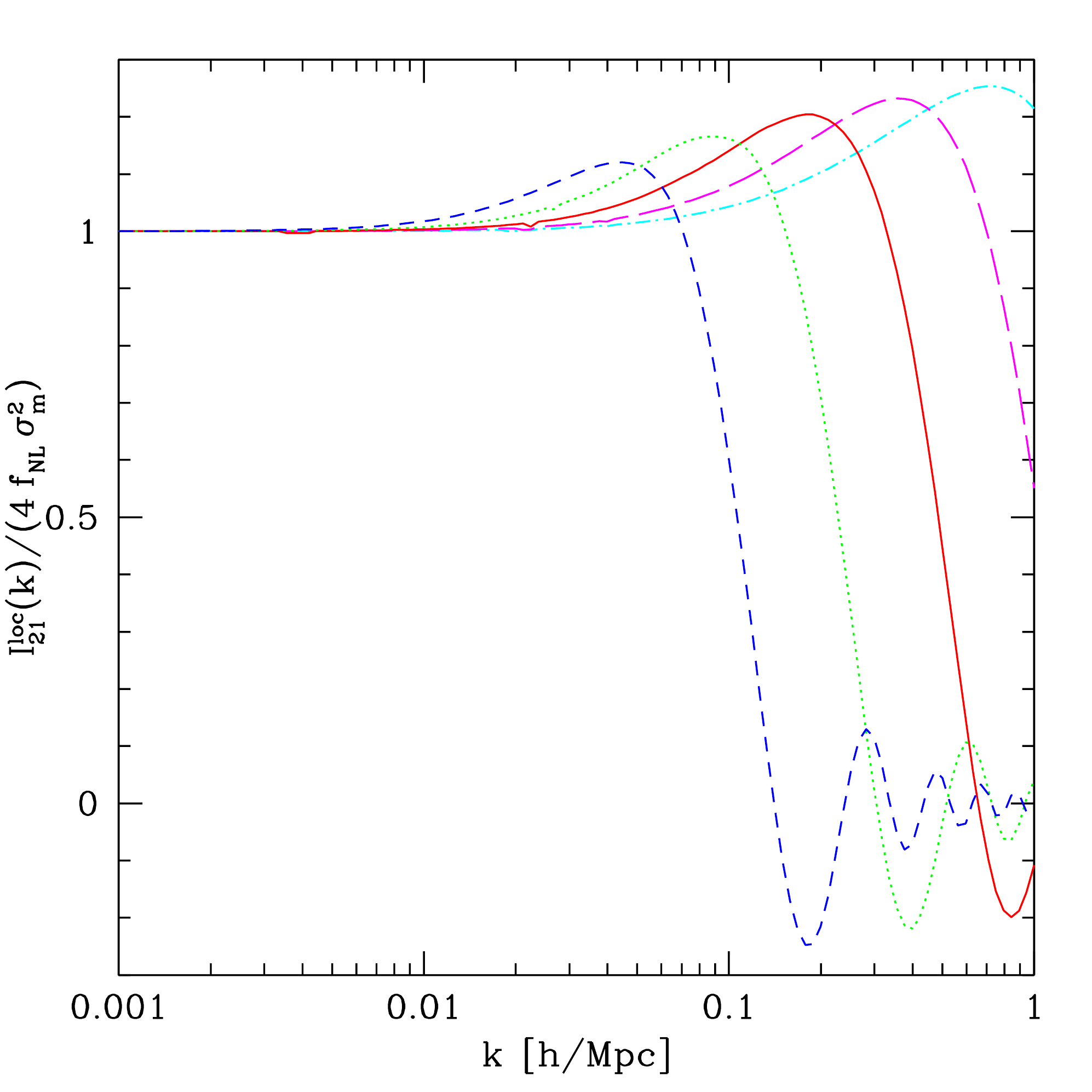}
 \caption{Full calculation of $I_{21}$, Eq.~(\ref{IBdef}), for local PNG normalized by its large-scale limit,  Eq.~(\protect\ref{IBls}), for halos of mass $\log_{10}m=16,15,14,13,12$ (dashed, dotted, solid, long-dashed, dot-dashed).}
 \label{IB}
 \end{figure}

To appreciate this latter point, In Figure~\ref{IB} we show the full calculation of $I_{21}(k,m)$ as a function of $k$ for different halo masses, normalized by it's large-scale limit, Eq.~(\ref{IBls}).  We use this numerical calculation of $I_{21}$ in all our computations below, except in Appendix~\ref{IBlocApp} where the different sources that contribute to scale-independent bias are separated by Taylor expansion.

 \begin{figure}[!t]
 \centering
 \includegraphics[ width=0.95\linewidth]{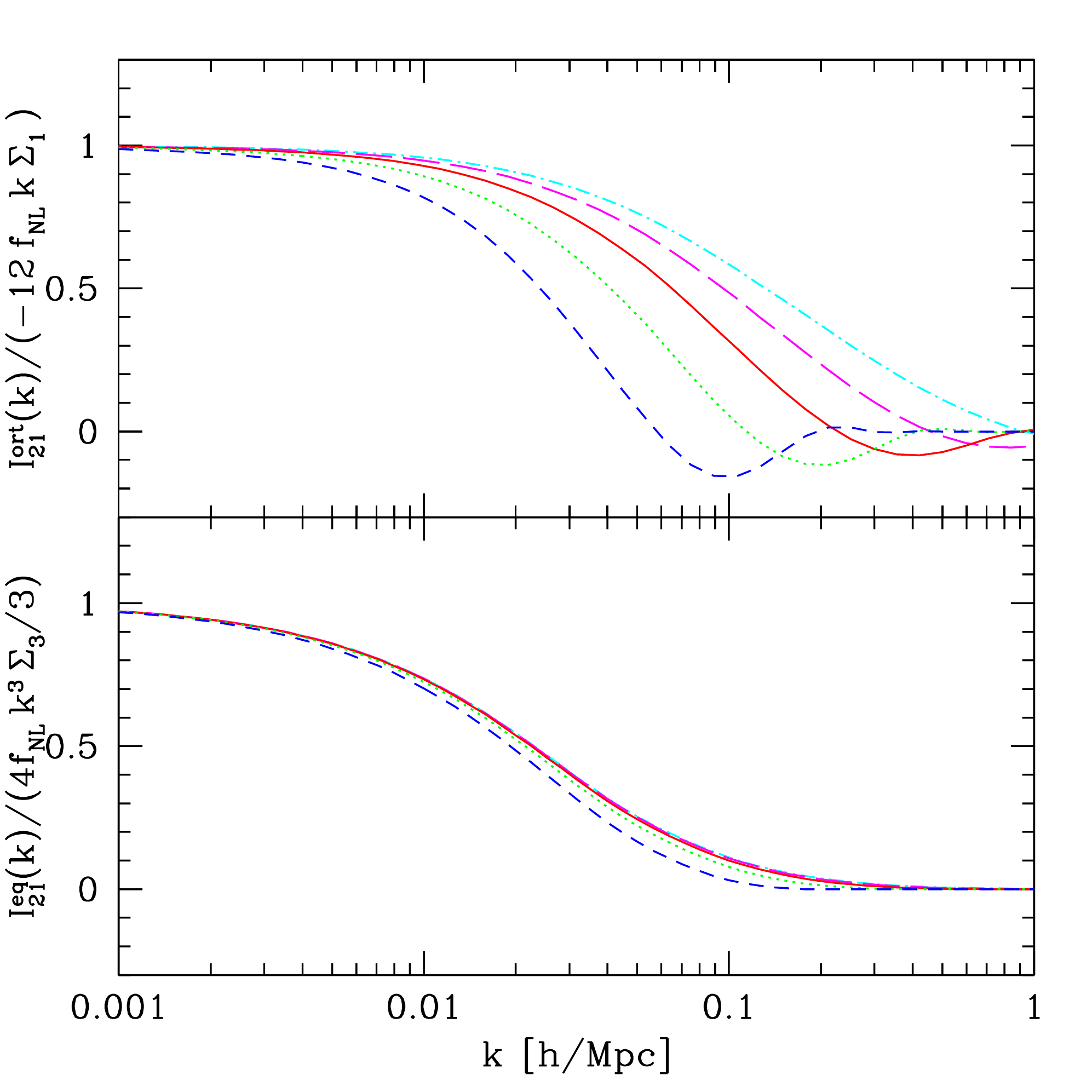}
 \caption{Same as Fig.~\ref{IB} but for non-local PNG of orthogonal  (top panel) and equilateral type (bottom panel). The $I_{21}$ are normalized by the large-scale values predicted by Eqs.~(\ref{I21ort}) and~(\ref{I21eq}), respectively.}
 \label{IBnonloc}
 \end{figure}

For non-local PNG, assuming Markovianity and universality we can rewrite Eq.~(\ref{DeltaB}) or its equivalent Eq.~(\ref{Bpeq2}) as
\beq
\Delta b_{1} = {I_{21}(k,m) \over 2\sigma^2_mM(k) }\delta_c \, b_{1L}^{(1)} + {1\over M(k)}\, \partial_{\ln \sigma^2_m} \Big({I_{21}(k,m)\over \sigma_m^2} \Big).
\label{SKcomp}
\eeq
The first term here agrees with the results of~\cite{2010PhRvD..82j3002S}, the second term (which vanishes for local models in the low-$k$ limit) has  been recently taken into account in~\cite{2011arXiv1105.3628D,2011arXiv1105.3476D}. Again, our results in Eq.~(\ref{DeltaB}) are more general than these as they incorporate the effects of non-universality of the mass function and non-Markovian behavior. Eq.~(\ref{SKcomp}) is of course nice in that it's analytic (aside from the
computation of $I_{21}$), whereas Eq.~(\ref{DeltaB}) requires knowing the
mass function. 
However, let us re-emphasize Eq.~(\ref{DeltaB}) requires only the
{\em Gaussian} mass function, and thus accurate fitting 
formula  could conceivably 
be used, though in this paper we use the measured
Gaussian mass function (see Eq.~\ref{implem} below for details of how we implement this).

Let us now specifically evaluate these results for the orthogonal and equilateral templates, in particular their scale dependence, which is governed by $I_{21}$. In the low-$k$ limit, we can expand the cross bispectrum in powers of $(k_\ell/k_s)$ with $k_\ell \ll k_s$, the so-called squeezed limit, which is described in detail in Appendix~\ref{sqz}. For the orthogonal template, using Eq.~(\ref{BORTSQZ}) for the squeezed limit of the cross bispectrum we obtain,
\beq
I_{21}^{\rm ort}(k\rightarrow0,m) \approx -12 \fnl \,k\, \Sigma_1(m),
\label{I21ort}
\eeq 
where we introduced (note that $\Sigma_0(m) = \sigma_m^2$)
\beq
\Sigma_n(m) \equiv \int d^3 q \, P(q) \, W^2_{\rm TH}(qR) \  q^{-n}
\label{sigmas}
\eeq
Equation~(\ref{I21ort}), when used into Eq.~(\ref{SKcomp}), gives the expected $k^{-1}$ scale-dependent bias\footnote{Recall, however, as mentioned in the introduction, that this is due to a pathology of the template we use, the true orthogonal PNG should scale as $I_{21} \sim k^2$ leading to scale-independent bias instead.}, while for the equilateral model, using Eq.~(\ref{BcrossSQZ3}) for the squeezed limit of the cross bispectrum we obtain instead,
\beq
I_{21}^{\rm eq}(k\rightarrow0,m) \approx {4\over 3} \fnl\, k^3\, \Sigma_3(m),
\label{I21eq}
\eeq 
which leads to $k^{+1}$ scale-dependent bias, although with a very small amplitude (percent level for $\fnl=-400$, see Fig.~\ref{BiasEq} below). This is not the expected result, but it arises because of cancellations when integrating over the cross bispectrum, while a naive use of the scaling of the kernel would predict a $k^{-1}$ scale-dependent bias, see Appendix~\ref{sqz} for details of the calculation. This different scaling is made possible in the first place because we used $\phi$ instead of $\Phi$ as the split variable, and for the equilateral template short $\phi$-modes can contribute to $\Phi$ at large distances through the singular squeezed limit of the $K^{(\ell)}$ kernel. See the next subsection and  Appendix~\ref{sqz} for more discussion. We will compare both approaches with simulations in Fig.~\ref{BiasEq} below.

Figure~\ref{IBnonloc} shows the results of the exact numerical integration of $I_{21}$ for these non-local models compared to their large-scale limits. We see that there are next-lo-leading order corrections to the large-scale result similar to the local PNG case shown in Fig.~\ref{IB}, that are particularly strong for the equilateral case (bottom panel, where $I_{21}^{\rm eq}$ drops to zero faster than in the other PNG models). 

\subsection{Using $\Phi$ as the large-scale constraint}
\label{ConstPhi}

Let us now go back and ask what would have happened if we imposed the large-scale constraint on $\Phi$ rather than $\phi$. The procedure is the same, but the expansion of the halo perturbation in Eq.~(\ref{deltaHL2}) is now in $\Phi_\ell(\k)$ rather than $\phi_\ell(\k)$ and we need to calculate the modulation of the cumulants of the small-scale density field $\delta_s$ by the large-scale field $\Phi_\ell$, which is  non-Gaussian. This modulation is described by the general result in Eq.~(\ref{AllDers}), of which Eq.~(\ref{RelBisp}) is the simplest version, corresponding to modulation of the variance by the large-scale field at ${\cal O}(\fnl)$. These results are {\em exact} for $\phi_\ell$ a Gaussian field, as one can easily check by doing a Taylor expansion of $c^{(p)}$ on $\phi_\ell$ and computing left- and right-hand sides. We need to extend these results to the non-Gaussian case when $\phi_\ell$ is replaced by $\Phi_\ell$. 

Let us consider the simplest case first, Eq.~(\ref{RelBisp}). It is easy to see that in the non-Gaussian case it gets modified to,
\begin{widetext}
\beq
\Big( {{\cal D} \sigma^2 \over {\cal D} \Phi_\ell(\k)} \Big)_0 =  { \langle \sigma^2\, \Phi_\ell^*(\k) \rangle_c \over P_\Phi(k)}
 - {1\over 2} \int d^3q { \langle \sigma^2\, \Phi_\ell^*(\q) \Phi_\ell^*(\k-\q) \rangle_c \over P_\Phi(k)\, P_\Phi(q)\, P_\Phi(|\k-\q|)}\ B_\Phi(\q,\k-\q,-\k) + \ldots
\label{1pt}
\eeq
where we neglected higher-order terms in PNG. The first term leads, to leading order, to the same Eq.~(\ref{RelBisp}) but where  $B_{\widehat{\delta}\widehat{\delta}\phi}$ is replaced by $B_{\widehat{\delta}\widehat{\delta}\Phi}$, while the second term is at least of ${\cal O}(\fnl^2)$ and can be dropped. This implies that the scaling of the halo bias in the case of the equilateral template {\em will be changed from $k^{+1}$ to scale-independent}. See Fig.~\ref{BiasEq} below for comparison of this prediction against simulations. For the other two templates (local and orthogonal) there are no significant changes in the predictions of the scale-dependent {\em linear} bias. 

The situation is a bit more interesting for the case of quadratic bias that will be discussed in the next subsection. For this we need in addition the modulation of the variance by two large-scale fields, which reads\footnote{These calculations are precisely equivalent to those in RPT that relate multi-point propagators to cross-correlations, see~\cite{2010PhRvD..82h3507B} for the non-Gaussian case. Equation~(\ref{2pt}) corresponds to their Eq.~(33), while Eq.~(\ref{1pt}) fixes a sign typo in their Eq.~(34).
}
\beq
\Big( {{\cal D}^2 \sigma^2 \over {\cal D} \Phi_\ell(\k_1) {\cal D} \Phi_\ell(\k_2)} \Big)_0 =  { \langle \sigma^2\, \Phi_\ell^*(\k_1) \Phi_\ell^*(\k_2) \rangle_c \over P_\Phi(k_1)\, P_\Phi(k_2)}
 -  { B_\Phi(\k_{12},-\k_1,-\k_2)\over  P_\Phi(k_1)\, P_\Phi(k_2)} \times 
 { \langle \sigma^2\, \Phi_\ell^*(\k_{12})  \rangle_c \over P_\Phi(k_{12})} + \ldots
\label{2pt}
\eeq
where now the first two terms are {\em of the same order}, i.e. ${\cal O}(\fnl^2)$ for $\fnl$-PNG. These results can be cast in terms of the $I_{pq}$'s defined in Eq.~(\ref{AllDers}), by saying that when $\Phi_\ell$ modes have been constrained we must change,
\beq
I_{21}(k) \rightarrow \tilde{I}_{21}(k), \ \ \ \ \ \ \ \ \ \ 
I_{22}(\k_1,\k_2) \rightarrow \tilde{I}_{22}(\k_1,\k_2) -  \tilde{I}_{21}(k_{12})\times { B_\Phi(\k_{12},-\k_1,-\k_2)\over  P_\Phi(k_1)\, P_\Phi(k_2)}
\label{transf}
\eeq
where the $\tilde{I}_{pq}$'s are calculated  in the same way as Eq.~(\ref{AllDers}) but with $T^{(p+q)}_{\widehat{\delta}\ldots \widehat{\delta}\, \phi \ldots \phi}$  replaced by $T^{(p+q)}_{\widehat{\delta}\ldots \widehat{\delta}\, \Phi \ldots \Phi}$. We now discuss the implications of these results for quadratic bias. 

\subsection{Quadratic Bias}
\label{QuadB}

Let us now go back to Eqs.~(\ref{deltaHL2}-\ref{PiSecond}) and collect the terms which lead to quadratic Lagrangian bias.  Of the five terms displayed in Eq.~(\ref{PiSecond}), one of them (the second on the first line) was already taken into account as it contributes to linear bias (restoring the non-Gaussian part of $\delta_\ell$). The remaining four contribute to quadratic bias. Because of Eq.~(\ref{AllDers}) all terms come with an $\x$-dependence that is just a plane wave with momentum $\k_{12}=\k_1+\k_2$, and thus the Fourier space biases are straightforward to compute. 
The first term, also present for Gaussian initial conditions, leads to the scale-independent quadratic bias of local form
\beq
b_{2L}^{(1)} = {\partial_m \int (\partial^2\Pi/\partial\delta_\ell^2)_0 \over \partial_m\int \Pi_0},
\label{b2Lsi}
\eeq
that is, it contributes a term $b_{2L}^{(1)} \delta_\ell^2/2$ in the expansion of $\delta_h$ given by Eq.~(\ref{deltaHL2}), or in Fourier space
\beq
b_{2L,{\bf k}}^{(1)}[\delta_\ell,\delta_\ell]\equiv b_{2L}^{(1)} \int \, [\dD] \, \delta_\ell(\k_1) \,  \delta_\ell(\k_2) \, d^3k_1 d^3k_2,
\label{b2local}
\eeq
where $[\dD] \equiv \dD(\k-\k_{12})$. 
Recall that $ \delta_\ell$ is the non-Gaussian large-scale density perturbation. While this contribution from Eq.~(\ref{PiSecond}) leads to only the Gaussian part of $\delta_\ell$, third and fourth derivative terms in Eq.~(\ref{deltaHL2}) with the same coefficients restore the non-Gaussian parts, as it is easy to check.

The other three terms are not local in real space (except for local PNG, see below), and they generically depend on scale. From the second line in Eq.~(\ref{PiSecond}) we obtain to ${\cal O}(\fnl)$
\beq
b_{2L,{\bf k}}^{(1,2)}[\phi_\ell,\delta_\ell] \equiv \int {\partial_m\Big[  [ {I_{21}(k_1,m) } +{I_{21}(k_2,m) }] \int (\partial^2\Pi/\partial\delta_\ell \partial\sigma_m^2)_0 \Big] \over \partial_m\int \Pi_0} \, [\dD] \, \phi_\ell(\k_1) \,  \delta_\ell(\k_2) \, d^3k_1 d^3k_2 
\label{b2L12} 
\eeq
while the other contributions are ${\cal O}(\fnl^2)$
\beq
b_{2L,{\bf k}}^{(2,2)}[\phi_\ell,\phi_\ell] \equiv \int {\partial_m\Big[   {I_{21}(k_1,m) } {I_{21}(k_2,m) } \int \partial^2\Pi_0/\partial(\sigma_m^2)^2 \Big] \over  \partial_m\int \Pi_0} \, [\dD] \, \phi_\ell(\k_1) \, \phi_\ell(\k_2) \, d^3k_1 d^3k_2 
\label{b2L22} 
\eeq
and  ${\cal O}(\gnl)$ (and  also ${\cal O}(\fnl^2)$)
\beq
b_{2L,{\bf k}}^{(2)}[\phi_\ell,\phi_\ell] \equiv \int {\partial_m\Big[   {I_{22}(\k_1,\k_2,m) } \int \partial\Pi_0/\partial\sigma_m^2 \Big] \over \partial_m\int \Pi_0} \, [\dD] \, \phi_\ell(\k_1) \, \phi_\ell(\k_2) \, d^3k_1 d^3k_2 
\label{b2L2} 
\eeq
Putting all these together in {\em Fourier space} we have
\beq
\delta_h^L(\k) = b_{1L}\, \delta(\k) + {1\over 2}b_{2L,{\bf k}}^{(1)}[\delta_\ell,\delta_\ell] +{1\over 2}b_{2L,{\bf k}}^{(1,2)}[\phi_\ell,\delta_\ell] + {1\over 2}b_{2L,{\bf k}}^{(2,2)}[\phi_\ell,\phi_\ell]  + {1\over 2}b_{2L,{\bf k}}^{(2)}[\phi_\ell,\phi_\ell] 
\label{deltahFS}
\eeq
\end{widetext}
The superscripts in $b_{2L}$ denote the physical origin of each term, e.g. $(1,2)$ in Eq.~(\ref{b2L12}) indicates that this contribution comes from modulation of the large-scale mean and small-scale variance, see also Table~\ref{BiasParam} for explanation of our bias parameter notation.  
Note that while $\phi_\ell$ is a Gaussian field, $\delta_\ell$ includes PNG contributions (which are higher-order in PNG because of the overall factors of $I_{pq}$). If we ignore these higher-order contributions, we can simplify the notation by defining quadratic bias kernels in Fourier space in terms of the  $\delta_\ell$'s, e.g.
\beqa
b_{2L,{\bf k}}^{(1,2)}[\phi_\ell,\delta_\ell] &\equiv &\int   b_{2L}^{(1,2)}(\k_1,\k_2)   \, \dD(\k-\k_{12}) \nonumber \\
& & \times\  \delta_\ell(\k_1) \,  \delta_\ell(\k_2) \, d^3k_1 d^3k_2,
\label{b2L12kernel} 
\eeqa
and similarly for the other two cases. The leading-order contribution ${\cal O}(\fnl)$ to scale-dependence of quadratic bias is thus given by Eq.~(\ref{b2L12}), which from Eq.~(\ref{Bpeq1}) can be rewritten in a compact form (compare to Eq.~\ref{Bpeq2})
\beq
 b_{2L}^{(1,2)}(k_1,k_2) = {\partial_{\sigma^2_m}  [I_{21}(k_1)\, b_{1L}^{(1)} \, {\cal F}_0 ]\over M(k_1)\, {\cal F}_0}  + k_1 \leftrightarrow k_2
\label{b2L12f0} 
\eeq
where again it is understood that any mass dependence inside the square brackets is rewritten in terms of $\sigma_m^2$, and we have suppressed the mass arguments of $I_{21}$, $b_{1L}^{(1)}$ and ${\cal F}_0$ for simplicity. The other two kernels can be similarly written,
\beq
 b_{2L}^{(2,2)}(k_1,k_2) = {\partial_{\sigma^2_m}  [I_{21}(k_1)\, I_{21}(k_2) \, \partial_{\sigma^2_m}{\cal F}_0 ]\over M(k_1)\, M(k_2)\, {\cal F}_0}  
\label{b2L22f0} 
\eeq

\beq
 b_{2L}^{(2)}(\k_1,\k_2) = {\partial_{\sigma^2_m}  [I_{22}(\k_1,\k_2) \, {\cal F}_0 ]\over M(k_1)\, M(k_2)\, {\cal F}_0}  
\label{b2L2f0} 
\eeq

These results can also be cast in terms of the mass function, in analogy to the linear bias scale-dependence, Eq.~(\ref{DeltaB}). For example, for the leading correction to ${\cal O}(\fnl)$ we have from Eq.~(\ref{b2L12f0})
\beqa
 b_{2L}^{(1,2)}(k_1,k_2) &=& {\partial_{m}  \Big[I_{21}(k_1)\, b_{1L}^{(1)} \, \Big({dn\over d\ln m}\Big)\, \Big({d\sigma_m^2\over dm}\Big)^{-1} \Big]\over M(k_1)\, \Big({dn\over d\ln m}\Big)}  \nonumber \\ & & + \quad k_1 \leftrightarrow k_2
\label{b2L12MF} 
\eeqa
which again can be implemented using measurements in {\em Gaussian} simulations as discussed above. Summarizing, the total quadratic bias kernel is simply the sum over all these contributions, i.e.
\beq
b_{2L} = b_{2L}^{(1)} +  b_{2L}^{(1,2)} +  b_{2L}^{(2,2)} +  b_{2L}^{(2)} + \ldots
\label{b2Lkernel}
\eeq
where only the first term is of local form for non-local PNG. 

Following the treatment in Section~\ref{CompRes}, we can take the low-$k$ limit and assume universality and Markovian evolution to simplify these expressions, and make contact with previous literature. For local PNG we have using Eq.~(\ref{IBls}) that  the leading-order scale-dependence, Eq.~(\ref{b2L12f0}), reads 
\beq
 b_{2L}^{(1,2)} \approx 2\fnl  (\delta_c\, b_{2L}^{(1)}-b_{1L}^{(1)}) \Big({1\over M(k_1)}+{1\over M(k_2)}\Big),
\label{b2L12UMlowk} 
\eeq
while for Eq.~(\ref{b2L22f0}) we have 
\beq
 b_{2L}^{(2,2)} \approx 4 \fnl^2 \delta_c \, {(\delta_c\, b_{2L}^{(1)}-3b_{1L}^{(1)}) \over M(k_1) M(k_2)}.
\label{b2L22UMlowk} 
\eeq
To give the low-$k$ limit of Eq.~(\ref{b2L2f0}) we need (from Eqs.~\ref{sigma-phi}-\ref{chisq} and Eq.~\ref{AllDers}) 
\beq
I_{22}^{\rm loc} (k_1,k_2 \rightarrow0,m) = (8\fnl^2+12\gnl)\, \sigma^2_m
\label{I22lowk}
\eeq
which gives (see Eq.~\ref{gnlPhi} for definition of $\gnl$)
\beq
 b_{2L}^{(2)} \approx {4 \fnl^2 \delta_c b_{1L}^{(1)}  + 6\gnl \delta_c b_{1L}^{(1)} \over M(k_1) M(k_2)}.
\label{b2L2UMlowk} 
\eeq
Equations~(\ref{b2L12UMlowk}-\ref{b2L22UMlowk}) and (\ref{b2L2UMlowk}) agree with previous results in the literature for local PNG~\cite{2010PhRvD..81f3530G,2011JCAP...04..006B}. There are no results on non-local PNG quadratic bias in the literature.

Let us now discuss how these results change if we impose the large-scale constraint on $\Phi_\ell$, rather than $\phi_\ell$. It's easy to check that the same results in Eqs.~(\ref{b2L12f0}-\ref{b2L12MF}) hold for quadratic biases, but with the ${I}_{pq}$'s changed by $\tilde{I}_{pq}$'s according to Eq.~(\ref{transf}).  For local PNG, is still true that Eq.~(\ref{IBls}) holds in the new variables, i.e. $\tilde{I}_{21}^{\rm loc}  = 4 \fnl \ \sigma_m^2 $, but Eq.~(\ref{I22lowk}) gets changed to 
\beqa
\tilde{I}_{22}^{\rm loc}  &=& 
4\fnl^2 \int d^3 q\, P_{\widehat{\delta}}(q) {P_\phi(|\k_{12}-\q|) \over P_\phi(k_1) P_\phi(k_2)} P_\phi(|\k_{1}-\q|)
\nonumber \\ & +& \k_1  \leftrightarrow \k_2 
\nonumber \\ &+&
 (8\fnl^2+ 6\gnl)  \, \Big[{ P_\phi(k_1)+P_\phi(k_{2}) \over P_\phi(k_1) \ P_\phi(k_{2}) } \Big]
\nonumber \\ &&
\times \int d^3 q\, P_{\widehat{\delta}}(q)  P_\phi(q) + 12 \, \gnl \, \sigma^2_m, 
 \label{I22lowkb}
\eeqa
where $P_{\widehat{\delta}}(q)$ is the power spectrum of the smoothed density field. 
In the low-$k$ limit, the factor in square brackets vanishes, therefore, only the first two terms give a modification of the ${\cal O}(\fnl^2)$ amplitude in Eq.~(\ref{I22lowk}), while the $\gnl$ amplitude remains the same in this limit. Therefore, we see that Eqs.~(\ref{b2L12UMlowk}-\ref{b2L22UMlowk}) are unchanged but Eq.~(\ref{b2L2UMlowk}) now reads,
\beqa
 b_{2L}^{(2)} &\approx & {6\gnl \delta_c b_{1L}^{(1)} \over M(k_1) M(k_2)}  + {2\fnl^2 \delta_c b_{1L}^{(1)}  \over  M(k_1) M(k_2)}   \nonumber \\ &&
\times \int d^3 q\, {P_{\widehat{\delta}}(q) \over \sigma^2_m} {P_\phi(|\k_{12}-\q|) \over P_\phi(k_1) P_\phi(k_2)} P_\phi(|\k_{1}-\q|)
\nonumber \\ &&+ \ \k_1  \leftrightarrow \k_2 
\label{b2L2UMlowkb} 
\eeqa
This modification of the ${\cal O}(\fnl^2)$ quadratic bias parameter may be probed through measurements of the halo bispectrum as a function of triangle shape. It arises from the same effect that can change the scaling in the low-$k$ limit for the linear bias, from contributions of the $K^{(\ell)}$ kernel that couples two short $\phi_s$ modes.

Finally, note that these results are for the {\em Lagrangian} quadratic bias parameters, what we need to compare against simulations is to compute their Eulerian counterparts. This is a standard procedure usually done in the spherical collapse approximation (see e.g.~\cite{2010PhRvD..81f3530G,2011JCAP...04..006B,EmiHaloBisp}) or, more accurately,  full perturbation theory. We leave this for an upcoming work where we implement these PBS predictions for the bispectrum and compare against simulations for halos and mock galaxy catalogs.

\section{Comparison with Simulations}
\label{NBcomp}

\begin{figure}
\center
\includegraphics[ width=0.95\linewidth]{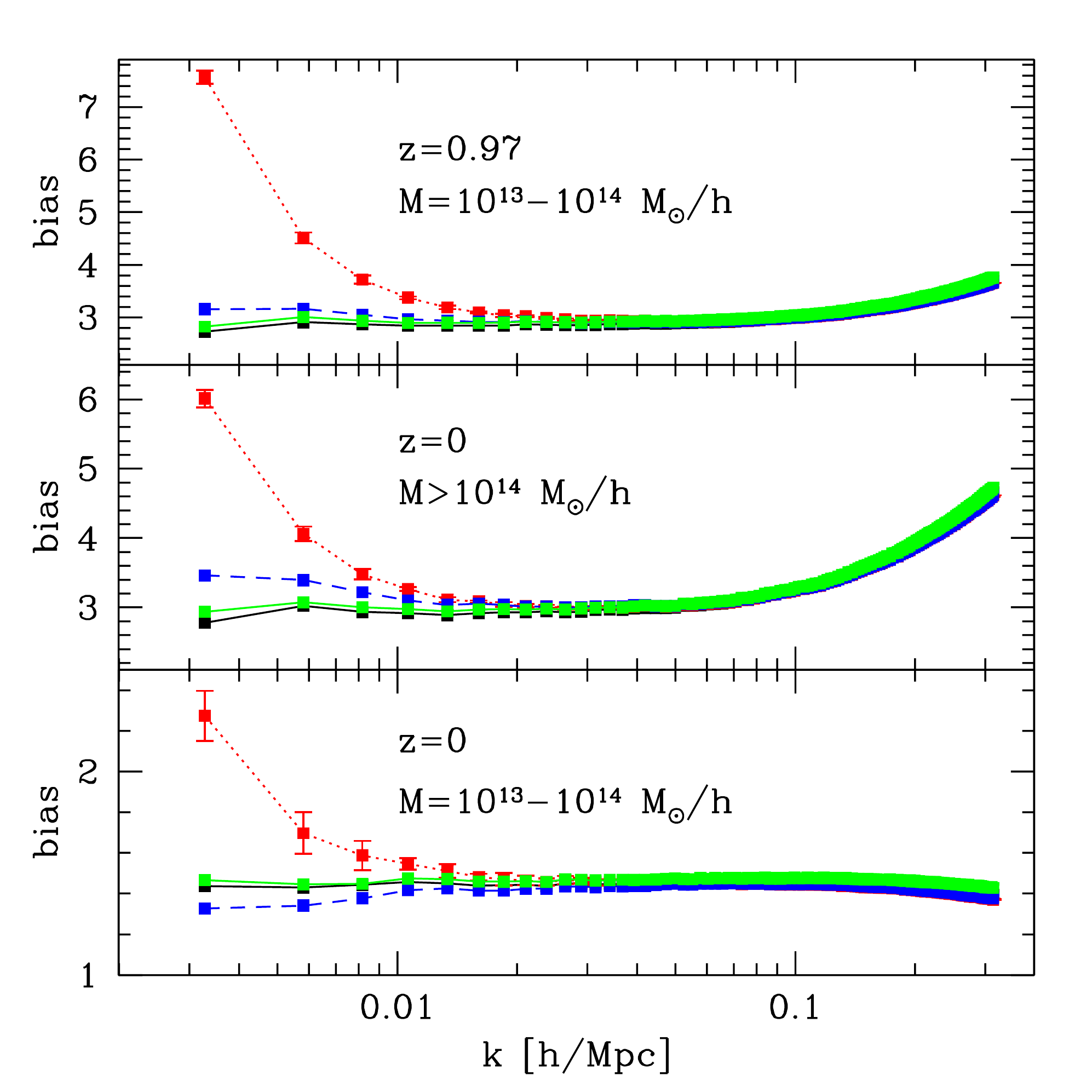}
\caption{ The bias for FOF$0.156$ halos as a function of scale for Gaussian and local, orthogonal and equilateral PNG initial conditions. Since cosmic variance is dominated by its Gaussian contribution, we only show error bars on the local PNG case for clarity. The equilateral and Gaussian case are very close to each other, whereas the orthogonal template ($\fnl=-400$, blue dashed lines) is in between them and local ($\fnl=100$, red dotted) for significantly biased objects (top two panels), but below the Gaussian (black solid) and equilateral ($\fnl=-400$, green solid) case for low-mass halos at $z=0$.}  
\label{plot_crossbias_many}
\end{figure}

We now contrast our predictions for large-scale linear bias with measurements in the simulations discussed in section~\ref{impl}. Since our predictions for the scale-dependent bias from PNG should be more widely valid than the standard results based on universality and Markovian evolution, our primary goal here is to test for the amplitude of this scale dependence. Previous results in the literature on this proceed by modeling the full bias factor, including scale-dependent and independent contributions, see e.g.~\cite{2009MNRAS.tmp..631D,2010MNRAS.402..191P,2010PhRvD..81f3530G,2010CQGra..27l4011D,2011arXiv1102.3229W,2011arXiv1105.3628D,2011arXiv1105.3476D,2010A&A...514A..46V}, and there is no consensus about whether a `fudge factor' is needed to properly account for the amplitude of scale-dependence for local PNG.

There are many reasons why this might be the case. First, not all works used the same halo definitions, we explore the dependence on halo definition below. Second, there is the impact on halo bias from from transients induced by setting up initial conditions in the simulations~\cite{2006MNRAS.373..369C}. For example, we find that using Zel'dovich initial conditions instead of 2LPT at $z=49$ for local PNG with $\fnl=100$ leads to a $z=1$ halo power spectrum ($M=10^{13}-10^{14} M_\odot/h$) that is larger by 14\% at $k=0.003 \kvecMpc$ and 3\%  at $k\ga0.05 \kvecMpc$. These transients also induce artificial violations of universality. 

From the theoretical point of view, deviations from the standard predictions are expected by violations of Markovianity and universality. While deviations from the former have not yet been established in a precise quantitative way, there is a significant body of work showing that universality of the mass function does not hold at the 5-10\% level~\cite{2002ApJS..143..241W,2007MNRAS.374....2R,2008ApJ...688..709T,2010MNRAS.402..589M,2009arXiv0912.0446M,2010MNRAS.403.1353C} for FOF halos, with more significant deviations for spherical overdensity (SO) halos~\cite{2008ApJ...688..709T}. In addition, the peak-background split calculations for Gaussian initial conditions show similar deviations ~\cite{2010MNRAS.402..589M,2009arXiv0912.0446M,2010ApJ...724..878T}. In this case, however, there is the extra complication in going from the bias parameters in the expansion of perturbations to the bias parameters that appear in the correlators such as the power spectrum, which will differ in general by renormalizations induced by loop corrections~\cite{1998MNRAS.301..797H,2006PhRvD..74j3512M,2007PhRvD..75f3512S}.

Figure~\ref{plot_crossbias_many} shows the bias computed from the halo-matter power spectrum for one of our choices of halo definition (FOF halos with linking length equal to $0.156$ times the interparticle separation) as a function of scale for Gaussian and local, orthogonal and equilateral PNG initial conditions. We see the expected scale dependence for the local case, a weaker dependence for the orthogonal template, and close to Gaussian bias in the equilateral model. Rather than performing a global fit for the scale-independent and dependent terms, our approach here is to look at the residual halo bias in simulations after the scale-dependent bias predicted by theory is substracted, i.e. (see Eq.~\ref{Db1Lapprox}  for definition of $\Delta b_1$)

\beq
b_{\rm res} \equiv \Big({P_{hm} \over P_{\rm mass}}\Big)_{\rm Nbody}  - \Delta b^{\rm theory}_{1},
\label{bres}
\eeq
where $P_{hm}$ is the cross-spectrum between halos and matter.  Note that the N-body quantities are for the PNG model under consideration, i.e. the mass power spectrum includes PNG. Simulations and perturbation theory calculations show that there are interesting PNG corrections for the mass power spectrum and bispectrum (see~\cite{2009PhRvD..80l3002S,2010PhRvD..82h3507B,2010MNRAS.406.1014S} and Fig.~\ref{dQics}), and even down to the nonlinear regime~\cite{2011PhRvD..83d3526S},  
but we won't explore those here.

If the theoretical model $\Delta b^{\rm theory}_{1}$ is correct, the residual bias $b_{\rm res}$ should be consistent with scale independence, whereas if the theoretical model does not predict the correct scale-dependent bias $b_{\rm res}$ will still show residual scale-dependence. Furthermore, {\em provided that  $b_{\rm res}$ is consistent with scale-independence}, we can look at the ratio of $b_{\rm res}$ to $b_{\rm G}$, the halo bias measured in our Gaussian simulations, to quantify  the magnitude of the PNG corrections to the scale-independent bias.

 \begin{figure}[!t]
 \centering
 \includegraphics[ width=0.95\linewidth]{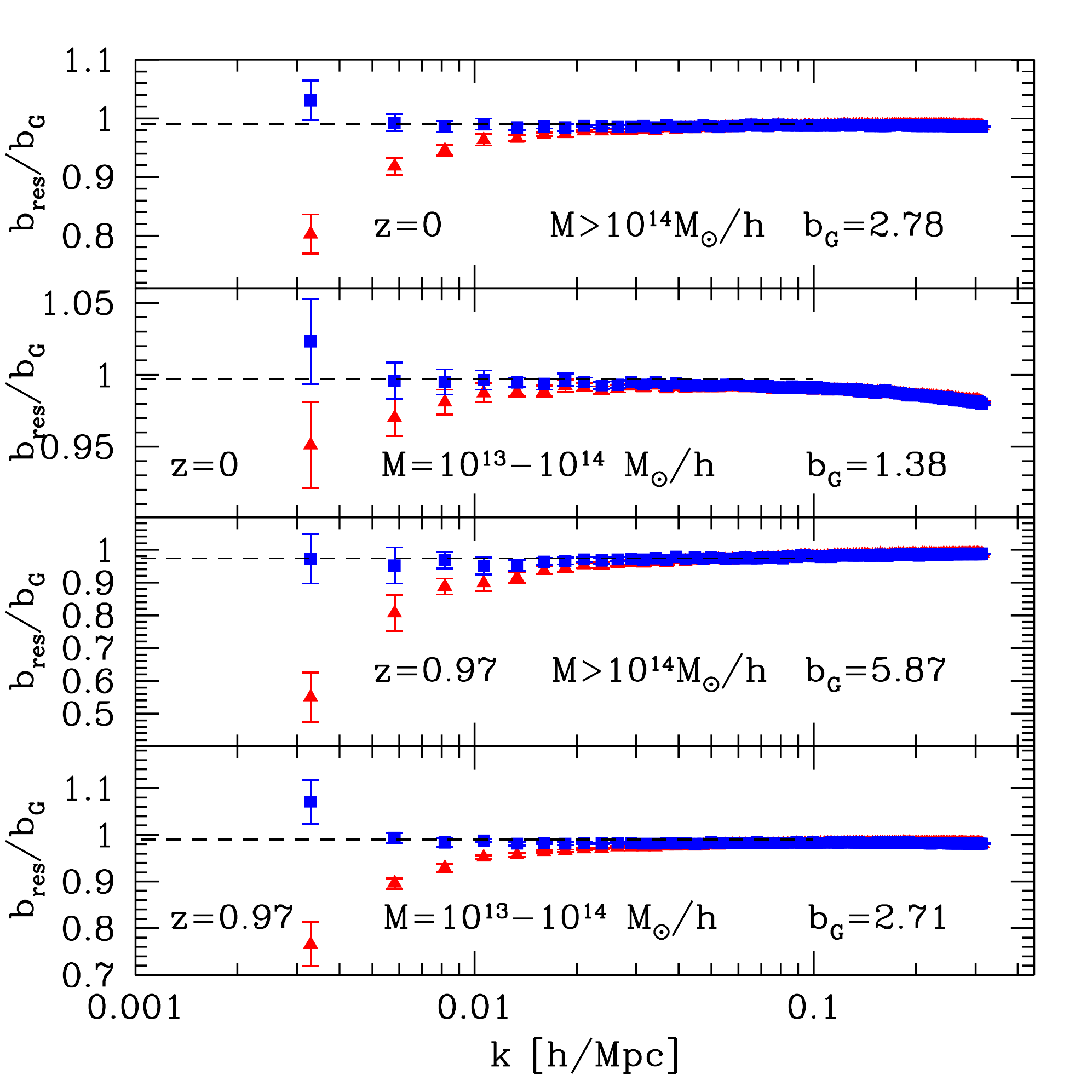} %{LocBiasHmassz0}
 \caption{The residual halo bias in local PNG with $\fnl=100$ (normalized by the bias measured in Gaussian simulations $b_G$) after the theoretical PBS scale-dependent bias is accounted for using two predictions: our result (Eq.~\ref{DeltaB}, blue squares), and the standard prediction (Eq.~\ref{Db1Lstd}, red triangles). This is for FOF$0.2$ halos and different halo masses and redshifts.  Our predictions are consistent with scale-independent residuals, while the standard prediction is not, more so for large-bias objects. The dashed lines show the expected ratio of scale-independent biases assuming universality plus Markovianity.}
 \label{BiasFOF0p2L}
 \end{figure} 

 \begin{figure}[!t]
 \centering
 \includegraphics[ width=0.95\linewidth]{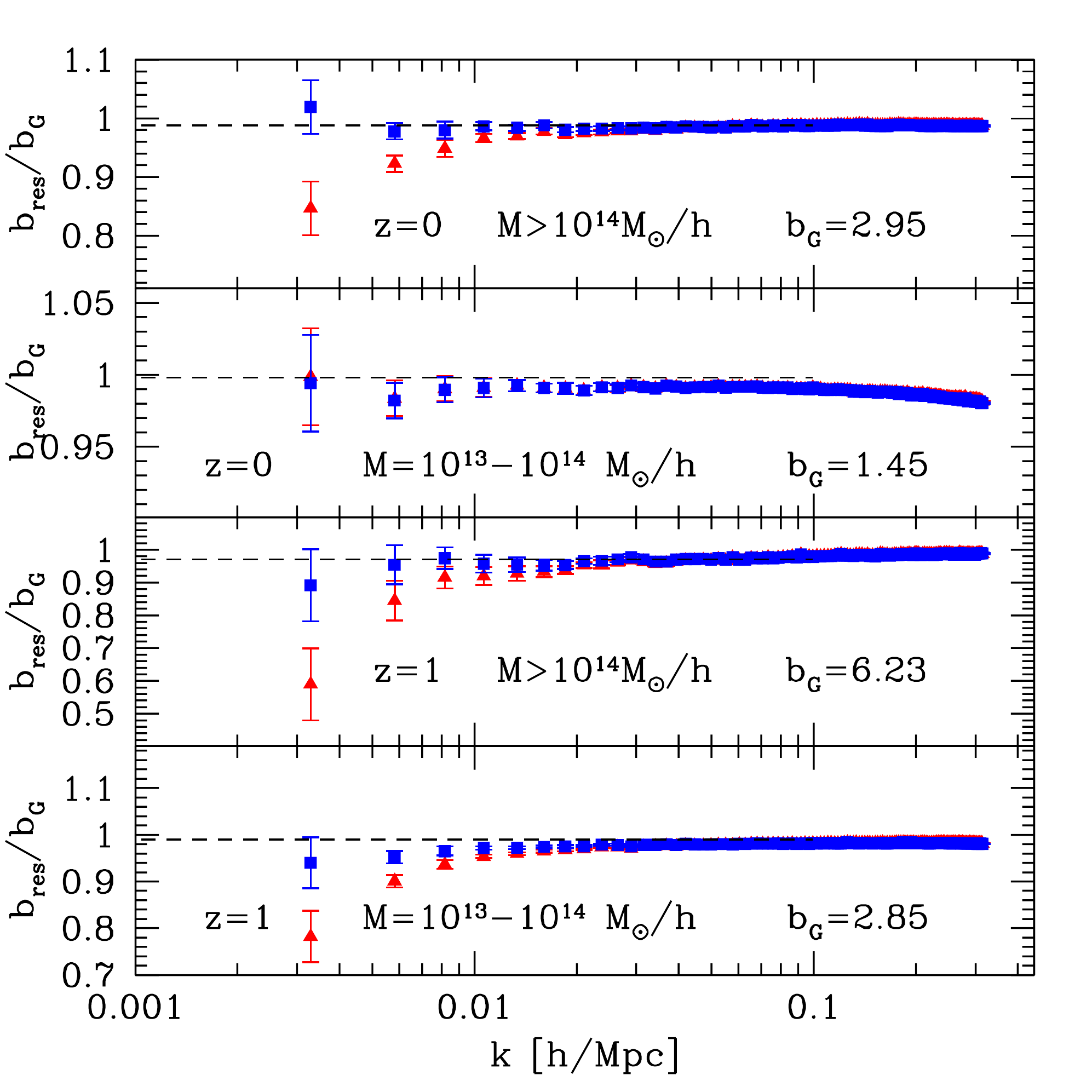} 
 \caption{Same as Fig.~\ref{BiasFOF0p2L} but for FOF$0.156$ halos. Similar results hold despite a very different halo definition.}
 \label{BiasFOF0p156L}
 \end{figure} 

To calculate our predictions for $\Delta b^{\rm theory}_{1}$, given by Eq.~(\ref{DeltaB}), a mass derivative of the Gaussian mass function is required. We  implement this by doing,
\beq
\Delta b_1 (k) = { \sum_i [I_{21}(k,m_i)N_h(i) (d\sigma^2_m/dm_i)^{-1} ]' \over M(k)\, N_h^{\rm tot}},
\label{implem}
\eeq
where $N_h(i)$ denote the number of halos in the Gaussian realizations in a bin of constant $d \ln m$, $N_h^{\rm tot}=\sum_i N_h(i)$, and the sum is over the mass bins belonging to the halo sample. The numerical derivative (denoted by a prime) is taken by doing central differences from neighboring bins. Care must be taken at low mass to have a smooth mass function, when the number of particles in a halo is smaller and binning effects  can induce artificial noise.

In Figure~\ref{BiasFOF0p2L}, we show $b_{\rm res}$ for FOF halos with linking length equal to 0.2 times the mean interparticle separation with local PNG with $\fnl=100$ normalized by the halo bias measured in our Gaussian simulations $b_{\rm G}$ (as labeled in each panel),  for different halo masses and redshifts. The symbols with error bars show $b_{\rm res}/b_{\rm G}$ for two different theoretical models, our prediction Eq.~(\ref{DeltaB}) shown by blue squares, and the standard prediction Eq.~(\ref{Db1Lstd}) denoted by red triangles. We see that  our  prediction for the scale-dependent bias performs better, as the residuals are consistent with scale-independence, whereas the standard prediction is not. The latter over-predicts the amplitude of the scale-dependent bias, as a result the residual bias $b_{\rm res}$ is suppressed at low-$k$. Figure~\ref{BiasFOF0p156L} shows the analogous results for FOF halos obtained from a linking length 0.156 times the mean interparticle separation, and shows a similar overprediction of the scale dependence by the standard formula. The magnitude of this deviation is somewhat larger for FOF$0.2$ halos, thus the details depend on halo definition. This is in qualitative agreement with previous studies that required a ``fudge factor" less than unity (typically $q\simeq 0.75$) on top of the standard prediction~\cite{2009MNRAS.398..321G,2010MNRAS.402..191P,2009MNRAS.tmp..631D,2010JCAP...07..002N}.

From the constancy of the ratio $b_{\rm res}/b_{\rm G}$ for our theoretical prediction (square symbols) we can read off that there is a PNG correction to scale-independent bias. The sign of the magnitude is expected as for local PNG with positive $\fnl$ the halo mass function is enhanced and the scale-independent bias is thus suppressed compared to the Gaussian $\fnl=0$ case. To be more specific, we show using dashed lines in Figs.~\ref{BiasFOF0p2L} and~\ref{BiasFOF0p156L} the expected {\em scale-independent} correction to halo bias assuming universality plus Markovianity, that is (see Eq.~\ref{b1LmarkovUniv})

\beq
\delta b_{1L}^{(1)} \equiv b_{1L}^{(1)}|_{\rm PNG} - b_{1L}^{(1)}|_{\rm G}
\label{dbPNG}
\eeq
where
\beq
b_{1L}^{(1)}|_{\rm PNG} = \Big({2\over \delta_c}\Big) \Big({d \ln \sigma^2 \over dm} \Big)^{-1} \partial_m \ln 
\Big( {
dn \over d \ln m}
\Big)_{\rm PNG}
\label{DbSI}
\eeq
and similarly for the Gaussian case~\cite{2008JCAP...08..031S,2008PhRvD..78l3507A,2009MNRAS.tmp..631D,2010MNRAS.402..191P,2010PhRvD..81f3530G}. Because this is for fixed mass, we integrate each expression for the bias at fixed $m$ weighted by the corresponding mass function over the desired mass bin. We see from Figs.~\ref{BiasFOF0p2L} and~\ref{BiasFOF0p156L}  that these predictions, for a wide set of halos (note the range in Gaussian bias parameters from 1.38 to 6.23) match rather well the residual bias from our theoretical prediction, although there are certainly deviations at the percent level.  This fact, together with the flatness of the residual bias as a function scale, tells us that  our improved treatment leads to a better description of the amplitude of scale-dependence in local PNG. 

In~\cite{2010CQGra..27l4011D} it is found that spherical overdensity (SO) halos obey the standard formula for scale-dependent bias more closely than FOF halos. Naively, since SO halos violate universality more strongly than FOF halos~\cite{2008ApJ...688..709T}, one would have expected the opposite (particularly at low mass, where the deviations from PBS bias plus Markovian and universality are stronger~\cite{2010ApJ...724..878T}). We don't currently have SO halos for the simulations we present here, but would be interesting to check our improved theoretical prediction against SO halos.

 \begin{figure}[!t]
 \centering
 \includegraphics[ width=0.95\linewidth]{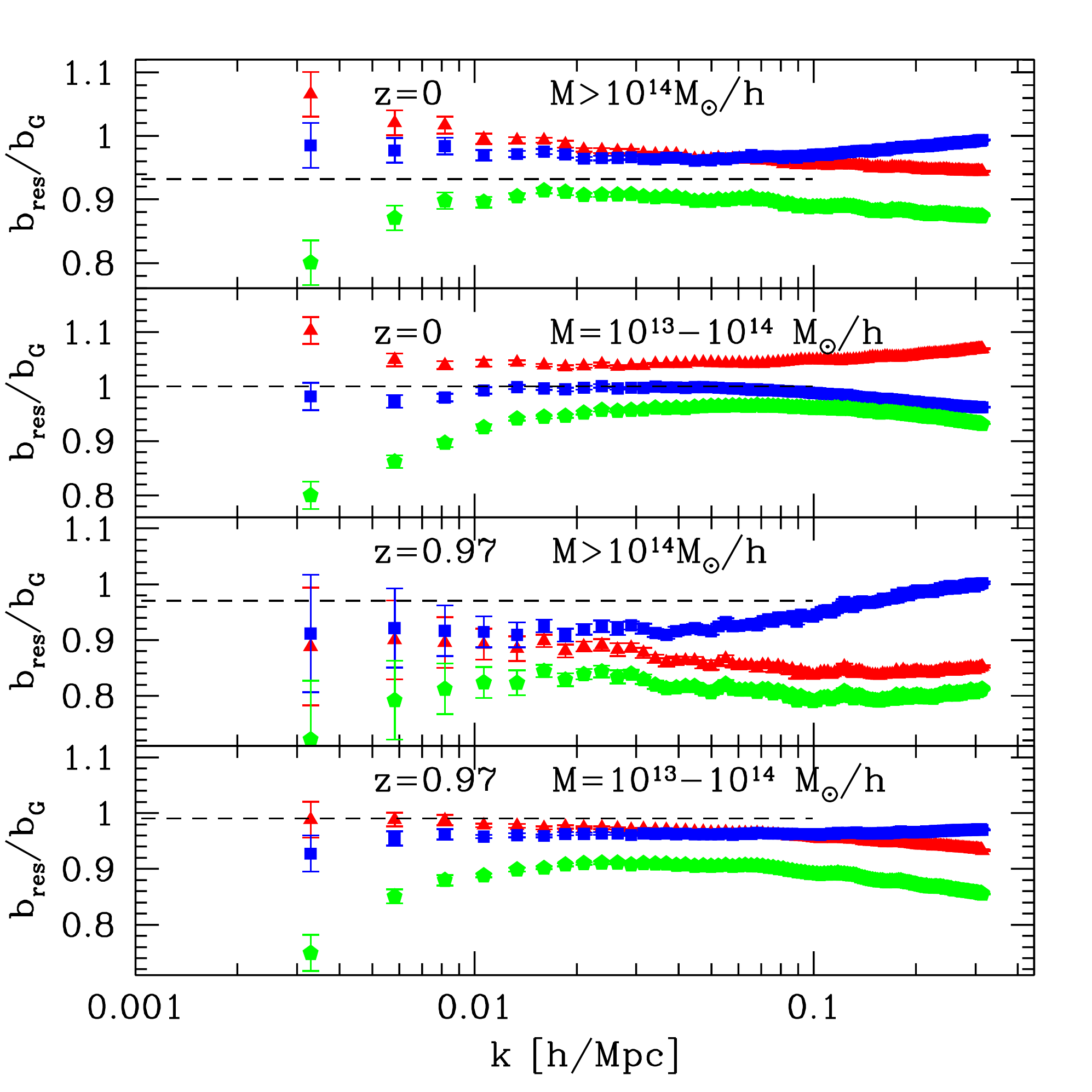} %{OrtBiasHmass0p2z0}
 \caption{Residual halo bias for orthogonal PNG with $\fnl=-400$ after our predictions for scale-dependent bias are included, Eq.~(\ref{DeltaB}), for FOF$0.156$ halos. The blue square symbols show our predictions, red triangles the predictions of Eq.~(\ref{SKcomp}), and green pentagons its first term only.  
 For our predictions, residuals are consistent with scale-independence at large scales for different halo masses, redshift and halo definitions.}
 \label{HMz0ORT}
 \end{figure} 

 \begin{figure}[!t]
 \centering
 \includegraphics[ width=0.95\linewidth]{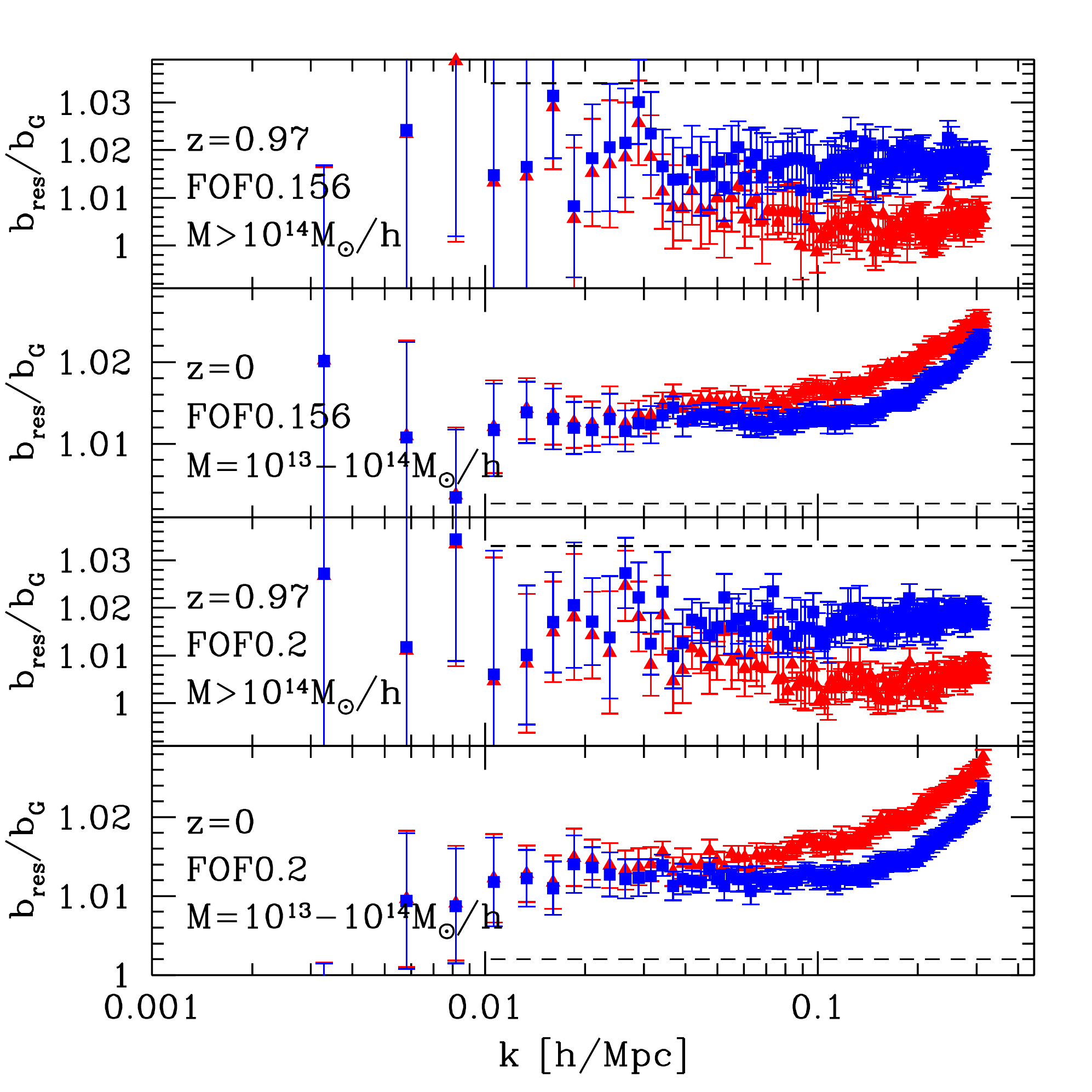} 
 \caption{Residual halo bias for equilateral PNG with $\fnl=-400$ after  predictions  for scale-dependent bias are included (blue squares), versus assuming no scale-dependent contributions as in the standard predictions (red triangles).  The different panels show two halo definitions, redshift and halo masses. Note the enhanced vertical scale in this figure.}
 \label{BiasEq}
 \end{figure}

Figure~\ref{HMz0ORT} shows the analogous results for the orthogonal template with $\fnl=-400$. Our prediction for the amplitude of the $k^{-1}$ bias  (blue squares)  leads again to a residual bias consistent with scale-independence at low-$k$. Note  from Fig.~\ref{plot_crossbias_many} that in our low-mass bin at $z=0$  (bottom panel) the scale-dependent bias changes sign, and our predictions correctly match this (second panel from top in Fig.~\ref{HMz0ORT}). At high-$k$, as nonlinear scales begin to be probed around $k\simeq 0.1 \kvecMpc$ there is significant evidence for scale-dependent non-Gaussian contributions unlike the local PNG case shown in Fig.~\ref{BiasFOF0p156L}. This must be due to the larger value of $\fnl$ in the orthogonal case ($\fnl=-400$ versus $\fnl=100$ in the local case).  We also show two other predictions, assuming Markovianity and universality (red triangles) which gives rise to Eq.~(\ref{SKcomp}) (as in~\cite{2011arXiv1105.3476D,2011arXiv1105.3628D}) and in green pentagons its first term only (corresponding to the predictions in~\cite{2010PhRvD..82j3002S}). While this is for FOF$0.156$ halos, we find very similar results for FOF$0.2$ halos. We conclude that our improved formula performs best compared to the alternatives. Note in this case that the residual bias predicted by Eq.~(\ref{DbSI}) (shown as dashed lines in Fig.~\ref{HMz0ORT}) shows larger deviations than for the local case. This might be due to  non-Markovian corrections proportional to $\fnl$~\cite{2010ApJ...717..526M} that are not included in Eq.~(\ref{DbSI}). 

	Finally, in Fig.~\ref{BiasEq} we present residual halo bias results for equilateral PNG for FOF$0.156$ halos. In this case we compare our prediction for residual bias (blue squares) based on subtracting the {\em scale-dependent} term given by the bottom panel in Fig.~\ref{IBnonloc} (and Eq.~\ref{I21eq} in the low-$k$ limit) and without substraction (red triangles) which correspond to the standard prediction (that includes only scale-independent corrections) and also the PBS prediction when the constraint is done on the $\Phi_\ell$ field (see Section~\ref{ConstPhi}). 
	
	We see that the differences are small, although the measurements are slightly more consistent with scale-independent residual bias for  the prediction based on the cross-bispectrum $B_{\Phi\Phi\phi}$ rather than $B_{\Phi}$, but we don't consider this statistically significant. Note that the sign of the scale-dependent effect in this case depends on halo mass (negative for high mass and positive for our low mass bin), and for high-mass at $z=0.97$ (with Gaussian bias $b_G\simeq 6$) the effect is only about 2\%,  thus for all practical purposes not very important.

\section{On local bias vs PBS, loops and effective theory of bias}
\label{PBSvsLOC}

\begin{figure*}[t!]
%\begin{center}
\centering
\includegraphics[ width=0.475\linewidth]{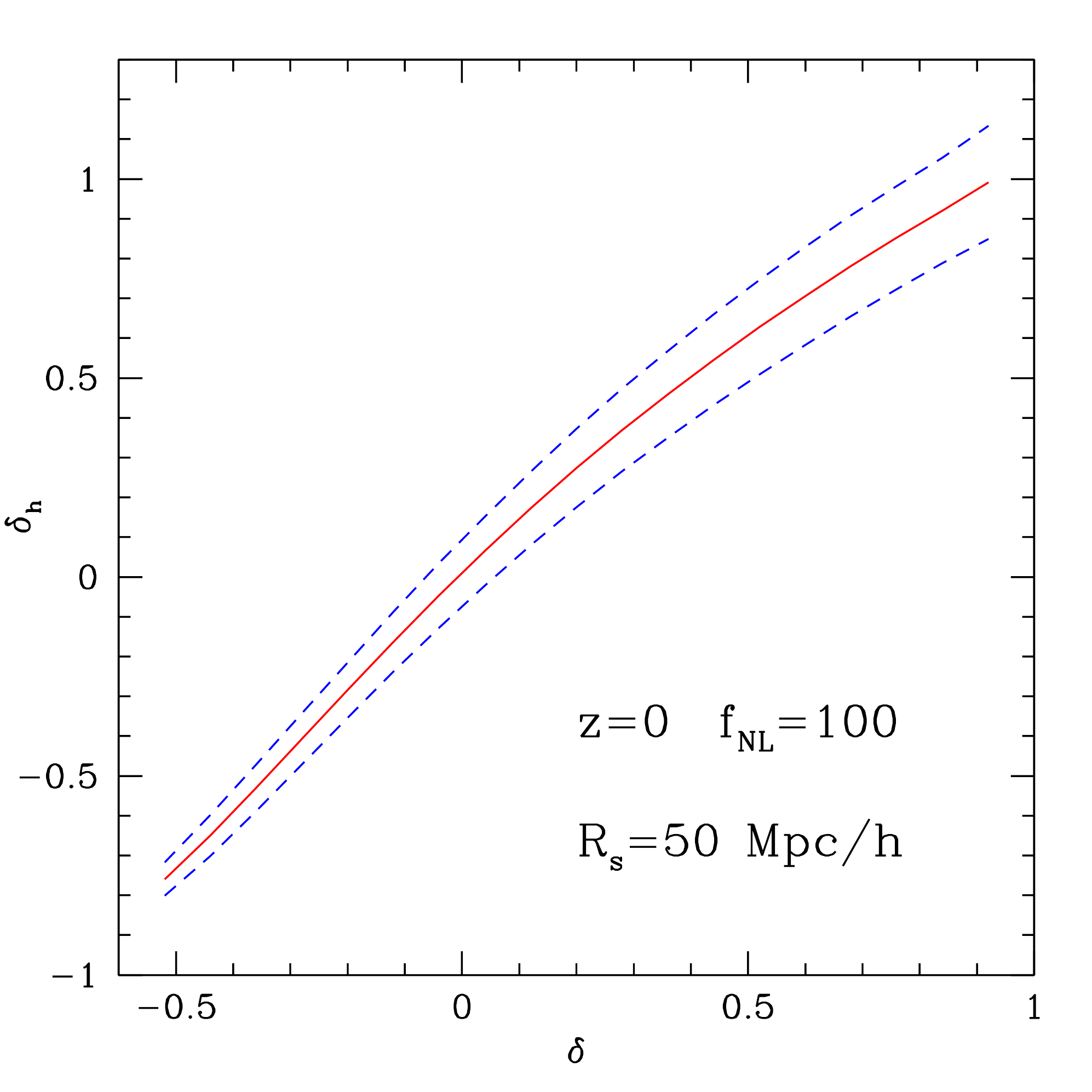}
\includegraphics[ width=0.475\linewidth]{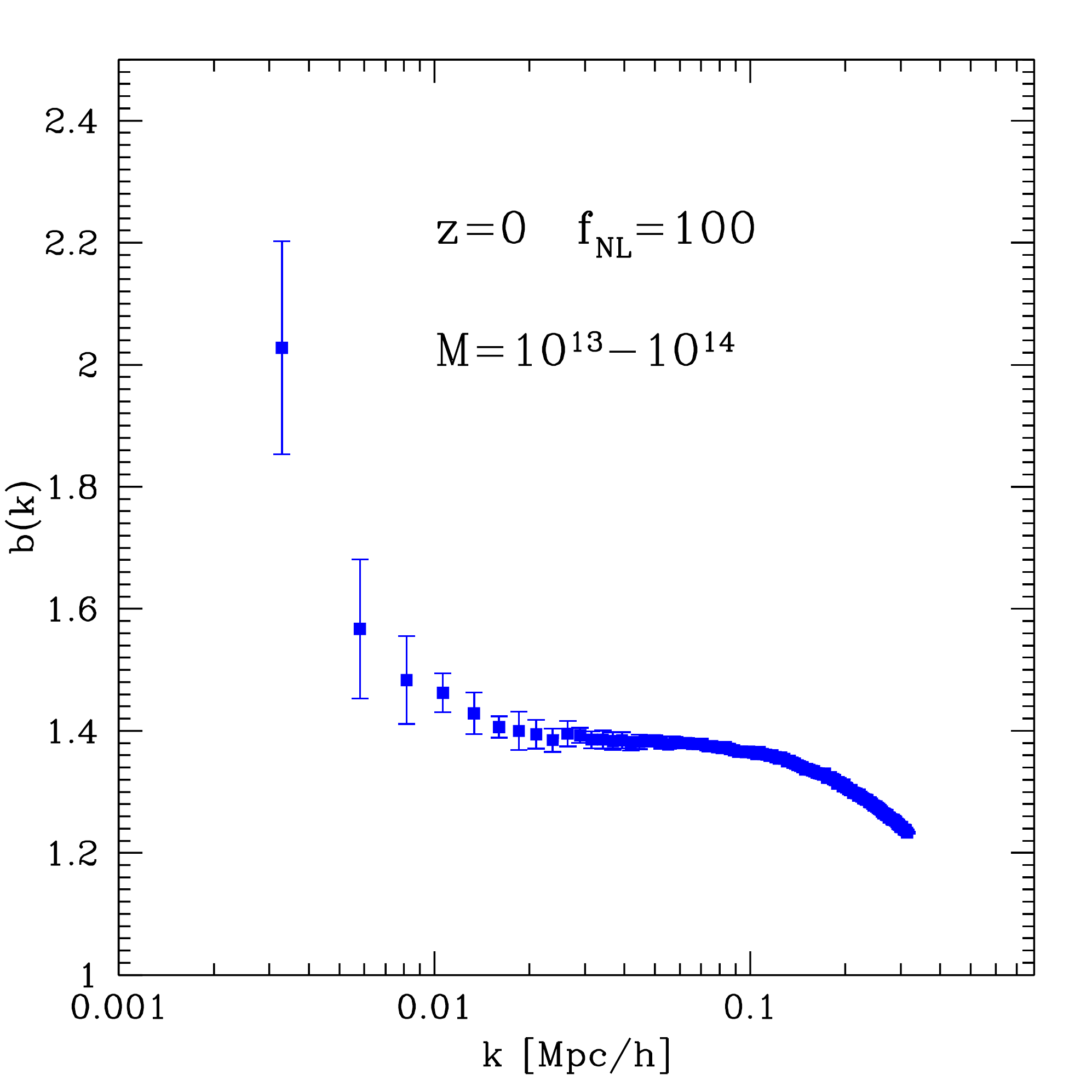}
\caption{{\em Left panel: }The $\delta_h$ vs $\delta$ relationship for local $\fnl=100$ for halos of mass in the $10^{13}-10^{14}$ $M_\odot/h$ range. The mean (solid line) is reasonably described by a local relation with bias parameters $b_1=1.34$ and $b_2=-0.58$. The standard deviation of $\delta_h$ given $\delta$ is shown by dashed lines. {\em Right panel: } The scale-dependence of the halo bias factor. Despite having $b_2<0$, these halos show a scale-dependent {\em enhancement} of their power spectrum at low-$k$, rather than the suppression predicted by the local bias model.}
\label{LocBiasNo}
%\end{center}
\end{figure*}

Let us now concentrate in this section in the simplest case, local PNG, where most results in the literature have been obtained. We will discuss first the large-scale behavior (scale-dependent bias at low-$k$) which arises from loop corrections in local bias,  and how this compares with simulations. We then discuss 
 what happens with bias towards the non-linear regime in the context of loops in the PBS, which provides a nice example of a formulation of an effective theory of bias that can be applied to the more complicated case of galaxies.

In Eulerian local bias model~\cite{FryGaz9308} the {\em smoothed} halo perturbation $\widehat{\delta}_h(\x)$ is assumed to be  a local function of the {\em smoothed} dark matter perturbation  
\beq
\widehat{\delta}_h(\x) \equiv f[\widehat{\delta}(\x)]
\label{TrueLoc}
\eeq
Note that both fields are smoothed on the same scale, say $R_s$, which is the scale one is interested in measuring things at, and must be large enough compared to the Lagrangian size of halos for local bias to make sense at all (operationally the function $f$ can be obtained from simulations by constructing a scatter plot such as the one presented in Fig.~\ref{LocBiasNo} below). At large smoothing scales, density perturbations are small so one can write a perturbative version of this
\beq
\widehat{\delta}_h = b_1\, \widehat{\delta} + {b_2\over 2}\, \widehat{\delta}^2 + \ldots
\label{FGexp}
\eeq
where we have suppressed a constant term $b_0$ that enforces $\langle \widehat{\delta}_h \rangle=0$. This leads to a halo-matter power spectrum,
\beq
P_{hm}(k)=b_1 P(k) + {b_2 \over 2} \int B(\k-\q,\q) \,d^3 q + \ldots
\label{PhmLB}
\eeq
where $B$ denotes the bispectrum of the density field, and we have deliberately left the UV cutoff in the loop momentum unspecified. Interestingly, as noted in~\cite{2008PhRvD..78l3534T} the primordial contribution to the density bispectrum gives rise to a $k^{-2}$ scale-dependent bias proportional to $\fnl$, i.e.
\beq
\Delta b_1 = {2\fnl \over M(k)} \, b_2 \sigma^2
\label{Db1LE}
\eeq
where $\sigma^2$ is the variance of the  smoothed density field. In such an approach, the sign of $\Delta b_1$ is precisely given by that of $b_2$. This result seems qualitatively similar to Eq.~(\ref{Db1Lstd}), but the amplitude depends on $b_2$ and $\sigma^2$ instead. The former, when measured from a scatter plot between $\delta_h$ and $\delta$, does not depend sensitively on the smoothing scale $R_s$ (for $R_s\ga 30 \Mpc$~\cite{2007PhRvD..75f3512S,2009arXiv0912.0446M}), but of course $\sigma^2$ is very sensitive to $R_s$, and thus the magnitude of this effect is uncertain\footnote{If we had followed the logic of Eq.~(\ref{FGexp}) the smoothing scale would be related to the observation scale $k$, as in Eq.~(\ref{TrueLoc}).  See later in this section for more discussion on this.}. However, {\em if} we take the rare event limit, corresponding to  halos (or peaks) of very large mass, {\em and} we take the smoothing length to be the Lagrangian radius of the halo $R_L$, then $ b_2 \sigma^2 \approx b_{1L}^{(1)}$ and Eq.(\ref{Db1LE}) agrees with Eq.~(\ref{Db1Lstd})~\cite{2009PhRvD..80l3002S}. Note however that this requires, effectively, to have taken the UV cutoff in the loop in Eq.~(\ref{PhmLB}) to be close to $R_L^{-1}$ where local bias is a dubious assumption, and even if it were to hold including such modes in $\widehat{\delta}$ would likely invalidate the perturbative expansion in Eq.~(\ref{FGexp}). 

It is also interesting that for halos away from the high-mass limit, where $b_2$ becomes negative, Eq.(\ref{Db1LE}) predicts that the halo power spectrum should be suppressed rather than enhanced (for $\fnl>0$) at low-$k$.  In Fig.~\ref{LocBiasNo} we test this prediction for FOF$0.2$ halos of mass in the $10^{13}-10^{14}$ $M_\odot/h$ range at $z=0$. The left panel shows a scatter plot of the halo versus matter perturbation smoothed on $R_s=50 \Mpc$ scales, showing that these halos have $b_2<0$ due to the negative curvature of the solid line, which  denotes the mean of the relation. The right panel shows however that the bias (defined from the cross spectrum) is enhanced rather than suppressed. Note that another way to see that $b_2<0$ for these halos is  that $b(k)$ decreases with scale for $k\ga 0.1 \kvecMpc$, as expected since for Gaussian initial conditions  the sign of $b_2$ controls the behavior of the scale-dependent bias at nonlinear scales~\cite{2007PhRvD..75f3512S}. One may argue that $b_2$ calculated at $R_s=50 \Mpc$ is not fair, since as we discussed Eq.~(\ref{Db1LE}) approximately agrees with the PBS result only when $\sigma^2$ is taken to be smoothed at the Lagrangian radius of the halo. However, evaluating $b_2$ at such small scale does not change the conclusions, as $b_2$ is driven to small values (but still negative) as the smoothing length decreases~\cite{2009arXiv0912.0446M}. In addition, one can see from scatter plots at such smoothing scales that local bias is not really a good representation of the data. As we discuss below, the PBS tells us that we shouldn't push the smoothing scale to the Lagrangian scale of halos as the perturbative expansion breaks down.

A generalization of Eq.~(\ref{Db1LE})  is available in other flavors of local biasing. In~\cite{1986ApJ...310...19G}, a general formula is given for biasing of exponential form, which reproduces Eq.~(\ref{Db1LE}) to lowest order. Similarly, in local Lagrangian bias, \cite{1986ApJ...310L..21M} derive the same formula as~\cite{1986ApJ...310...19G}, and Eq.~(\ref{Db1LE}) follows with $b_2$ replaced by its Lagrangian version $b_{2L}^{(1)}$ for clustering in Lagrangian space (see e.g.~\cite{2010PhRvD..82j3002S} for recent discussion on this). For high-peaks, the Lagrangian to Eulerian mapping does not matter (it is subleading in $\nu^2$), but for the halos  considered in Fig.~\ref{LocBiasNo} the contributions from the mapping {\em are not} negligible. If the mapping is done in the usual spherical approximation, then the same Eq.~(\ref{Db1LE}) follows for the Eulerian (as measured in the simulations) scale-dependent bias where $b_{2}=b_{2L}^{(1)}+(8/21) b_{1L}^{(1)}$, which means that $b_{2L}^{(1)}$ is even more negative than the measured $b_2$. In any case, the conclusion is that these local bias models (at least in their perturbative form) do not match the halo clustering behavior seen in Fig.~\ref{LocBiasNo} where the (Eulerian linear) bias is enhanced at low-$k$ while the quadratic Eulerian bias is negative. It would be interesting to do a more rigorous peak-type calculation that goes beyond a perturbative expansion to see whether the predictions differ from a quadratic bias effect for low bias tracers.

Another proposal for dealing with bias in PNG models is presented in~\cite{2008PhRvD..78l3519M}, based on earlier work for Gaussian initial conditions~\cite{2006PhRvD..74j3512M}. In this case the starting point is the Eulerian local bias expansion in Eq.~(\ref{FGexp}), but rather than being restricted to smoothed fields, $\delta$ is formally assumed to be unsmoothed, leading to a scale-dependent bias at low-$k$ for local PNG  from Eq.~(\ref{Db1LE}) that is formally ultraviolet (UV) divergent. To cure this divergence, a new (non-local in density) term proportional to $\phi$ with arbitrary amplitude is added to Eq.~(\ref{FGexp}). Since this has the same effect in the low-$k$ power spectrum (leading to a $k^{-2}$ bias) one can tune its amplitude to cancel the UV divergence from Eq.~(\ref{Db1LE}). In this way, the model is left with a ``renormalized" finite amplitude of scale-dependent bias that can be fit to simulations. In a sense, in this approach the UV sensitivities are used as a guide to select the appropriate modification of the local bias model for a given PNG type, and after the renormalization procedure is complete the theory is no longer UV sensitive. By contrast, in the PBS calculation we presented in this paper, the effective {\em large-scale} degrees of freedom appear naturally after implementing the split and there are UV sensitivities left (although properly suppressed by inverse powers of the UV cutoff of the theory, see below), which alert us to when the perturbative expansion breaks down.

In the PBS, there is also a quadratic bias parameter (see Section~\ref{QuadB}). Does this mean that one gets, in addition to the already discussed PBS results,  {\em another} contribution to scale-dependent bias from the effect leading to Eq.~(\ref{Db1LE})? The answer is `yes', but since the expansion in the PBS is in the {\em large-scale} modes $\delta_\ell$ (with $\sigma_\ell^2 \ll \sigma_m^2$) the contributions from quadratic bias are highly suppressed. That is, the contribution from the effect leading to Eq.~(\ref{Db1LE}) reads in PBS (compare to Eq.~\ref{Db1Lstd})
\beqa
\Delta b_1 & \supset & {2\fnl \over M(k)} \, b_2 \sigma_\ell^2 = {2\fnl \over M(k)} \, [(b_{2L}^{(1)}+(8/21) b_{1L}^{(1)}] \, \sigma_\ell^2
\nonumber \\ &
\stackrel{\nu\gg 1}{\approx} & {2\fnl \over M(k)} \, \delta_c\, b_{1L}^{(1)} \, \Big({\sigma_\ell^2 \over \sigma^2_m}\Big)
\label{Db1fromb2PBS}
\eeqa
where for simplicity in the last equality we have assumed the high-peak limit. Therefore, we see that as long as $\sigma_\ell^2 \ll \sigma_m^2$ these extra corrections are well under control. Since this correction arises as a loop integration over the momenta of large-scale modes, here is where the details of the split enter. In our formulation, we take the split right at the scale $k$ we are computing the power spectrum (i.e. we choose a sharp-$k$ filter, as explained in Eqs.~\ref{SharpPhi}-\ref{Sharp}), thus $\sigma_\ell^2$ is $k$-dependent (see Eq.~\ref{sigmaell}).  
 For example, at $k=0.01 \kvecMpc$, for a $10^{14}$ ($10^{15}$) $M_\odot/h$ halo, $(\sigma_\ell^2 /\sigma_m^2)$ is 0.0005 (0.0015).  At $k=0.03 \kvecMpc$, where the scale-dependent contributions are already subleading (see Fig.~\ref{plot_crossbias_many}), the large-scale to small-scale variance ratios become 0.015 and 0.045, respectively. That loops lead to $\sigma_\ell^2(k)$ factors was already used in ~\cite{2000ApJ...542....1S} to implement the perturbative bias expansion for PNG models of $\chi^2$ type.
 
 This shows that the PBS calculation we have presented before is self-consistent, i.e. insensitive to the inclusion of non-linear bias loops (they vanish as $k\rightarrow 0$). In fact, as $(\sigma_\ell^2 /\sigma_m^2)$ becomes non-negligible, one should go back and include such corrections in  Eq.~(\ref{Piapprox1}),  which correspond to changing $\sigma^2_m \rightarrow \sigma_m^2-\sigma_\ell^2$ for a sharp-$k$ filter (see Eq.~\ref{sigma-phi}), well-known from conditional mass functions~\cite{1991ApJ...379..440B,1993MNRAS.262..627L}. In this regime, however, one is trying to study bias at scales comparable to the Lagrangian size of the halo, thus  finite-size effects become important and bias ceases to be local (in the UV sense, i.e. terms with positive powers of $k$ show up). Indeed as $k$ approaches the inverse Lagrangian size $R_L$ of halos, $(\sigma_\ell^2 /\sigma_m^2) \propto (kR_L)^{n+3}$, where $n$ is the effective spectral index at $k$.  At these scales, the perturbative expansion of PBS breaks down, and it is replaced by a non-perturbative, non-local description  with exclusion effects that  drive the halo autocorrelation function to $-1$, see~\cite{1999MNRAS.304..767S} (physically, what happens is that since $\delta_\ell$ is approaching $\delta_c$ the region of interest will correspond to a halo, thus proper account of the `cloud in cloud' problem is essential). In the case of galaxies, it is at these scales where the non-local bias due to halo profiles (in the language of the halo model) takes over. 
 
In~\cite{2010PhRvD..81f3530G} a bivariate expansion of the halo bias in terms of the {\em smoothed} $\phi$ and $\delta$ is developed and it is argued that the quadratic effect must be small based on choosing a smoothing scale $R=10 \Mpc$. As we mentioned in Section~\ref{modbias} this type of approach arises from using a real-space top-hat filter to make the split between low-$k$ and high-$k$ modes, which is more appropriate to calculate counts-in-cells statistics than Fourier space correlators. Our conclusion here is similar, but with a key difference: rather than choose an arbitrary {\em fixed} smoothing scale $R$, we instead use a {\em varying splitting scale} between small and large-scale modes that follows the $k$-mode we are interested in. Since the Lagrangian halo bias parameters can be written as $b_{nL} = {\cal B}_n(\nu)/\sigma_m^n$ for some dimensionless ${\cal B}_n(\nu)$~\cite{ScoSheHui01} the perturbative expansion of PBS reads (here we ignore terms depending on $\phi_\ell$ since we are interested in the UV limit),
\beq
\delta_h^L = \sum_{n=1} b_{nL}\, \delta_\ell^n = \sum_{n=1} {\cal B}_n(\nu) \, \Big({\delta_\ell \over \sigma_m}\Big)^n
\label{PBSpert}
\eeq
that is, all loop corrections coming from nonlinear bias (such as Eq.~\ref{Db1fromb2PBS}) will be suppressed by ratios of $\sigma^2_\ell(k)/\sigma^2_m$, and thus the PBS perturbative expansion is well-behaved as long as we study scales larger than the Lagrangian size of the halos (which acts as the UV cutoff of the theory). In fact the resulting expansion is like in effective field theory, where the corrections are given by powers of the scale $k$ over the cutoff $R_L^{-1}$. 

A different view on loops in PBS is presented in~\cite{2011JCAP...04..006B}, where they argue that the PBS bias parameters correspond to the renormalized bias parameters of~\cite{2008PhRvD..78l3519M,2006PhRvD..74j3512M} and thus when computing loops one should discard terms which are UV sensitive. In this proposal, the running due to loops controlled by $({\sigma_\ell^2(k) / \sigma_m^2})$ is absent, so this makes distinct predictions from our approach. 
Detailed calculations of correlation functions including loops and comparison  against simulations will be presented elsewhere.

\section{Conclusions}

In this paper we have studied two major issues, the generation of initial conditions in N-body simulations with primordial non-Gaussian (PNG) non-local models motivated by inflation and the peak-background split (PBS) method for predicting the scale-dependent bias in these models. The main results in each case are,

\begin{enumerate}

\item[i)] using as an example the factorizable templates motivated by inflationary bispectra, we showed how to construct initial conditions for nonlocal PNG models in an efficient way. Our algorithm is only slightly slower for nonlocal PNG  than for the local case, and scales with particle number in the same way ($N_{\rm par} \ln N_{\rm par}$) as    Gaussian initial conditions. The resulting implementation is given in Eqs.~(\ref{KEQ}-\ref{KORT}) for equilateral and orthogonal templates. Appendix~\ref{NSI} generalizes these results beyond scale-invariance and Appendix~\ref{ICsSDfnl} gives the general prescription for more general templates (including scale-dependent $\fnl$ models) and non-local $\gnl$ PNG.

\item[ii)] we derived predictions for the scale-dependent bias from the PBS in the context of the excursion-set approach to halos for general, non-local, PNG to describe scale-dependent bias. Our main result for the scale-dependent bias contribution in generic $\fnl$ PNG is given by Eq.~(\ref{DeltaB}), which generalizes previous results in the literature to go beyond the assumption of universality of the mass function, Markovian evolution of the random walk, and includes beyond leading-order effects due to the transfer function, filter effects and gradients. Our results, which only require an accurate knowledge of the {\em Gaussian} mass function, show improved agreement against numerical simulations when compared to the predictions previously available in the literature (see Figs.~\ref{BiasFOF0p2L}-\ref{BiasEq}). We also present, for the first time, results for the quadratic bias parameters for non-local PNG.

\end{enumerate}

Let us now expand on each of these  in more detail. 

Regarding the generation of initial conditions, we showed that there are multiple solutions for the non-local kernel given a desired bispectrum, but our general method does lead to a (sum of) factorizable kernel for a given factorizable bispectrum, preserving this advantage for a fast numerical generation of initial conditions. We discussed in detail how to restrict the linear combination of solutions using the one-loop corrections to the power spectrum to preserve the large-scale limit given by  primordial correlations. Even after these restrictions, there is still a one-parameter family of solutions for the equilateral and orthogonal templates. This lack of uniqueness may even show up in the squeezed limit (see Appendix~\ref{sqz}). This freedom is actually welcomed, as it should help reproduce the correct ``snake topology" (proportional to  $\tnl$) of the trispectrum  in these models, something that is left for future work.

On the other hand, we stressed that while the freedom may be removed by choosing the reduced bispectrum as a kernel~\cite{2010PhRvD..82j3002S}, doing so imposes a restrictive symmetry (under permutations of $k_1,k_2,k_3$) that is not clearly required, it imposes a non-trivial constraint between the power spectrum, bispectrum and snake amplitude of the four-point function that is unlikely to hold, and making the kernel non-factorizable  complicates substantially the generation of initial conditions.

We then used our algorithm to generate initial conditions to run a large suite of numerical simulations (12 realizations of $1280^3$ particles in a $2.4 \Gpc$ box) with PNG initial conditions of local, equilateral and orthogonal type and study the scale-dependent bias of dark matter halos.

Concerning the implementation of PBS for generic PNG, we find that our predictions for the amplitude of the scale dependence are in excellent agreement with the measurements in our numerical simulations for a variety of halo masses, redshifts and halo definitions. Our prediction for $\fnl$ PNG, Eq.~(\ref{DeltaB}), is valid for {\em any non-local} PNG and is simple to implement (requiring only knowing the mass function for Gaussian simulations with the same cosmology) and we demonstrated it to be more accurate than previous  results in the literature that assume universality (Eq.~\ref{Univ}), Markovian evolution (Eq.~\ref{Piapprox2}) and the large-scale limit ($k\rightarrow 0$). Therefore, with our improved predictions, there is no need to include any fudge factors to describe scale-dependent bias, and the result can be written for $\fnl$ PNG in terms of the Gaussian mass function independent of the dynamics assumed for halo formation. 
In addition, we extended our PBS formalism to deal with non-local $\gnl$ PNG.

We also stressed that in doing the PBS the choice of the split variable (the Gaussian $\phi$ vs the non-Gaussian $\Phi$) may lead to different theoretical predictions. For local and orthogonal models  constraining the large-scale modes using $\phi_\ell$ or $\Phi_\ell$ gives consistent answers at ${\cal O}(\fnl)$ at large scales (though they lead to sub-leading differences at small scales described by the difference between the cross bispectrum $B_{\Phi\Phi\phi}$ and $B_{\Phi\Phi\Phi}$). However,  under special circumstances (when the long-mode kernel is sufficiently singular) these two predictions may differ even  in their scalings in the squeezed limit (see Appendix~\ref{sqz}) and lead to different predictions for the scale-dependent bias at ${\cal O}(\fnl)$. However this is not surprising, since in this case $\Phi_\ell$ cannot be considered a function of just $\phi_\ell$, as  $\phi_s$-modes can give an important contribution to $\Phi_\ell$, and thus constraining the large-scale $\phi$ and $\Phi$ can differ significantly.

An example of this situation is given by our implementation of the equilateral template kernel, where a $k^{+1}$ bias is predicted in the PBS in $\phi$ while a scale-independent bias is predicted by doing the split in $\Phi$. In practice, the difference in the predictions is however very small  (2\% in the most biased samples) and thus unlikely to be important in practice and inconclusive on which approach is the preferred one. We pointed out, in addition, that the choice of the split variable can however be important for the halo bispectrum even in models where the kernels are regular (as in {\em local} PNG), as imposing the large-scale constraint in $\phi$ vs the non-Gaussian $\Phi$ leads to {\em large-scale} different predictions for  quadratic bias parameters at ${\cal O}(\fnl^2)$. This can be probed by measuring the halo bispectrum as a function of triangle shape, which we will address elsewhere.

In addition, in Section~\ref{PBSvsLOC} we showed that the prediction of perturbative local bias models that the amplitude of the scale-dependent bias is controlled by the quadratic bias $b_2$ does not agree with simulations, where halos with negative $b_2$ have an enhancement of bias at low-$k$ rather than suppression. 

Finally, we considered what happens with loop corrections due to nonlinear bias in the PBS, and showed that these are suppressed by the large- to small-scale variance $\sigma^2_\ell(k)/\sigma_m^2$, and thus they induced runnings with $k$ away from the tree-level predictions that are well under control as long as $k R_L \ll1$, where $R_L$ is the Lagrangian size of halos of mass $m$. As this scale is approached, bias ceases to be local (in UV sense) and the perturbative expansion breaks down due to the effects of halo exclusion. However, this expansion should be well-behaved when studying halos and galaxies at large-scales, which is where most of the constraining power on PNG lies. We will report on this in detail in the near future.

\vskip 1pc

While this work was being prepared for submission, the paper~\cite{RegSchShe1108} appeared in which the algorithm we present here (in the generic form given in Appendix~\ref{ICsSDfnl}) is also put forward to construct non-local PNG fields. They, however, used a different method to generate the kernel (the symmetrized kernel given by the reduced bispectrum).

\vskip 1 pc 

The parallel code using 2LPT initial conditions for non-local PNG models developed for this work is available\footnote{see {\tt http://cosmo.nyu.edu/roman/2LPT}}.

 \acknowledgements 

We thank F.~Bernardeau, P.~Creminelli, N.~Dalal, G.~d'Amico, V.~Desjacques, D.~Huterer, D.~Jeong, M.~LoVerde, P.~McDonald, C.~Porciani, F.~Schmidt, R.~Sheth, E. Sefusatti, L.~Senatore, S.~Shandera, K.~Smith, R.~Smith and M.~Zaldarriaga for comments and discussions. The simulations presented here are part of the LasDamas collaboration suite\footnote{\tt http://lss.phy.vanderbilt.edu/lasdamas} extended to non-Gaussian initial conditions and were run thanks to a Teragrid allocation and the use of RPI and NYU computing resources. The parallel 2LPT initial conditions code developed for this work is based on a Zel'dovich Gaussian initial conditions code originally written by Volker Springel. R.S. was partially supported by grants NSF AST-1109432 and NASA NNA10A171G. LH is supported by the DOE and NASA under contracts DE-FG02-92-ER40699 and NNX10AN14G, and thanks HKU and the IAS at HKUST for hospitality. MM acknowledges funds from the European Research Council. KCC acknowledges the support of James Arthur Graduate Assistantship and Mark Leslie Graduate Assistantship.

\bibliography{masterbiblio}

\begin{thebibliography}{121}%
\makeatletter
\providecommand \@ifxundefined [1]{%
 \@ifx{#1\undefined}
}%
\providecommand \@ifnum [1]{%
 \ifnum #1\expandafter \@firstoftwo
 \else \expandafter \@secondoftwo
 \fi
}%
\providecommand \@ifx [1]{%
 \ifx #1\expandafter \@firstoftwo
 \else \expandafter \@secondoftwo
 \fi
}%
\providecommand \natexlab [1]{#1}%
\providecommand \enquote  [1]{``#1''}%
\providecommand \bibnamefont  [1]{#1}%
\providecommand \bibfnamefont [1]{#1}%
\providecommand \citenamefont [1]{#1}%
\providecommand \href@noop [0]{\@secondoftwo}%
\providecommand \href [0]{\begingroup \@sanitize@url \@href}%
\providecommand \@href[1]{\@@startlink{#1}\@@href}%
\providecommand \@@href[1]{\endgroup#1\@@endlink}%
\providecommand \@sanitize@url [0]{\catcode `\\12\catcode `\$12\catcode
  `\&12\catcode `\#12\catcode `\^12\catcode `\_12\catcode `\%12\relax}%
\providecommand \@@startlink[1]{}%
\providecommand \@@endlink[0]{}%
\providecommand \url  [0]{\begingroup\@sanitize@url \@url }%
\providecommand \@url [1]{\endgroup\@href {#1}{\urlprefix }}%
\providecommand \urlprefix  [0]{URL }%
\providecommand \Eprint [0]{\href }%
\providecommand \doibase [0]{http://dx.doi.org/}%
\providecommand \selectlanguage [0]{\@gobble}%
\providecommand \bibinfo  [0]{\@secondoftwo}%
\providecommand \bibfield  [0]{\@secondoftwo}%
\providecommand \translation [1]{[#1]}%
\providecommand \BibitemOpen [0]{}%
\providecommand \bibitemStop [0]{}%
\providecommand \bibitemNoStop [0]{.\EOS\space}%
\providecommand \EOS [0]{\spacefactor3000\relax}%
\providecommand \BibitemShut  [1]{\csname bibitem#1\endcsname}%
\let\auto@bib@innerbib\@empty
%</preamble>
\bibitem [{\citenamefont {{Maldacena}}(2003)}]{2003JHEP...05..013M}%
  \BibitemOpen
  \bibfield  {author} {\bibinfo {author} {\bibfnamefont {J.}~\bibnamefont
  {{Maldacena}}},\ }\bibfield  {title} {\enquote {\bibinfo {title}
  {{Non-gaussian features of primordial fluctuations in single field
  inflationary models}},}\ }\href {\doibase 10.1088/1126-6708/2003/05/013}
  {\bibfield  {journal} {\bibinfo  {journal} {Journal of High Energy Physics}\
  }\textbf {\bibinfo {volume} {5}},\ \bibinfo {pages} {13--+} (\bibinfo {year}
  {2003})},\ \Eprint {http://arxiv.org/abs/arXiv:astro-ph/0210603}
  {arXiv:astro-ph/0210603} \BibitemShut {NoStop}%
\bibitem [{\citenamefont {{Komatsu}}\ \emph {et~al.}(2010)\citenamefont
  {{Komatsu}}, \citenamefont {{Smith}}, \citenamefont {{Dunkley}},
  \citenamefont {{Bennett}}, \citenamefont {{Gold}}, \citenamefont {{Hinshaw}},
  \citenamefont {{Jarosik}}, \citenamefont {{Larson}}, \citenamefont {{Nolta}},
  \citenamefont {{Page}}, \citenamefont {{Spergel}}, \citenamefont {{Halpern}},
  \citenamefont {{Hill}}, \citenamefont {{Kogut}}, \citenamefont {{Limon}},
  \citenamefont {{Meyer}}, \citenamefont {{Odegard}}, \citenamefont {{Tucker}},
  \citenamefont {{Weiland}}, \citenamefont {{Wollack}},\ and\ \citenamefont
  {{Wright}}}]{2010arXiv1001.4538K}%
  \BibitemOpen
  \bibfield  {author} {\bibinfo {author} {\bibfnamefont {E.}~\bibnamefont
  {{Komatsu}}}, \bibinfo {author} {\bibfnamefont {K.~M.}\ \bibnamefont
  {{Smith}}}, \bibinfo {author} {\bibfnamefont {J.}~\bibnamefont {{Dunkley}}},
  \bibinfo {author} {\bibfnamefont {C.~L.}\ \bibnamefont {{Bennett}}}, \bibinfo
  {author} {\bibfnamefont {B.}~\bibnamefont {{Gold}}}, \bibinfo {author}
  {\bibfnamefont {G.}~\bibnamefont {{Hinshaw}}}, \bibinfo {author}
  {\bibfnamefont {N.}~\bibnamefont {{Jarosik}}}, \bibinfo {author}
  {\bibfnamefont {D.}~\bibnamefont {{Larson}}}, \bibinfo {author}
  {\bibfnamefont {M.~R.}\ \bibnamefont {{Nolta}}}, \bibinfo {author}
  {\bibfnamefont {L.}~\bibnamefont {{Page}}}, \bibinfo {author} {\bibfnamefont
  {D.~N.}\ \bibnamefont {{Spergel}}}, \bibinfo {author} {\bibfnamefont
  {M.}~\bibnamefont {{Halpern}}}, \bibinfo {author} {\bibfnamefont {R.~S.}\
  \bibnamefont {{Hill}}}, \bibinfo {author} {\bibfnamefont {A.}~\bibnamefont
  {{Kogut}}}, \bibinfo {author} {\bibfnamefont {M.}~\bibnamefont {{Limon}}},
  \bibinfo {author} {\bibfnamefont {S.~S.}\ \bibnamefont {{Meyer}}}, \bibinfo
  {author} {\bibfnamefont {N.}~\bibnamefont {{Odegard}}}, \bibinfo {author}
  {\bibfnamefont {G.~S.}\ \bibnamefont {{Tucker}}}, \bibinfo {author}
  {\bibfnamefont {J.~L.}\ \bibnamefont {{Weiland}}}, \bibinfo {author}
  {\bibfnamefont {E.}~\bibnamefont {{Wollack}}}, \ and\ \bibinfo {author}
  {\bibfnamefont {E.~L.}\ \bibnamefont {{Wright}}},\ }\bibfield  {title}
  {\enquote {\bibinfo {title} {{Seven-Year Wilkinson Microwave Anisotropy Probe
  (WMAP) Observations: Cosmological Interpretation}},}\ }\href@noop {}
  {\bibfield  {journal} {\bibinfo  {journal} {ArXiv e-prints}\ } (\bibinfo
  {year} {2010})},\ \Eprint {http://arxiv.org/abs/1001.4538} {arXiv:1001.4538
  [astro-ph.CO]} \BibitemShut {NoStop}%
\bibitem [{\citenamefont {{Komatsu}}\ \emph {et~al.}(2009)\citenamefont
  {{Komatsu}}, \citenamefont {{Afshordi}}, \citenamefont {{Bartolo}},
  \citenamefont {{Baumann}}, \citenamefont {{Bond}}, \citenamefont
  {{Buchbinder}}, \citenamefont {{Byrnes}}, \citenamefont {{Chen}},
  \citenamefont {{Chung}}, \citenamefont {{Cooray}}, \citenamefont
  {{Creminelli}}, \citenamefont {{Dalal}}, \citenamefont {{Dore}},
  \citenamefont {{Easther}}, \citenamefont {{Frolov}}, \citenamefont
  {{Khoury}}, \citenamefont {{Kinney}}, \citenamefont {{Kofman}}, \citenamefont
  {{Koyama}}, \citenamefont {{Leblond}}, \citenamefont {{Lehners}},
  \citenamefont {{Lidsey}}, \citenamefont {{Liguori}}, \citenamefont {{Lim}},
  \citenamefont {{Linde}}, \citenamefont {{Lyth}}, \citenamefont {{Maldacena}},
  \citenamefont {{Matarrese}}, \citenamefont {{McAllister}}, \citenamefont
  {{McDonald}}, \citenamefont {{Mukohyama}}, \citenamefont {{Ovrut}},
  \citenamefont {{Peiris}}, \citenamefont {{Riotto}}, \citenamefont
  {{Rodrigues}}, \citenamefont {{Sasaki}}, \citenamefont {{Scoccimarro}},
  \citenamefont {{Seery}}, \citenamefont {{Sefusatti}}, \citenamefont
  {{Smith}}, \citenamefont {{Starobinsky}}, \citenamefont {{Steinhardt}},
  \citenamefont {{Takahashi}}, \citenamefont {{Tegmark}}, \citenamefont
  {{Tolley}}, \citenamefont {{Verde}}, \citenamefont {{Wandelt}}, \citenamefont
  {{Wands}}, \citenamefont {{Weinberg}}, \citenamefont {{Wyman}}, \citenamefont
  {{Yadav}},\ and\ \citenamefont {{Zaldarriaga}}}]{2009astro2010S.158K}%
  \BibitemOpen
  \bibfield  {author} {\bibinfo {author} {\bibfnamefont {E.}~\bibnamefont
  {{Komatsu}}}, \bibinfo {author} {\bibfnamefont {N.}~\bibnamefont
  {{Afshordi}}}, \bibinfo {author} {\bibfnamefont {N.}~\bibnamefont
  {{Bartolo}}}, \bibinfo {author} {\bibfnamefont {D.}~\bibnamefont
  {{Baumann}}}, \bibinfo {author} {\bibfnamefont {J.~R.}\ \bibnamefont
  {{Bond}}}, \bibinfo {author} {\bibfnamefont {E.~I.}\ \bibnamefont
  {{Buchbinder}}}, \bibinfo {author} {\bibfnamefont {C.~T.}\ \bibnamefont
  {{Byrnes}}}, \bibinfo {author} {\bibfnamefont {X.}~\bibnamefont {{Chen}}},
  \bibinfo {author} {\bibfnamefont {D.~J.~H.}\ \bibnamefont {{Chung}}},
  \bibinfo {author} {\bibfnamefont {A.}~\bibnamefont {{Cooray}}}, \bibinfo
  {author} {\bibfnamefont {P.}~\bibnamefont {{Creminelli}}}, \bibinfo {author}
  {\bibfnamefont {N.}~\bibnamefont {{Dalal}}}, \bibinfo {author} {\bibfnamefont
  {O.}~\bibnamefont {{Dore}}}, \bibinfo {author} {\bibfnamefont
  {R.}~\bibnamefont {{Easther}}}, \bibinfo {author} {\bibfnamefont {A.~V.}\
  \bibnamefont {{Frolov}}}, \bibinfo {author} {\bibfnamefont {J.}~\bibnamefont
  {{Khoury}}}, \bibinfo {author} {\bibfnamefont {W.~H.}\ \bibnamefont
  {{Kinney}}}, \bibinfo {author} {\bibfnamefont {L.}~\bibnamefont {{Kofman}}},
  \bibinfo {author} {\bibfnamefont {K.}~\bibnamefont {{Koyama}}}, \bibinfo
  {author} {\bibfnamefont {L.}~\bibnamefont {{Leblond}}}, \bibinfo {author}
  {\bibfnamefont {J.-L.}\ \bibnamefont {{Lehners}}}, \bibinfo {author}
  {\bibfnamefont {J.~E.}\ \bibnamefont {{Lidsey}}}, \bibinfo {author}
  {\bibfnamefont {M.}~\bibnamefont {{Liguori}}}, \bibinfo {author}
  {\bibfnamefont {E.~A.}\ \bibnamefont {{Lim}}}, \bibinfo {author}
  {\bibfnamefont {A.}~\bibnamefont {{Linde}}}, \bibinfo {author} {\bibfnamefont
  {D.~H.}\ \bibnamefont {{Lyth}}}, \bibinfo {author} {\bibfnamefont
  {J.}~\bibnamefont {{Maldacena}}}, \bibinfo {author} {\bibfnamefont
  {S.}~\bibnamefont {{Matarrese}}}, \bibinfo {author} {\bibfnamefont
  {L.}~\bibnamefont {{McAllister}}}, \bibinfo {author} {\bibfnamefont
  {P.}~\bibnamefont {{McDonald}}}, \bibinfo {author} {\bibfnamefont
  {S.}~\bibnamefont {{Mukohyama}}}, \bibinfo {author} {\bibfnamefont
  {B.}~\bibnamefont {{Ovrut}}}, \bibinfo {author} {\bibfnamefont {H.~V.}\
  \bibnamefont {{Peiris}}}, \bibinfo {author} {\bibfnamefont {A.}~\bibnamefont
  {{Riotto}}}, \bibinfo {author} {\bibfnamefont {Y.}~\bibnamefont
  {{Rodrigues}}}, \bibinfo {author} {\bibfnamefont {M.}~\bibnamefont
  {{Sasaki}}}, \bibinfo {author} {\bibfnamefont {R.}~\bibnamefont
  {{Scoccimarro}}}, \bibinfo {author} {\bibfnamefont {D.}~\bibnamefont
  {{Seery}}}, \bibinfo {author} {\bibfnamefont {A.}~\bibnamefont
  {{Sefusatti}}}, \bibinfo {author} {\bibfnamefont {K.~M.}\ \bibnamefont
  {{Smith}}}, \bibinfo {author} {\bibfnamefont {A.~A.}\ \bibnamefont
  {{Starobinsky}}}, \bibinfo {author} {\bibfnamefont {P.~J.}\ \bibnamefont
  {{Steinhardt}}}, \bibinfo {author} {\bibfnamefont {F.}~\bibnamefont
  {{Takahashi}}}, \bibinfo {author} {\bibfnamefont {M.}~\bibnamefont
  {{Tegmark}}}, \bibinfo {author} {\bibfnamefont {A.~J.}\ \bibnamefont
  {{Tolley}}}, \bibinfo {author} {\bibfnamefont {L.}~\bibnamefont {{Verde}}},
  \bibinfo {author} {\bibfnamefont {B.~D.}\ \bibnamefont {{Wandelt}}}, \bibinfo
  {author} {\bibfnamefont {D.}~\bibnamefont {{Wands}}}, \bibinfo {author}
  {\bibfnamefont {S.}~\bibnamefont {{Weinberg}}}, \bibinfo {author}
  {\bibfnamefont {M.}~\bibnamefont {{Wyman}}}, \bibinfo {author} {\bibfnamefont
  {A.~P.~S.}\ \bibnamefont {{Yadav}}}, \ and\ \bibinfo {author} {\bibfnamefont
  {M.}~\bibnamefont {{Zaldarriaga}}},\ }\bibfield  {title} {\enquote {\bibinfo
  {title} {{Non-Gaussianity as a Probe of the Physics of the Primordial
  Universe and the Astrophysics of the Low Redshift Universe}},}\ }in\
  \href@noop {} {\emph {\bibinfo {booktitle} {astro2010: The Astronomy and
  Astrophysics Decadal Survey}}},\ \bibinfo {series} {ArXiv Astrophysics
  e-prints}, Vol.\ \bibinfo {volume} {2010}\ (\bibinfo {year} {2009})\ pp.\
  \bibinfo {pages} {158--+},\ \Eprint {http://arxiv.org/abs/0902.4759}
  {arXiv:0902.4759 [astro-ph.CO]} \BibitemShut {NoStop}%
\bibitem [{\citenamefont {{Linde}}\ and\ \citenamefont
  {{Mukhanov}}(1997)}]{1997PhRvD..56..535L}%
  \BibitemOpen
  \bibfield  {author} {\bibinfo {author} {\bibfnamefont {A.}~\bibnamefont
  {{Linde}}}\ and\ \bibinfo {author} {\bibfnamefont {V.}~\bibnamefont
  {{Mukhanov}}},\ }\bibfield  {title} {\enquote {\bibinfo {title}
  {{Non-Gaussian isocurvature perturbations from inflation}},}\ }\href@noop {}
  {\bibfield  {journal} {\bibinfo  {journal} {\prd}\ }\textbf {\bibinfo
  {volume} {56}},\ \bibinfo {pages} {535--+} (\bibinfo {year} {1997})},\
  \Eprint {http://arxiv.org/abs/arXiv:astro-ph/9610219}
  {arXiv:astro-ph/9610219} \BibitemShut {NoStop}%
\bibitem [{\citenamefont {{Lyth}}\ and\ \citenamefont
  {{Wands}}(2002)}]{2002PhLB..524....5L}%
  \BibitemOpen
  \bibfield  {author} {\bibinfo {author} {\bibfnamefont {D.~H.}\ \bibnamefont
  {{Lyth}}}\ and\ \bibinfo {author} {\bibfnamefont {D.}~\bibnamefont
  {{Wands}}},\ }\bibfield  {title} {\enquote {\bibinfo {title} {{Generating the
  curvature perturbation without an inflaton}},}\ }\href {\doibase
  10.1016/S0370-2693(01)01366-1} {\bibfield  {journal} {\bibinfo  {journal}
  {Physics Letters B}\ }\textbf {\bibinfo {volume} {524}},\ \bibinfo {pages}
  {5--14} (\bibinfo {year} {2002})},\ \Eprint
  {http://arxiv.org/abs/arXiv:hep-ph/0110002} {arXiv:hep-ph/0110002}
  \BibitemShut {NoStop}%
\bibitem [{\citenamefont {{Enqvist}}\ and\ \citenamefont
  {{Sloth}}(2002)}]{EnqSlo0204}%
  \BibitemOpen
  \bibfield  {author} {\bibinfo {author} {\bibfnamefont {K.}~\bibnamefont
  {{Enqvist}}}\ and\ \bibinfo {author} {\bibfnamefont {M.~S.}\ \bibnamefont
  {{Sloth}}},\ }\bibfield  {title} {\enquote {\bibinfo {title} {{Adiabatic CMB
  perturbations in pre-Big-Bang string cosmology}},}\ }\href {\doibase
  10.1016/S0550-3213(02)00043-3} {\bibfield  {journal} {\bibinfo  {journal}
  {Nuclear Physics B}\ }\textbf {\bibinfo {volume} {626}},\ \bibinfo {pages}
  {395--409} (\bibinfo {year} {2002})},\ \Eprint
  {http://arxiv.org/abs/arXiv:hep-ph/0109214} {arXiv:hep-ph/0109214}
  \BibitemShut {NoStop}%
\bibitem [{\citenamefont {{Bernardeau}}\ and\ \citenamefont
  {{Uzan}}(2003)}]{2003PhRvD..67l1301B}%
  \BibitemOpen
  \bibfield  {author} {\bibinfo {author} {\bibfnamefont {F.}~\bibnamefont
  {{Bernardeau}}}\ and\ \bibinfo {author} {\bibfnamefont {J.}~\bibnamefont
  {{Uzan}}},\ }\bibfield  {title} {\enquote {\bibinfo {title} {{Inflationary
  models inducing non-Gaussian metric fluctuations}},}\ }\href@noop {}
  {\bibfield  {journal} {\bibinfo  {journal} {\prd}\ }\textbf {\bibinfo
  {volume} {67}},\ \bibinfo {pages} {121301} (\bibinfo {year}
  {2003})}\BibitemShut {NoStop}%
\bibitem [{\citenamefont {{Kofman}}(2003)}]{2003astro.ph..3614K}%
  \BibitemOpen
  \bibfield  {author} {\bibinfo {author} {\bibfnamefont {L.}~\bibnamefont
  {{Kofman}}},\ }\bibfield  {title} {\enquote {\bibinfo {title} {{Probing
  String Theory with Modulated Cosmological Fluctuations}},}\ }\href@noop {}
  {\bibfield  {journal} {\bibinfo  {journal} {ArXiv Astrophysics e-prints}\ }
  (\bibinfo {year} {2003})},\ \Eprint
  {http://arxiv.org/abs/arXiv:astro-ph/0303614} {arXiv:astro-ph/0303614}
  \BibitemShut {NoStop}%
\bibitem [{\citenamefont {{Dvali}}\ \emph
  {et~al.}(2004{\natexlab{a}})\citenamefont {{Dvali}}, \citenamefont
  {{Gruzinov}},\ and\ \citenamefont {{Zaldarriaga}}}]{2004PhRvD..69b3505D}%
  \BibitemOpen
  \bibfield  {author} {\bibinfo {author} {\bibfnamefont {G.}~\bibnamefont
  {{Dvali}}}, \bibinfo {author} {\bibfnamefont {A.}~\bibnamefont {{Gruzinov}}},
  \ and\ \bibinfo {author} {\bibfnamefont {M.}~\bibnamefont {{Zaldarriaga}}},\
  }\bibfield  {title} {\enquote {\bibinfo {title} {{New mechanism for
  generating density perturbations from inflation}},}\ }\href@noop {}
  {\bibfield  {journal} {\bibinfo  {journal} {\prd}\ }\textbf {\bibinfo
  {volume} {69}},\ \bibinfo {pages} {023505--+} (\bibinfo {year}
  {2004}{\natexlab{a}})}\BibitemShut {NoStop}%
\bibitem [{\citenamefont {{Bernardeau}}\ \emph {et~al.}(2004)\citenamefont
  {{Bernardeau}}, \citenamefont {{Kofman}},\ and\ \citenamefont
  {{Uzan}}}]{2004PhRvD..70h3004B}%
  \BibitemOpen
  \bibfield  {author} {\bibinfo {author} {\bibfnamefont {F.}~\bibnamefont
  {{Bernardeau}}}, \bibinfo {author} {\bibfnamefont {L.}~\bibnamefont
  {{Kofman}}}, \ and\ \bibinfo {author} {\bibfnamefont {J.-P.}\ \bibnamefont
  {{Uzan}}},\ }\bibfield  {title} {\enquote {\bibinfo {title} {{Modulated
  fluctuations from hybrid inflation}},}\ }\href {\doibase
  10.1103/PhysRevD.70.083004} {\bibfield  {journal} {\bibinfo  {journal}
  {\prd}\ }\textbf {\bibinfo {volume} {70}},\ \bibinfo {pages} {083004--+}
  (\bibinfo {year} {2004})},\ \Eprint
  {http://arxiv.org/abs/arXiv:astro-ph/0403315} {arXiv:astro-ph/0403315}
  \BibitemShut {NoStop}%
\bibitem [{\citenamefont {{Alabidi}}\ \emph {et~al.}(2010)\citenamefont
  {{Alabidi}}, \citenamefont {{Malik}}, \citenamefont {{Byrnes}},\ and\
  \citenamefont {{Choi}}}]{2010JCAP...11..037A}%
  \BibitemOpen
  \bibfield  {author} {\bibinfo {author} {\bibfnamefont {L.}~\bibnamefont
  {{Alabidi}}}, \bibinfo {author} {\bibfnamefont {K.}~\bibnamefont {{Malik}}},
  \bibinfo {author} {\bibfnamefont {C.~T.}\ \bibnamefont {{Byrnes}}}, \ and\
  \bibinfo {author} {\bibfnamefont {K.-Y.}\ \bibnamefont {{Choi}}},\ }\bibfield
   {title} {\enquote {\bibinfo {title} {{How the curvaton scenario, modulated
  reheating and an inhomogeneous end of inflation are related}},}\ }\href
  {\doibase 10.1088/1475-7516/2010/11/037} {\bibfield  {journal} {\bibinfo
  {journal} {\jcap}\ }\textbf {\bibinfo {volume} {11}},\ \bibinfo {pages}
  {37--+} (\bibinfo {year} {2010})},\ \Eprint {http://arxiv.org/abs/1002.1700}
  {arXiv:1002.1700 [astro-ph.CO]} \BibitemShut {NoStop}%
\bibitem [{\citenamefont {{Salopek}}\ and\ \citenamefont
  {{Bond}}(1991)}]{SalBon9102}%
  \BibitemOpen
  \bibfield  {author} {\bibinfo {author} {\bibfnamefont {D.~S.}\ \bibnamefont
  {{Salopek}}}\ and\ \bibinfo {author} {\bibfnamefont {J.~R.}\ \bibnamefont
  {{Bond}}},\ }\bibfield  {title} {\enquote {\bibinfo {title} {{Stochastic
  inflation and nonlinear gravity}},}\ }\href {\doibase
  10.1103/PhysRevD.43.1005} {\bibfield  {journal} {\bibinfo  {journal} {\prd}\
  }\textbf {\bibinfo {volume} {43}},\ \bibinfo {pages} {1005--1031} (\bibinfo
  {year} {1991})}\BibitemShut {NoStop}%
\bibitem [{\citenamefont {{Salopek}}\ and\ \citenamefont
  {{Bond}}(1990)}]{SalBon9012}%
  \BibitemOpen
  \bibfield  {author} {\bibinfo {author} {\bibfnamefont {D.~S.}\ \bibnamefont
  {{Salopek}}}\ and\ \bibinfo {author} {\bibfnamefont {J.~R.}\ \bibnamefont
  {{Bond}}},\ }\bibfield  {title} {\enquote {\bibinfo {title} {{Nonlinear
  evolution of long-wavelength metric fluctuations in inflationary models}},}\
  }\href {\doibase 10.1103/PhysRevD.42.3936} {\bibfield  {journal} {\bibinfo
  {journal} {\prd}\ }\textbf {\bibinfo {volume} {42}},\ \bibinfo {pages}
  {3936--3962} (\bibinfo {year} {1990})}\BibitemShut {NoStop}%
\bibitem [{\citenamefont {{Dvali}}\ \emph
  {et~al.}(2004{\natexlab{b}})\citenamefont {{Dvali}}, \citenamefont
  {{Gruzinov}},\ and\ \citenamefont {{Zaldarriaga}}}]{2004PhRvD..69h3505D}%
  \BibitemOpen
  \bibfield  {author} {\bibinfo {author} {\bibfnamefont {G.}~\bibnamefont
  {{Dvali}}}, \bibinfo {author} {\bibfnamefont {A.}~\bibnamefont {{Gruzinov}}},
  \ and\ \bibinfo {author} {\bibfnamefont {M.}~\bibnamefont {{Zaldarriaga}}},\
  }\bibfield  {title} {\enquote {\bibinfo {title} {{Cosmological perturbations
  from inhomogeneous reheating, freeze-out, and mass domination}},}\ }\href
  {\doibase 10.1103/PhysRevD.69.083505} {\bibfield  {journal} {\bibinfo
  {journal} {\prd}\ }\textbf {\bibinfo {volume} {69}},\ \bibinfo {pages}
  {083505--+} (\bibinfo {year} {2004}{\natexlab{b}})}\BibitemShut {NoStop}%
\bibitem [{\citenamefont {{Zaldarriaga}}(2004)}]{2004PhRvD..69d3508Z}%
  \BibitemOpen
  \bibfield  {author} {\bibinfo {author} {\bibfnamefont {M.}~\bibnamefont
  {{Zaldarriaga}}},\ }\bibfield  {title} {\enquote {\bibinfo {title}
  {{Non-Gaussianities in models with a varying inflaton decay rate}},}\
  }\href@noop {} {\bibfield  {journal} {\bibinfo  {journal} {\prd}\ }\textbf
  {\bibinfo {volume} {69}},\ \bibinfo {pages} {043508--+} (\bibinfo {year}
  {2004})}\BibitemShut {NoStop}%
\bibitem [{\citenamefont {{Slosar}}\ \emph {et~al.}(2008)\citenamefont
  {{Slosar}}, \citenamefont {{Hirata}}, \citenamefont {{Seljak}}, \citenamefont
  {{Ho}},\ and\ \citenamefont {{Padmanabhan}}}]{2008JCAP...08..031S}%
  \BibitemOpen
  \bibfield  {author} {\bibinfo {author} {\bibfnamefont {A.}~\bibnamefont
  {{Slosar}}}, \bibinfo {author} {\bibfnamefont {C.}~\bibnamefont {{Hirata}}},
  \bibinfo {author} {\bibfnamefont {U.}~\bibnamefont {{Seljak}}}, \bibinfo
  {author} {\bibfnamefont {S.}~\bibnamefont {{Ho}}}, \ and\ \bibinfo {author}
  {\bibfnamefont {N.}~\bibnamefont {{Padmanabhan}}},\ }\bibfield  {title}
  {\enquote {\bibinfo {title} {{Constraints on local primordial non-Gaussianity
  from large scale structure}},}\ }\href {\doibase
  10.1088/1475-7516/2008/08/031} {\bibfield  {journal} {\bibinfo  {journal}
  {\jcap}\ }\textbf {\bibinfo {volume} {8}},\ \bibinfo {pages} {31--+}
  (\bibinfo {year} {2008})},\ \Eprint {http://arxiv.org/abs/0805.3580}
  {arXiv:0805.3580} \BibitemShut {NoStop}%
\bibitem [{\citenamefont {{Dalal}}\ \emph {et~al.}(2008)\citenamefont
  {{Dalal}}, \citenamefont {{Dor{\'e}}}, \citenamefont {{Huterer}},\ and\
  \citenamefont {{Shirokov}}}]{2008PhRvD..77l3514D}%
  \BibitemOpen
  \bibfield  {author} {\bibinfo {author} {\bibfnamefont {N.}~\bibnamefont
  {{Dalal}}}, \bibinfo {author} {\bibfnamefont {O.}~\bibnamefont {{Dor{\'e}}}},
  \bibinfo {author} {\bibfnamefont {D.}~\bibnamefont {{Huterer}}}, \ and\
  \bibinfo {author} {\bibfnamefont {A.}~\bibnamefont {{Shirokov}}},\ }\bibfield
   {title} {\enquote {\bibinfo {title} {{Imprints of primordial
  non-Gaussianities on large-scale structure: Scale-dependent bias and
  abundance of virialized objects}},}\ }\href {\doibase
  10.1103/PhysRevD.77.123514} {\bibfield  {journal} {\bibinfo  {journal}
  {\prd}\ }\textbf {\bibinfo {volume} {77}},\ \bibinfo {pages} {123514--+}
  (\bibinfo {year} {2008})},\ \Eprint {http://arxiv.org/abs/0710.4560}
  {arXiv:0710.4560} \BibitemShut {NoStop}%
\bibitem [{\citenamefont {{Grinstein}}\ and\ \citenamefont
  {{Wise}}(1986)}]{1986ApJ...310...19G}%
  \BibitemOpen
  \bibfield  {author} {\bibinfo {author} {\bibfnamefont {B.}~\bibnamefont
  {{Grinstein}}}\ and\ \bibinfo {author} {\bibfnamefont {M.~B.}\ \bibnamefont
  {{Wise}}},\ }\bibfield  {title} {\enquote {\bibinfo {title} {{Non-Gaussian
  fluctuations and the correlations of galaxies or rich clusters of
  galaxies}},}\ }\href {\doibase 10.1086/164660} {\bibfield  {journal}
  {\bibinfo  {journal} {\apj}\ }\textbf {\bibinfo {volume} {310}},\ \bibinfo
  {pages} {19--22} (\bibinfo {year} {1986})}\BibitemShut {NoStop}%
\bibitem [{\citenamefont {{Alishahiha}}\ \emph {et~al.}(2004)\citenamefont
  {{Alishahiha}}, \citenamefont {{Silverstein}},\ and\ \citenamefont
  {{Tong}}}]{2004PhRvD..70l3505A}%
  \BibitemOpen
  \bibfield  {author} {\bibinfo {author} {\bibfnamefont {M.}~\bibnamefont
  {{Alishahiha}}}, \bibinfo {author} {\bibfnamefont {E.}~\bibnamefont
  {{Silverstein}}}, \ and\ \bibinfo {author} {\bibfnamefont {D.}~\bibnamefont
  {{Tong}}},\ }\bibfield  {title} {\enquote {\bibinfo {title} {{DBI in the sky:
  Non-Gaussianity from inflation with a speed limit}},}\ }\href {\doibase
  10.1103/PhysRevD.70.123505} {\bibfield  {journal} {\bibinfo  {journal}
  {\prd}\ }\textbf {\bibinfo {volume} {70}},\ \bibinfo {pages} {123505--+}
  (\bibinfo {year} {2004})},\ \Eprint
  {http://arxiv.org/abs/arXiv:hep-th/0404084} {arXiv:hep-th/0404084}
  \BibitemShut {NoStop}%
\bibitem [{\citenamefont {{Silverstein}}\ and\ \citenamefont
  {{Tong}}(2004)}]{2004PhRvD..70j3505S}%
  \BibitemOpen
  \bibfield  {author} {\bibinfo {author} {\bibfnamefont {E.}~\bibnamefont
  {{Silverstein}}}\ and\ \bibinfo {author} {\bibfnamefont {D.}~\bibnamefont
  {{Tong}}},\ }\bibfield  {title} {\enquote {\bibinfo {title} {{Scalar speed
  limits and cosmology: Acceleration from D-cceleration}},}\ }\href {\doibase
  10.1103/PhysRevD.70.103505} {\bibfield  {journal} {\bibinfo  {journal}
  {\prd}\ }\textbf {\bibinfo {volume} {70}},\ \bibinfo {pages} {103505--+}
  (\bibinfo {year} {2004})},\ \Eprint
  {http://arxiv.org/abs/arXiv:hep-th/0310221} {arXiv:hep-th/0310221}
  \BibitemShut {NoStop}%
\bibitem [{\citenamefont {{Armend{\'a}riz-Pic{\'o}n}}\ \emph
  {et~al.}(1999)\citenamefont {{Armend{\'a}riz-Pic{\'o}n}}, \citenamefont
  {{Damour}},\ and\ \citenamefont {{Mukhanov}}}]{1999PhLB..458..209A}%
  \BibitemOpen
  \bibfield  {author} {\bibinfo {author} {\bibfnamefont {C.}~\bibnamefont
  {{Armend{\'a}riz-Pic{\'o}n}}}, \bibinfo {author} {\bibfnamefont
  {T.}~\bibnamefont {{Damour}}}, \ and\ \bibinfo {author} {\bibfnamefont
  {V.}~\bibnamefont {{Mukhanov}}},\ }\bibfield  {title} {\enquote {\bibinfo
  {title} {{k-Inflation}},}\ }\href {\doibase 10.1016/S0370-2693(99)00603-6}
  {\bibfield  {journal} {\bibinfo  {journal} {Physics Letters B}\ }\textbf
  {\bibinfo {volume} {458}},\ \bibinfo {pages} {209--218} (\bibinfo {year}
  {1999})},\ \Eprint {http://arxiv.org/abs/arXiv:hep-th/9904075}
  {arXiv:hep-th/9904075} \BibitemShut {NoStop}%
\bibitem [{\citenamefont {{Li}}\ \emph {et~al.}(2008)\citenamefont {{Li}},
  \citenamefont {{Wang}},\ and\ \citenamefont {{Wang}}}]{2008JCAP...03..028L}%
  \BibitemOpen
  \bibfield  {author} {\bibinfo {author} {\bibfnamefont {M.}~\bibnamefont
  {{Li}}}, \bibinfo {author} {\bibfnamefont {T.}~\bibnamefont {{Wang}}}, \ and\
  \bibinfo {author} {\bibfnamefont {Y.}~\bibnamefont {{Wang}}},\ }\bibfield
  {title} {\enquote {\bibinfo {title} {{General single field inflation with
  large positive non-Gaussianity}},}\ }\href {\doibase
  10.1088/1475-7516/2008/03/028} {\bibfield  {journal} {\bibinfo  {journal}
  {\jcap}\ }\textbf {\bibinfo {volume} {3}},\ \bibinfo {pages} {28--+}
  (\bibinfo {year} {2008})},\ \Eprint {http://arxiv.org/abs/0801.0040}
  {arXiv:0801.0040} \BibitemShut {NoStop}%
\bibitem [{\citenamefont {{Tolley}}\ and\ \citenamefont
  {{Wyman}}(2010)}]{2010PhRvD..81d3502T}%
  \BibitemOpen
  \bibfield  {author} {\bibinfo {author} {\bibfnamefont {A.~J.}\ \bibnamefont
  {{Tolley}}}\ and\ \bibinfo {author} {\bibfnamefont {M.}~\bibnamefont
  {{Wyman}}},\ }\bibfield  {title} {\enquote {\bibinfo {title} {{Equilateral
  non-Gaussianity from multifield dynamics}},}\ }\href {\doibase
  10.1103/PhysRevD.81.043502} {\bibfield  {journal} {\bibinfo  {journal}
  {\prd}\ }\textbf {\bibinfo {volume} {81}},\ \bibinfo {pages} {043502--+}
  (\bibinfo {year} {2010})},\ \Eprint {http://arxiv.org/abs/0910.1853}
  {arXiv:0910.1853 [hep-th]} \BibitemShut {NoStop}%
\bibitem [{\citenamefont {{Chen}}\ \emph {et~al.}(2007)\citenamefont {{Chen}},
  \citenamefont {{Huang}}, \citenamefont {{Kachru}},\ and\ \citenamefont
  {{Shiu}}}]{2007JCAP...01..002C}%
  \BibitemOpen
  \bibfield  {author} {\bibinfo {author} {\bibfnamefont {X.}~\bibnamefont
  {{Chen}}}, \bibinfo {author} {\bibfnamefont {{M.-x.}}\ \bibnamefont
  {{Huang}}}, \bibinfo {author} {\bibfnamefont {S.}~\bibnamefont {{Kachru}}}, \
  and\ \bibinfo {author} {\bibfnamefont {G.}~\bibnamefont {{Shiu}}},\
  }\bibfield  {title} {\enquote {\bibinfo {title} {{Observational signatures
  and non-Gaussianities of general single-field inflation}},}\ }\href {\doibase
  10.1088/1475-7516/2007/01/002} {\bibfield  {journal} {\bibinfo  {journal}
  {\jcap}\ }\textbf {\bibinfo {volume} {1}},\ \bibinfo {pages} {2--+} (\bibinfo
  {year} {2007})},\ \Eprint {http://arxiv.org/abs/arXiv:hep-th/0605045}
  {arXiv:hep-th/0605045} \BibitemShut {NoStop}%
\bibitem [{\citenamefont {{Senatore}}\ \emph {et~al.}(2010)\citenamefont
  {{Senatore}}, \citenamefont {{Smith}},\ and\ \citenamefont
  {{Zaldarriaga}}}]{2010JCAP...01..028S}%
  \BibitemOpen
  \bibfield  {author} {\bibinfo {author} {\bibfnamefont {L.}~\bibnamefont
  {{Senatore}}}, \bibinfo {author} {\bibfnamefont {K.~M.}\ \bibnamefont
  {{Smith}}}, \ and\ \bibinfo {author} {\bibfnamefont {M.}~\bibnamefont
  {{Zaldarriaga}}},\ }\bibfield  {title} {\enquote {\bibinfo {title}
  {{Non-Gaussianities in single field inflation and their optimal limits from
  the WMAP 5-year data}},}\ }\href {\doibase 10.1088/1475-7516/2010/01/028}
  {\bibfield  {journal} {\bibinfo  {journal} {\jcap}\ }\textbf {\bibinfo
  {volume} {1}},\ \bibinfo {pages} {28--+} (\bibinfo {year} {2010})},\ \Eprint
  {http://arxiv.org/abs/0905.3746} {arXiv:0905.3746 [astro-ph.CO]} \BibitemShut
  {NoStop}%
\bibitem [{\citenamefont {{Creminelli}}\ \emph
  {et~al.}(2011{\natexlab{a}})\citenamefont {{Creminelli}}, \citenamefont
  {{D'Amico}}, \citenamefont {{Musso}}, \citenamefont {{Nore{\~n}a}},\ and\
  \citenamefont {{Trincherini}}}]{2011JCAP...02..006C}%
  \BibitemOpen
  \bibfield  {author} {\bibinfo {author} {\bibfnamefont {P.}~\bibnamefont
  {{Creminelli}}}, \bibinfo {author} {\bibfnamefont {G.}~\bibnamefont
  {{D'Amico}}}, \bibinfo {author} {\bibfnamefont {M.}~\bibnamefont {{Musso}}},
  \bibinfo {author} {\bibfnamefont {J.}~\bibnamefont {{Nore{\~n}a}}}, \ and\
  \bibinfo {author} {\bibfnamefont {E.}~\bibnamefont {{Trincherini}}},\
  }\bibfield  {title} {\enquote {\bibinfo {title} {{Galilean symmetry in the
  effective theory of inflation: new shapes of non-Gaussianity}},}\ }\href
  {\doibase 10.1088/1475-7516/2011/02/006} {\bibfield  {journal} {\bibinfo
  {journal} {\jcap}\ }\textbf {\bibinfo {volume} {2}},\ \bibinfo {pages} {6--+}
  (\bibinfo {year} {2011}{\natexlab{a}})},\ \Eprint
  {http://arxiv.org/abs/1011.3004} {arXiv:1011.3004 [hep-th]} \BibitemShut
  {NoStop}%
\bibitem [{\citenamefont {{Chen}}\ and\ \citenamefont
  {{Wang}}(2010{\natexlab{a}})}]{2010JCAP...04..027C}%
  \BibitemOpen
  \bibfield  {author} {\bibinfo {author} {\bibfnamefont {X.}~\bibnamefont
  {{Chen}}}\ and\ \bibinfo {author} {\bibfnamefont {Y.}~\bibnamefont
  {{Wang}}},\ }\bibfield  {title} {\enquote {\bibinfo {title} {{Quasi-single
  field inflation and non-Gaussianities}},}\ }\href {\doibase
  10.1088/1475-7516/2010/04/027} {\bibfield  {journal} {\bibinfo  {journal}
  {\jcap}\ }\textbf {\bibinfo {volume} {4}},\ \bibinfo {pages} {27--+}
  (\bibinfo {year} {2010}{\natexlab{a}})},\ \Eprint
  {http://arxiv.org/abs/0911.3380} {arXiv:0911.3380 [hep-th]} \BibitemShut
  {NoStop}%
\bibitem [{\citenamefont {{Chen}}\ and\ \citenamefont
  {{Wang}}(2010{\natexlab{b}})}]{2010PhRvD..81f3511C}%
  \BibitemOpen
  \bibfield  {author} {\bibinfo {author} {\bibfnamefont {X.}~\bibnamefont
  {{Chen}}}\ and\ \bibinfo {author} {\bibfnamefont {Y.}~\bibnamefont
  {{Wang}}},\ }\bibfield  {title} {\enquote {\bibinfo {title} {{Large
  non-Gaussianities with intermediate shapes from quasi-single-field
  inflation}},}\ }\href {\doibase 10.1103/PhysRevD.81.063511} {\bibfield
  {journal} {\bibinfo  {journal} {\prd}\ }\textbf {\bibinfo {volume} {81}},\
  \bibinfo {pages} {063511--+} (\bibinfo {year} {2010}{\natexlab{b}})},\
  \Eprint {http://arxiv.org/abs/0909.0496} {arXiv:0909.0496 [astro-ph.CO]}
  \BibitemShut {NoStop}%
\bibitem [{\citenamefont {{Creminelli}}\ \emph
  {et~al.}(2011{\natexlab{b}})\citenamefont {{Creminelli}}, \citenamefont
  {{D'Amico}}, \citenamefont {{Musso}},\ and\ \citenamefont
  {{Nore{\~n}a}}}]{2011arXiv1106.1462C}%
  \BibitemOpen
  \bibfield  {author} {\bibinfo {author} {\bibfnamefont {P.}~\bibnamefont
  {{Creminelli}}}, \bibinfo {author} {\bibfnamefont {G.}~\bibnamefont
  {{D'Amico}}}, \bibinfo {author} {\bibfnamefont {M.}~\bibnamefont {{Musso}}},
  \ and\ \bibinfo {author} {\bibfnamefont {J.}~\bibnamefont {{Nore{\~n}a}}},\
  }\bibfield  {title} {\enquote {\bibinfo {title} {{The (not so) squeezed limit
  of the primordial 3-point function}},}\ }\href@noop {} {\bibfield  {journal}
  {\bibinfo  {journal} {ArXiv e-prints}\ } (\bibinfo {year}
  {2011}{\natexlab{b}})},\ \Eprint {http://arxiv.org/abs/1106.1462}
  {arXiv:1106.1462 [astro-ph.CO]} \BibitemShut {NoStop}%
\bibitem [{\citenamefont {{Xia}}\ \emph {et~al.}(2011)\citenamefont {{Xia}},
  \citenamefont {{Baccigalupi}}, \citenamefont {{Matarrese}}, \citenamefont
  {{Verde}},\ and\ \citenamefont {{Viel}}}]{2011arXiv1104.5015X}%
  \BibitemOpen
  \bibfield  {author} {\bibinfo {author} {\bibfnamefont {J.-Q.}\ \bibnamefont
  {{Xia}}}, \bibinfo {author} {\bibfnamefont {C.}~\bibnamefont
  {{Baccigalupi}}}, \bibinfo {author} {\bibfnamefont {S.}~\bibnamefont
  {{Matarrese}}}, \bibinfo {author} {\bibfnamefont {L.}~\bibnamefont
  {{Verde}}}, \ and\ \bibinfo {author} {\bibfnamefont {M.}~\bibnamefont
  {{Viel}}},\ }\bibfield  {title} {\enquote {\bibinfo {title} {{Constraints on
  Primordial Non-Gaussianity from Large Scale Structure Probes}},}\ }\href@noop
  {} {\bibfield  {journal} {\bibinfo  {journal} {ArXiv e-prints}\ } (\bibinfo
  {year} {2011})},\ \Eprint {http://arxiv.org/abs/1104.5015} {arXiv:1104.5015
  [astro-ph.CO]} \BibitemShut {NoStop}%
\bibitem [{\citenamefont {{Holman}}\ and\ \citenamefont
  {{Tolley}}(2008)}]{2008JCAP...05..001H}%
  \BibitemOpen
  \bibfield  {author} {\bibinfo {author} {\bibfnamefont {R.}~\bibnamefont
  {{Holman}}}\ and\ \bibinfo {author} {\bibfnamefont {A.~J.}\ \bibnamefont
  {{Tolley}}},\ }\bibfield  {title} {\enquote {\bibinfo {title} {{Enhanced
  non-Gaussianity from excited initial states}},}\ }\href {\doibase
  10.1088/1475-7516/2008/05/001} {\bibfield  {journal} {\bibinfo  {journal}
  {\jcap}\ }\textbf {\bibinfo {volume} {5}},\ \bibinfo {pages} {1--+} (\bibinfo
  {year} {2008})},\ \Eprint {http://arxiv.org/abs/0710.1302} {arXiv:0710.1302
  [hep-th]} \BibitemShut {NoStop}%
\bibitem [{\citenamefont {{Meerburg}}\ \emph {et~al.}(2009)\citenamefont
  {{Meerburg}}, \citenamefont {{van der Schaar}},\ and\ \citenamefont {{Stefano
  Corasaniti}}}]{2009JCAP...05..018M}%
  \BibitemOpen
  \bibfield  {author} {\bibinfo {author} {\bibfnamefont {P.~D.}\ \bibnamefont
  {{Meerburg}}}, \bibinfo {author} {\bibfnamefont {J.~P.}\ \bibnamefont {{van
  der Schaar}}}, \ and\ \bibinfo {author} {\bibfnamefont {P.}~\bibnamefont
  {{Stefano Corasaniti}}},\ }\bibfield  {title} {\enquote {\bibinfo {title}
  {{Signatures of initial state modifications on bispectrum statistics}},}\
  }\href {\doibase 10.1088/1475-7516/2009/05/018} {\bibfield  {journal}
  {\bibinfo  {journal} {\jcap}\ }\textbf {\bibinfo {volume} {5}},\ \bibinfo
  {pages} {18--+} (\bibinfo {year} {2009})},\ \Eprint
  {http://arxiv.org/abs/0901.4044} {arXiv:0901.4044 [hep-th]} \BibitemShut
  {NoStop}%
\bibitem [{\citenamefont {{Meerburg}}\ \emph {et~al.}(2010)\citenamefont
  {{Meerburg}}, \citenamefont {{van der Schaar}},\ and\ \citenamefont
  {{Jackson}}}]{2010JCAP...02..001M}%
  \BibitemOpen
  \bibfield  {author} {\bibinfo {author} {\bibfnamefont {P.~D.}\ \bibnamefont
  {{Meerburg}}}, \bibinfo {author} {\bibfnamefont {J.~P.}\ \bibnamefont {{van
  der Schaar}}}, \ and\ \bibinfo {author} {\bibfnamefont {M.~G.}\ \bibnamefont
  {{Jackson}}},\ }\bibfield  {title} {\enquote {\bibinfo {title} {{Bispectrum
  signatures of a modified vacuum in single field inflation with a small speed
  of sound}},}\ }\href {\doibase 10.1088/1475-7516/2010/02/001} {\bibfield
  {journal} {\bibinfo  {journal} {\jcap}\ }\textbf {\bibinfo {volume} {2}},\
  \bibinfo {pages} {1--+} (\bibinfo {year} {2010})},\ \Eprint
  {http://arxiv.org/abs/0910.4986} {arXiv:0910.4986 [hep-th]} \BibitemShut
  {NoStop}%
\bibitem [{\citenamefont {{Ashoorioon}}\ and\ \citenamefont
  {{Shiu}}(2011)}]{AshShi1103}%
  \BibitemOpen
  \bibfield  {author} {\bibinfo {author} {\bibfnamefont {A.}~\bibnamefont
  {{Ashoorioon}}}\ and\ \bibinfo {author} {\bibfnamefont {G.}~\bibnamefont
  {{Shiu}}},\ }\bibfield  {title} {\enquote {\bibinfo {title} {{A note on calm
  excited states of inflation}},}\ }\href {\doibase
  10.1088/1475-7516/2011/03/025} {\bibfield  {journal} {\bibinfo  {journal}
  {\jcap}\ }\textbf {\bibinfo {volume} {3}},\ \bibinfo {pages} {25} (\bibinfo
  {year} {2011})},\ \Eprint {http://arxiv.org/abs/1012.3392} {arXiv:1012.3392
  [astro-ph.CO]} \BibitemShut {NoStop}%
\bibitem [{\citenamefont {{Flauger}}\ and\ \citenamefont
  {{Pajer}}(2010)}]{2010arXiv1002.0833F}%
  \BibitemOpen
  \bibfield  {author} {\bibinfo {author} {\bibfnamefont {R.}~\bibnamefont
  {{Flauger}}}\ and\ \bibinfo {author} {\bibfnamefont {E.}~\bibnamefont
  {{Pajer}}},\ }\bibfield  {title} {\enquote {\bibinfo {title} {{Resonant
  Non-Gaussianity}},}\ }\href@noop {} {\bibfield  {journal} {\bibinfo
  {journal} {ArXiv e-prints}\ } (\bibinfo {year} {2010})},\ \Eprint
  {http://arxiv.org/abs/1002.0833} {arXiv:1002.0833 [hep-th]} \BibitemShut
  {NoStop}%
\bibitem [{\citenamefont {{Cheung}}\ \emph {et~al.}(2008)\citenamefont
  {{Cheung}}, \citenamefont {{Fitzpatrick}}, \citenamefont {{Kaplan}},
  \citenamefont {{Senatore}},\ and\ \citenamefont
  {{Creminelli}}}]{2008JHEP...03..014C}%
  \BibitemOpen
  \bibfield  {author} {\bibinfo {author} {\bibfnamefont {C.}~\bibnamefont
  {{Cheung}}}, \bibinfo {author} {\bibfnamefont {A.~L.}\ \bibnamefont
  {{Fitzpatrick}}}, \bibinfo {author} {\bibfnamefont {J.}~\bibnamefont
  {{Kaplan}}}, \bibinfo {author} {\bibfnamefont {L.}~\bibnamefont
  {{Senatore}}}, \ and\ \bibinfo {author} {\bibfnamefont {P.}~\bibnamefont
  {{Creminelli}}},\ }\bibfield  {title} {\enquote {\bibinfo {title} {{The
  effective field theory of inflation}},}\ }\href {\doibase
  10.1088/1126-6708/2008/03/014} {\bibfield  {journal} {\bibinfo  {journal}
  {Journal of High Energy Physics}\ }\textbf {\bibinfo {volume} {3}},\ \bibinfo
  {pages} {14--014} (\bibinfo {year} {2008})},\ \Eprint
  {http://arxiv.org/abs/0709.0293} {arXiv:0709.0293 [hep-th]} \BibitemShut
  {NoStop}%
\bibitem [{\citenamefont {{Enqvist}}\ \emph {et~al.}(2005)\citenamefont
  {{Enqvist}}, \citenamefont {{Jokinen}}, \citenamefont {{Mazumdar}},
  \citenamefont {{Multam{\"a}ki}},\ and\ \citenamefont
  {{V{\"a}ihk{\"o}nen}}}]{EnqJokMaz0504}%
  \BibitemOpen
  \bibfield  {author} {\bibinfo {author} {\bibfnamefont {K.}~\bibnamefont
  {{Enqvist}}}, \bibinfo {author} {\bibfnamefont {A.}~\bibnamefont
  {{Jokinen}}}, \bibinfo {author} {\bibfnamefont {A.}~\bibnamefont
  {{Mazumdar}}}, \bibinfo {author} {\bibfnamefont {T.}~\bibnamefont
  {{Multam{\"a}ki}}}, \ and\ \bibinfo {author} {\bibfnamefont {A.}~\bibnamefont
  {{V{\"a}ihk{\"o}nen}}},\ }\bibfield  {title} {\enquote {\bibinfo {title}
  {{Non-Gaussianity from preheating}},}\ }\href {\doibase
  10.1103/PhysRevLett.94.161301} {\bibfield  {journal} {\bibinfo  {journal}
  {Physical Review Letters}\ }\textbf {\bibinfo {volume} {94}},\ \bibinfo {eid}
  {161301} (\bibinfo {year} {2005})},\ \Eprint
  {http://arxiv.org/abs/arXiv:astro-ph/0411394} {arXiv:astro-ph/0411394}
  \BibitemShut {NoStop}%
\bibitem [{\citenamefont {{Barnaby}}\ and\ \citenamefont
  {{Cline}}(2006)}]{BarCli0605}%
  \BibitemOpen
  \bibfield  {author} {\bibinfo {author} {\bibfnamefont {N.}~\bibnamefont
  {{Barnaby}}}\ and\ \bibinfo {author} {\bibfnamefont {J.~M.}\ \bibnamefont
  {{Cline}}},\ }\bibfield  {title} {\enquote {\bibinfo {title} {{Non-Gaussian
  and nonscale-invariant perturbations from tachyonic preheating in hybrid
  inflation}},}\ }\href {\doibase 10.1103/PhysRevD.73.106012} {\bibfield
  {journal} {\bibinfo  {journal} {\prd}\ }\textbf {\bibinfo {volume} {73}},\
  \bibinfo {eid} {106012} (\bibinfo {year} {2006})},\ \Eprint
  {http://arxiv.org/abs/arXiv:astro-ph/0601481} {arXiv:astro-ph/0601481}
  \BibitemShut {NoStop}%
\bibitem [{\citenamefont {{Barnaby}}\ and\ \citenamefont
  {{Cline}}(2007)}]{BarCli0704}%
  \BibitemOpen
  \bibfield  {author} {\bibinfo {author} {\bibfnamefont {N.}~\bibnamefont
  {{Barnaby}}}\ and\ \bibinfo {author} {\bibfnamefont {J.~M.}\ \bibnamefont
  {{Cline}}},\ }\bibfield  {title} {\enquote {\bibinfo {title}
  {{Non-Gaussianity from tachyonic preheating in hybrid inflation}},}\ }\href
  {\doibase 10.1103/PhysRevD.75.086004} {\bibfield  {journal} {\bibinfo
  {journal} {\prd}\ }\textbf {\bibinfo {volume} {75}},\ \bibinfo {eid} {086004}
  (\bibinfo {year} {2007})},\ \Eprint
  {http://arxiv.org/abs/arXiv:astro-ph/0611750} {arXiv:astro-ph/0611750}
  \BibitemShut {NoStop}%
\bibitem [{\citenamefont {{Chambers}}\ and\ \citenamefont
  {{Rajantie}}(2008{\natexlab{a}})}]{ChaRaj0808}%
  \BibitemOpen
  \bibfield  {author} {\bibinfo {author} {\bibfnamefont {A.}~\bibnamefont
  {{Chambers}}}\ and\ \bibinfo {author} {\bibfnamefont {A.}~\bibnamefont
  {{Rajantie}}},\ }\bibfield  {title} {\enquote {\bibinfo {title}
  {{Non-Gaussianity from massless preheating}},}\ }\href {\doibase
  10.1088/1475-7516/2008/08/002} {\bibfield  {journal} {\bibinfo  {journal}
  {\jcap}\ }\textbf {\bibinfo {volume} {8}},\ \bibinfo {pages} {2} (\bibinfo
  {year} {2008}{\natexlab{a}})},\ \Eprint {http://arxiv.org/abs/0805.4795}
  {arXiv:0805.4795} \BibitemShut {NoStop}%
\bibitem [{\citenamefont {{Chambers}}\ and\ \citenamefont
  {{Rajantie}}(2008{\natexlab{b}})}]{ChaRaj0802}%
  \BibitemOpen
  \bibfield  {author} {\bibinfo {author} {\bibfnamefont {A.}~\bibnamefont
  {{Chambers}}}\ and\ \bibinfo {author} {\bibfnamefont {A.}~\bibnamefont
  {{Rajantie}}},\ }\bibfield  {title} {\enquote {\bibinfo {title} {{Lattice
  Calculation of Non-Gaussian Density Perturbations from the Massless
  Preheating Inflationary Model}},}\ }\href {\doibase
  10.1103/PhysRevLett.100.041302} {\bibfield  {journal} {\bibinfo  {journal}
  {Physical Review Letters}\ }\textbf {\bibinfo {volume} {100}},\ \bibinfo
  {eid} {041302} (\bibinfo {year} {2008}{\natexlab{b}})},\ \Eprint
  {http://arxiv.org/abs/0710.4133} {arXiv:0710.4133} \BibitemShut {NoStop}%
\bibitem [{\citenamefont {{Bond}}\ \emph {et~al.}(2009)\citenamefont {{Bond}},
  \citenamefont {{Frolov}}, \citenamefont {{Huang}},\ and\ \citenamefont
  {{Kofman}}}]{BonFroHua0908}%
  \BibitemOpen
  \bibfield  {author} {\bibinfo {author} {\bibfnamefont {J.~R.}\ \bibnamefont
  {{Bond}}}, \bibinfo {author} {\bibfnamefont {A.~V.}\ \bibnamefont
  {{Frolov}}}, \bibinfo {author} {\bibfnamefont {Z.}~\bibnamefont {{Huang}}}, \
  and\ \bibinfo {author} {\bibfnamefont {L.}~\bibnamefont {{Kofman}}},\
  }\bibfield  {title} {\enquote {\bibinfo {title} {{Non-Gaussian Curvature
  Spikes from Chaotic Billiards in Inflation Preheating}},}\ }\href {\doibase
  10.1103/PhysRevLett.103.071301} {\bibfield  {journal} {\bibinfo  {journal}
  {Physical Review Letters}\ }\textbf {\bibinfo {volume} {103}},\ \bibinfo
  {eid} {071301} (\bibinfo {year} {2009})},\ \Eprint
  {http://arxiv.org/abs/0903.3407} {arXiv:0903.3407 [astro-ph.CO]} \BibitemShut
  {NoStop}%
\bibitem [{\citenamefont {{Figueroa}}\ \emph {et~al.}(2010)\citenamefont
  {{Figueroa}}, \citenamefont {{Caldwell}},\ and\ \citenamefont
  {{Kamionkowski}}}]{FigCalKam1006}%
  \BibitemOpen
  \bibfield  {author} {\bibinfo {author} {\bibfnamefont {D.~G.}\ \bibnamefont
  {{Figueroa}}}, \bibinfo {author} {\bibfnamefont {R.~R.}\ \bibnamefont
  {{Caldwell}}}, \ and\ \bibinfo {author} {\bibfnamefont {M.}~\bibnamefont
  {{Kamionkowski}}},\ }\bibfield  {title} {\enquote {\bibinfo {title}
  {{Non-Gaussianity from self-ordering scalar fields}},}\ }\href {\doibase
  10.1103/PhysRevD.81.123504} {\bibfield  {journal} {\bibinfo  {journal}
  {\prd}\ }\textbf {\bibinfo {volume} {81}},\ \bibinfo {eid} {123504} (\bibinfo
  {year} {2010})},\ \Eprint {http://arxiv.org/abs/1003.0672} {arXiv:1003.0672
  [astro-ph.CO]} \BibitemShut {NoStop}%
\bibitem [{\citenamefont {{Regan}}\ and\ \citenamefont
  {{Shellard}}(2010)}]{RegShe1009}%
  \BibitemOpen
  \bibfield  {author} {\bibinfo {author} {\bibfnamefont {D.~M.}\ \bibnamefont
  {{Regan}}}\ and\ \bibinfo {author} {\bibfnamefont {E.~P.~S.}\ \bibnamefont
  {{Shellard}}},\ }\bibfield  {title} {\enquote {\bibinfo {title} {{Cosmic
  string power spectrum, bispectrum, and trispectrum}},}\ }\href {\doibase
  10.1103/PhysRevD.82.063527} {\bibfield  {journal} {\bibinfo  {journal}
  {\prd}\ }\textbf {\bibinfo {volume} {82}},\ \bibinfo {eid} {063527} (\bibinfo
  {year} {2010})},\ \Eprint {http://arxiv.org/abs/0911.2491} {arXiv:0911.2491
  [astro-ph.CO]} \BibitemShut {NoStop}%
\bibitem [{\citenamefont {{Matarrese}}\ and\ \citenamefont
  {{Verde}}(2008)}]{2008ApJ...677L..77M}%
  \BibitemOpen
  \bibfield  {author} {\bibinfo {author} {\bibfnamefont {S.}~\bibnamefont
  {{Matarrese}}}\ and\ \bibinfo {author} {\bibfnamefont {L.}~\bibnamefont
  {{Verde}}},\ }\bibfield  {title} {\enquote {\bibinfo {title} {{The Effect of
  Primordial Non-Gaussianity on Halo Bias}},}\ }\href {\doibase 10.1086/587840}
  {\bibfield  {journal} {\bibinfo  {journal} {\apjl}\ }\textbf {\bibinfo
  {volume} {677}},\ \bibinfo {pages} {L77--L80} (\bibinfo {year} {2008})},\
  \Eprint {http://arxiv.org/abs/0801.4826} {arXiv:0801.4826} \BibitemShut
  {NoStop}%
\bibitem [{\citenamefont {{Taruya}}\ \emph {et~al.}(2008)\citenamefont
  {{Taruya}}, \citenamefont {{Koyama}},\ and\ \citenamefont
  {{Matsubara}}}]{2008PhRvD..78l3534T}%
  \BibitemOpen
  \bibfield  {author} {\bibinfo {author} {\bibfnamefont {A.}~\bibnamefont
  {{Taruya}}}, \bibinfo {author} {\bibfnamefont {K.}~\bibnamefont {{Koyama}}},
  \ and\ \bibinfo {author} {\bibfnamefont {T.}~\bibnamefont {{Matsubara}}},\
  }\bibfield  {title} {\enquote {\bibinfo {title} {{Signature of primordial
  non-Gaussianity on the matter power spectrum}},}\ }\href {\doibase
  10.1103/PhysRevD.78.123534} {\bibfield  {journal} {\bibinfo  {journal}
  {\prd}\ }\textbf {\bibinfo {volume} {78}},\ \bibinfo {pages} {123534--+}
  (\bibinfo {year} {2008})},\ \Eprint {http://arxiv.org/abs/0808.4085}
  {arXiv:0808.4085} \BibitemShut {NoStop}%
\bibitem [{\citenamefont {{Afshordi}}\ and\ \citenamefont
  {{Tolley}}(2008)}]{2008PhRvD..78l3507A}%
  \BibitemOpen
  \bibfield  {author} {\bibinfo {author} {\bibfnamefont {N.}~\bibnamefont
  {{Afshordi}}}\ and\ \bibinfo {author} {\bibfnamefont {A.~J.}\ \bibnamefont
  {{Tolley}}},\ }\bibfield  {title} {\enquote {\bibinfo {title} {{Primordial
  non-Gaussianity, statistics of collapsed objects, and the integrated
  Sachs-Wolfe effect}},}\ }\href {\doibase 10.1103/PhysRevD.78.123507}
  {\bibfield  {journal} {\bibinfo  {journal} {\prd}\ }\textbf {\bibinfo
  {volume} {78}},\ \bibinfo {pages} {123507--+} (\bibinfo {year} {2008})},\
  \Eprint {http://arxiv.org/abs/0806.1046} {arXiv:0806.1046} \BibitemShut
  {NoStop}%
\bibitem [{\citenamefont {{Sefusatti}}\ and\ \citenamefont
  {{Komatsu}}(2007)}]{2007PhRvD..76h3004S}%
  \BibitemOpen
  \bibfield  {author} {\bibinfo {author} {\bibfnamefont {E.}~\bibnamefont
  {{Sefusatti}}}\ and\ \bibinfo {author} {\bibfnamefont {E.}~\bibnamefont
  {{Komatsu}}},\ }\bibfield  {title} {\enquote {\bibinfo {title} {{Bispectrum
  of galaxies from high-redshift galaxy surveys: Primordial non-Gaussianity and
  nonlinear galaxy bias}},}\ }\href {\doibase 10.1103/PhysRevD.76.083004}
  {\bibfield  {journal} {\bibinfo  {journal} {\prd}\ }\textbf {\bibinfo
  {volume} {76}},\ \bibinfo {pages} {083004--+} (\bibinfo {year} {2007})},\
  \Eprint {http://arxiv.org/abs/0705.0343} {arXiv:0705.0343} \BibitemShut
  {NoStop}%
\bibitem [{\citenamefont {{Jeong}}\ and\ \citenamefont
  {{Komatsu}}(2009)}]{2009ApJ...703.1230J}%
  \BibitemOpen
  \bibfield  {author} {\bibinfo {author} {\bibfnamefont {D.}~\bibnamefont
  {{Jeong}}}\ and\ \bibinfo {author} {\bibfnamefont {E.}~\bibnamefont
  {{Komatsu}}},\ }\bibfield  {title} {\enquote {\bibinfo {title} {{Primordial
  Non-Gaussianity, Scale-dependent Bias, and the Bispectrum of Galaxies}},}\
  }\href {\doibase 10.1088/0004-637X/703/2/1230} {\bibfield  {journal}
  {\bibinfo  {journal} {\apj}\ }\textbf {\bibinfo {volume} {703}},\ \bibinfo
  {pages} {1230--1248} (\bibinfo {year} {2009})},\ \Eprint
  {http://arxiv.org/abs/0904.0497} {arXiv:0904.0497 [astro-ph.CO]} \BibitemShut
  {NoStop}%
\bibitem [{\citenamefont {{Nishimichi}}\ \emph {et~al.}(2010)\citenamefont
  {{Nishimichi}}, \citenamefont {{Taruya}}, \citenamefont {{Koyama}},\ and\
  \citenamefont {{Sabiu}}}]{2010JCAP...07..002N}%
  \BibitemOpen
  \bibfield  {author} {\bibinfo {author} {\bibfnamefont {T.}~\bibnamefont
  {{Nishimichi}}}, \bibinfo {author} {\bibfnamefont {A.}~\bibnamefont
  {{Taruya}}}, \bibinfo {author} {\bibfnamefont {K.}~\bibnamefont {{Koyama}}},
  \ and\ \bibinfo {author} {\bibfnamefont {C.}~\bibnamefont {{Sabiu}}},\
  }\bibfield  {title} {\enquote {\bibinfo {title} {{Scale dependence of halo
  bispectrum from non-Gaussian initial conditions in cosmological N-body
  simulations}},}\ }\href {\doibase 10.1088/1475-7516/2010/07/002} {\bibfield
  {journal} {\bibinfo  {journal} {\jcap}\ }\textbf {\bibinfo {volume} {7}},\
  \bibinfo {pages} {2--+} (\bibinfo {year} {2010})},\ \Eprint
  {http://arxiv.org/abs/0911.4768} {arXiv:0911.4768 [astro-ph.CO]} \BibitemShut
  {NoStop}%
\bibitem [{\citenamefont {{Baldauf}}\ \emph {et~al.}(2011)\citenamefont
  {{Baldauf}}, \citenamefont {{Seljak}},\ and\ \citenamefont
  {{Senatore}}}]{2011JCAP...04..006B}%
  \BibitemOpen
  \bibfield  {author} {\bibinfo {author} {\bibfnamefont {T.}~\bibnamefont
  {{Baldauf}}}, \bibinfo {author} {\bibfnamefont {U.}~\bibnamefont {{Seljak}}},
  \ and\ \bibinfo {author} {\bibfnamefont {L.}~\bibnamefont {{Senatore}}},\
  }\bibfield  {title} {\enquote {\bibinfo {title} {{Primordial non-Gaussianity
  in the bispectrum of the halo density field}},}\ }\href {\doibase
  10.1088/1475-7516/2011/04/006} {\bibfield  {journal} {\bibinfo  {journal}
  {\jcap}\ }\textbf {\bibinfo {volume} {4}},\ \bibinfo {pages} {6--+} (\bibinfo
  {year} {2011})},\ \Eprint {http://arxiv.org/abs/1011.1513} {arXiv:1011.1513
  [astro-ph.CO]} \BibitemShut {NoStop}%
\bibitem [{\citenamefont {{Sefusatti}}\ \emph {et~al.}(2011)\citenamefont
  {{Sefusatti}}, \citenamefont {{Crocce}},\ and\ \citenamefont
  {{Desjacques}}}]{EmiHaloBisp}%
  \BibitemOpen
  \bibfield  {author} {\bibinfo {author} {\bibfnamefont {E.}~\bibnamefont
  {{Sefusatti}}}, \bibinfo {author} {\bibfnamefont {M.}~\bibnamefont
  {{Crocce}}}, \ and\ \bibinfo {author} {\bibfnamefont {V.}~\bibnamefont
  {{Desjacques}}},\ }\bibfield  {title} {\enquote {\bibinfo {title} {{The halo
  bispectrum in N-body simulations with non-Gaussian initial conditions}},}\
  }\href@noop {} {\bibfield  {journal} {\bibinfo  {journal} {In preparation}\ }
  (\bibinfo {year} {2011})}\BibitemShut {NoStop}%
\bibitem [{\citenamefont {{Sefusatti}}\ \emph {et~al.}(2009)\citenamefont
  {{Sefusatti}}, \citenamefont {{Liguori}}, \citenamefont {{Yadav}},
  \citenamefont {{Jackson}},\ and\ \citenamefont
  {{Pajer}}}]{2009JCAP...12..022S}%
  \BibitemOpen
  \bibfield  {author} {\bibinfo {author} {\bibfnamefont {E.}~\bibnamefont
  {{Sefusatti}}}, \bibinfo {author} {\bibfnamefont {M.}~\bibnamefont
  {{Liguori}}}, \bibinfo {author} {\bibfnamefont {A.~P.~S.}\ \bibnamefont
  {{Yadav}}}, \bibinfo {author} {\bibfnamefont {M.~G.}\ \bibnamefont
  {{Jackson}}}, \ and\ \bibinfo {author} {\bibfnamefont {E.}~\bibnamefont
  {{Pajer}}},\ }\bibfield  {title} {\enquote {\bibinfo {title} {{Constraining
  running non-gaussianity}},}\ }\href {\doibase 10.1088/1475-7516/2009/12/022}
  {\bibfield  {journal} {\bibinfo  {journal} {\jcap}\ }\textbf {\bibinfo
  {volume} {12}},\ \bibinfo {pages} {22--+} (\bibinfo {year} {2009})},\ \Eprint
  {http://arxiv.org/abs/0906.0232} {arXiv:0906.0232 [astro-ph.CO]} \BibitemShut
  {NoStop}%
\bibitem [{\citenamefont {{Byrnes}}\ \emph
  {et~al.}(2010{\natexlab{a}})\citenamefont {{Byrnes}}, \citenamefont
  {{Nurmi}}, \citenamefont {{Tasinato}},\ and\ \citenamefont
  {{Wands}}}]{2010JCAP...02..034B}%
  \BibitemOpen
  \bibfield  {author} {\bibinfo {author} {\bibfnamefont {C.~T.}\ \bibnamefont
  {{Byrnes}}}, \bibinfo {author} {\bibfnamefont {S.}~\bibnamefont {{Nurmi}}},
  \bibinfo {author} {\bibfnamefont {G.}~\bibnamefont {{Tasinato}}}, \ and\
  \bibinfo {author} {\bibfnamefont {D.}~\bibnamefont {{Wands}}},\ }\bibfield
  {title} {\enquote {\bibinfo {title} {{Scale dependence of local $\fnl$}},}\
  }\href {\doibase 10.1088/1475-7516/2010/02/034} {\bibfield  {journal}
  {\bibinfo  {journal} {\jcap}\ }\textbf {\bibinfo {volume} {2}},\ \bibinfo
  {pages} {34--+} (\bibinfo {year} {2010}{\natexlab{a}})},\ \Eprint
  {http://arxiv.org/abs/0911.2780} {arXiv:0911.2780 [astro-ph.CO]} \BibitemShut
  {NoStop}%
\bibitem [{\citenamefont {{Byrnes}}\ \emph
  {et~al.}(2010{\natexlab{b}})\citenamefont {{Byrnes}}, \citenamefont
  {{Gerstenlauer}}, \citenamefont {{Nurmi}}, \citenamefont {{Tasinato}},\ and\
  \citenamefont {{Wands}}}]{2010JCAP...10..004B}%
  \BibitemOpen
  \bibfield  {author} {\bibinfo {author} {\bibfnamefont {C.~T.}\ \bibnamefont
  {{Byrnes}}}, \bibinfo {author} {\bibfnamefont {M.}~\bibnamefont
  {{Gerstenlauer}}}, \bibinfo {author} {\bibfnamefont {S.}~\bibnamefont
  {{Nurmi}}}, \bibinfo {author} {\bibfnamefont {G.}~\bibnamefont {{Tasinato}}},
  \ and\ \bibinfo {author} {\bibfnamefont {D.}~\bibnamefont {{Wands}}},\
  }\bibfield  {title} {\enquote {\bibinfo {title} {{Scale-dependent
  non-Gaussianity probes inflationary physics}},}\ }\href {\doibase
  10.1088/1475-7516/2010/10/004} {\bibfield  {journal} {\bibinfo  {journal}
  {\jcap}\ }\textbf {\bibinfo {volume} {10}},\ \bibinfo {pages} {4--+}
  (\bibinfo {year} {2010}{\natexlab{b}})},\ \Eprint
  {http://arxiv.org/abs/1007.4277} {arXiv:1007.4277 [astro-ph.CO]} \BibitemShut
  {NoStop}%
\bibitem [{\citenamefont {{Becker}}\ \emph {et~al.}(2011)\citenamefont
  {{Becker}}, \citenamefont {{Huterer}},\ and\ \citenamefont
  {{Kadota}}}]{2011JCAP...01..006B}%
  \BibitemOpen
  \bibfield  {author} {\bibinfo {author} {\bibfnamefont {A.}~\bibnamefont
  {{Becker}}}, \bibinfo {author} {\bibfnamefont {D.}~\bibnamefont {{Huterer}}},
  \ and\ \bibinfo {author} {\bibfnamefont {K.}~\bibnamefont {{Kadota}}},\
  }\bibfield  {title} {\enquote {\bibinfo {title} {{Scale-dependent
  non-Gaussianity as a generalization of the local model}},}\ }\href {\doibase
  10.1088/1475-7516/2011/01/006} {\bibfield  {journal} {\bibinfo  {journal}
  {\jcap}\ }\textbf {\bibinfo {volume} {1}},\ \bibinfo {pages} {6--+} (\bibinfo
  {year} {2011})},\ \Eprint {http://arxiv.org/abs/1009.4189} {arXiv:1009.4189
  [astro-ph.CO]} \BibitemShut {NoStop}%
\bibitem [{\citenamefont {{Huang}}(2010)}]{Hua1012}%
  \BibitemOpen
  \bibfield  {author} {\bibinfo {author} {\bibfnamefont {Q.-G.}\ \bibnamefont
  {{Huang}}},\ }\bibfield  {title} {\enquote {\bibinfo {title} {{Scale
  dependence of f$_{NL}$ in N-flation}},}\ }\href {\doibase
  10.1088/1475-7516/2010/12/017} {\bibfield  {journal} {\bibinfo  {journal}
  {\jcap}\ }\textbf {\bibinfo {volume} {12}},\ \bibinfo {pages} {17} (\bibinfo
  {year} {2010})},\ \Eprint {http://arxiv.org/abs/1009.3326} {arXiv:1009.3326
  [astro-ph.CO]} \BibitemShut {NoStop}%
\bibitem [{\citenamefont {{Shandera}}\ \emph {et~al.}(2011)\citenamefont
  {{Shandera}}, \citenamefont {{Dalal}},\ and\ \citenamefont
  {{Huterer}}}]{2011JCAP...03..017S}%
  \BibitemOpen
  \bibfield  {author} {\bibinfo {author} {\bibfnamefont {S.}~\bibnamefont
  {{Shandera}}}, \bibinfo {author} {\bibfnamefont {N.}~\bibnamefont {{Dalal}}},
  \ and\ \bibinfo {author} {\bibfnamefont {D.}~\bibnamefont {{Huterer}}},\
  }\bibfield  {title} {\enquote {\bibinfo {title} {{A generalized local ansatz
  and its effect on halo bias}},}\ }\href {\doibase
  10.1088/1475-7516/2011/03/017} {\bibfield  {journal} {\bibinfo  {journal}
  {\jcap}\ }\textbf {\bibinfo {volume} {3}},\ \bibinfo {pages} {17--+}
  (\bibinfo {year} {2011})},\ \Eprint {http://arxiv.org/abs/1010.3722}
  {arXiv:1010.3722 [astro-ph.CO]} \BibitemShut {NoStop}%
\bibitem [{\citenamefont {{Schmidt}}\ and\ \citenamefont
  {{Kamionkowski}}(2010)}]{2010PhRvD..82j3002S}%
  \BibitemOpen
  \bibfield  {author} {\bibinfo {author} {\bibfnamefont {F.}~\bibnamefont
  {{Schmidt}}}\ and\ \bibinfo {author} {\bibfnamefont {M.}~\bibnamefont
  {{Kamionkowski}}},\ }\bibfield  {title} {\enquote {\bibinfo {title} {{Halo
  clustering with nonlocal non-Gaussianity}},}\ }\href {\doibase
  10.1103/PhysRevD.82.103002} {\bibfield  {journal} {\bibinfo  {journal}
  {\prd}\ }\textbf {\bibinfo {volume} {82}},\ \bibinfo {pages} {103002--+}
  (\bibinfo {year} {2010})},\ \Eprint {http://arxiv.org/abs/1008.0638}
  {arXiv:1008.0638 [astro-ph.CO]} \BibitemShut {NoStop}%
\bibitem [{\citenamefont {{Byrnes}}\ \emph {et~al.}(2006)\citenamefont
  {{Byrnes}}, \citenamefont {{Sasaki}},\ and\ \citenamefont
  {{Wands}}}]{2006PhRvD..74l3519B}%
  \BibitemOpen
  \bibfield  {author} {\bibinfo {author} {\bibfnamefont {C.~T.}\ \bibnamefont
  {{Byrnes}}}, \bibinfo {author} {\bibfnamefont {M.}~\bibnamefont {{Sasaki}}},
  \ and\ \bibinfo {author} {\bibfnamefont {D.}~\bibnamefont {{Wands}}},\
  }\bibfield  {title} {\enquote {\bibinfo {title} {{Primordial trispectrum from
  inflation}},}\ }\href {\doibase 10.1103/PhysRevD.74.123519} {\bibfield
  {journal} {\bibinfo  {journal} {\prd}\ }\textbf {\bibinfo {volume} {74}},\
  \bibinfo {pages} {123519--+} (\bibinfo {year} {2006})},\ \Eprint
  {http://arxiv.org/abs/arXiv:astro-ph/0611075} {arXiv:astro-ph/0611075}
  \BibitemShut {NoStop}%
\bibitem [{\citenamefont {{Bernardeau}}\ \emph {et~al.}(2002)\citenamefont
  {{Bernardeau}}, \citenamefont {{Colombi}}, \citenamefont {{Gaztanaga}},\ and\
  \citenamefont {{Scoccimarro}}}]{BerColGaz02}%
  \BibitemOpen
  \bibfield  {author} {\bibinfo {author} {\bibfnamefont {F.}~\bibnamefont
  {{Bernardeau}}}, \bibinfo {author} {\bibfnamefont {S.}~\bibnamefont
  {{Colombi}}}, \bibinfo {author} {\bibfnamefont {E.}~\bibnamefont
  {{Gaztanaga}}}, \ and\ \bibinfo {author} {\bibfnamefont {R.}~\bibnamefont
  {{Scoccimarro}}},\ }\bibfield  {title} {\enquote {\bibinfo {title}
  {{Large-scale structure of the Universe and cosmological perturbation
  theory.}}}\ }\href@noop {} {\bibfield  {journal} {\bibinfo  {journal}
  {\physrep}\ }\textbf {\bibinfo {volume} {367}},\ \bibinfo {pages} {1--128}
  (\bibinfo {year} {2002})}\BibitemShut {NoStop}%
\bibitem [{\citenamefont {{Smith}}\ and\ \citenamefont
  {{Zaldarriaga}}(2006)}]{2006astro.ph.12571S}%
  \BibitemOpen
  \bibfield  {author} {\bibinfo {author} {\bibfnamefont {K.~M.}\ \bibnamefont
  {{Smith}}}\ and\ \bibinfo {author} {\bibfnamefont {M.}~\bibnamefont
  {{Zaldarriaga}}},\ }\bibfield  {title} {\enquote {\bibinfo {title}
  {{Algorithms for bispectra: forecasting, optimal analysis, and
  simulation}},}\ }\href@noop {} {\bibfield  {journal} {\bibinfo  {journal}
  {ArXiv Astrophysics e-prints}\ } (\bibinfo {year} {2006})},\ \Eprint
  {http://arxiv.org/abs/astro-ph/0612571} {astro-ph/0612571} \BibitemShut
  {NoStop}%
\bibitem [{\citenamefont {{Fergusson}}\ \emph
  {et~al.}(2010{\natexlab{a}})\citenamefont {{Fergusson}}, \citenamefont
  {{Liguori}},\ and\ \citenamefont {{Shellard}}}]{2010PhRvD..82b3502F}%
  \BibitemOpen
  \bibfield  {author} {\bibinfo {author} {\bibfnamefont {J.~R.}\ \bibnamefont
  {{Fergusson}}}, \bibinfo {author} {\bibfnamefont {M.}~\bibnamefont
  {{Liguori}}}, \ and\ \bibinfo {author} {\bibfnamefont {E.~P.~S.}\
  \bibnamefont {{Shellard}}},\ }\bibfield  {title} {\enquote {\bibinfo {title}
  {{General CMB and primordial bispectrum estimation: Mode expansion, map
  making, and measures of $\fnl$}},}\ }\href {\doibase
  10.1103/PhysRevD.82.023502} {\bibfield  {journal} {\bibinfo  {journal}
  {\prd}\ }\textbf {\bibinfo {volume} {82}},\ \bibinfo {pages} {023502--+}
  (\bibinfo {year} {2010}{\natexlab{a}})},\ \Eprint
  {http://arxiv.org/abs/0912.5516} {arXiv:0912.5516 [astro-ph.CO]} \BibitemShut
  {NoStop}%
\bibitem [{\citenamefont {{Wagner}}\ \emph {et~al.}(2010)\citenamefont
  {{Wagner}}, \citenamefont {{Verde}},\ and\ \citenamefont
  {{Boubekeur}}}]{2010JCAP...10..022W}%
  \BibitemOpen
  \bibfield  {author} {\bibinfo {author} {\bibfnamefont {C.}~\bibnamefont
  {{Wagner}}}, \bibinfo {author} {\bibfnamefont {L.}~\bibnamefont {{Verde}}}, \
  and\ \bibinfo {author} {\bibfnamefont {L.}~\bibnamefont {{Boubekeur}}},\
  }\bibfield  {title} {\enquote {\bibinfo {title} {{N-body simulations with
  generic non-Gaussian initial conditions I: power spectrum and halo mass
  function}},}\ }\href {\doibase 10.1088/1475-7516/2010/10/022} {\bibfield
  {journal} {\bibinfo  {journal} {\jcap}\ }\textbf {\bibinfo {volume} {10}},\
  \bibinfo {pages} {22--+} (\bibinfo {year} {2010})},\ \Eprint
  {http://arxiv.org/abs/1006.5793} {arXiv:1006.5793 [astro-ph.CO]} \BibitemShut
  {NoStop}%
\bibitem [{\citenamefont {{Fergusson}}\ \emph
  {et~al.}(2010{\natexlab{b}})\citenamefont {{Fergusson}}, \citenamefont
  {{Regan}},\ and\ \citenamefont {{Shellard}}}]{2010arXiv1008.1730F}%
  \BibitemOpen
  \bibfield  {author} {\bibinfo {author} {\bibfnamefont {J.~R.}\ \bibnamefont
  {{Fergusson}}}, \bibinfo {author} {\bibfnamefont {D.~M.}\ \bibnamefont
  {{Regan}}}, \ and\ \bibinfo {author} {\bibfnamefont {E.~P.~S.}\ \bibnamefont
  {{Shellard}}},\ }\bibfield  {title} {\enquote {\bibinfo {title} {{Rapid
  Separable Analysis of Higher Order Correlators in Large Scale Structure}},}\
  }\href@noop {} {\bibfield  {journal} {\bibinfo  {journal} {ArXiv e-prints}\ }
  (\bibinfo {year} {2010}{\natexlab{b}})},\ \Eprint
  {http://arxiv.org/abs/1008.1730} {arXiv:1008.1730 [astro-ph.CO]} \BibitemShut
  {NoStop}%
\bibitem [{\citenamefont {{Wagner}}\ and\ \citenamefont
  {{Verde}}(2011)}]{2011arXiv1102.3229W}%
  \BibitemOpen
  \bibfield  {author} {\bibinfo {author} {\bibfnamefont {C.}~\bibnamefont
  {{Wagner}}}\ and\ \bibinfo {author} {\bibfnamefont {L.}~\bibnamefont
  {{Verde}}},\ }\bibfield  {title} {\enquote {\bibinfo {title} {{N-body
  simulations with generic non-Gaussian initial conditions II: Halo bias}},}\
  }\href@noop {} {\bibfield  {journal} {\bibinfo  {journal} {ArXiv e-prints}\ }
  (\bibinfo {year} {2011})},\ \Eprint {http://arxiv.org/abs/1102.3229}
  {arXiv:1102.3229 [astro-ph.CO]} \BibitemShut {NoStop}%
\bibitem [{\citenamefont {{Scoccimarro}}(1998)}]{Sco98}%
  \BibitemOpen
  \bibfield  {author} {\bibinfo {author} {\bibfnamefont {R.}~\bibnamefont
  {{Scoccimarro}}},\ }\bibfield  {title} {\enquote {\bibinfo {title}
  {{Transients from initial conditions: a perturbative analysis}},}\
  }\href@noop {} {\bibfield  {journal} {\bibinfo  {journal} {\mnras}\ }\textbf
  {\bibinfo {volume} {299}},\ \bibinfo {pages} {1097--1118} (\bibinfo {year}
  {1998})}\BibitemShut {NoStop}%
\bibitem [{\citenamefont {{Sirko}}(2005)}]{Sir0511}%
  \BibitemOpen
  \bibfield  {author} {\bibinfo {author} {\bibfnamefont {E.}~\bibnamefont
  {{Sirko}}},\ }\bibfield  {title} {\enquote {\bibinfo {title} {{Initial
  Conditions to Cosmological N-Body Simulations, or, How to Run an Ensemble of
  Simulations}},}\ }\href {\doibase 10.1086/497090} {\bibfield  {journal}
  {\bibinfo  {journal} {\apj}\ }\textbf {\bibinfo {volume} {634}},\ \bibinfo
  {pages} {728--743} (\bibinfo {year} {2005})},\ \Eprint
  {http://arxiv.org/abs/arXiv:astro-ph/0503106} {arXiv:astro-ph/0503106}
  \BibitemShut {NoStop}%
\bibitem [{\citenamefont {{Crocce}}\ \emph {et~al.}(2006)\citenamefont
  {{Crocce}}, \citenamefont {{Pueblas}},\ and\ \citenamefont
  {{Scoccimarro}}}]{2006MNRAS.373..369C}%
  \BibitemOpen
  \bibfield  {author} {\bibinfo {author} {\bibfnamefont {M.}~\bibnamefont
  {{Crocce}}}, \bibinfo {author} {\bibfnamefont {S.}~\bibnamefont {{Pueblas}}},
  \ and\ \bibinfo {author} {\bibfnamefont {R.}~\bibnamefont {{Scoccimarro}}},\
  }\bibfield  {title} {\enquote {\bibinfo {title} {{Transients from initial
  conditions in cosmological simulations}},}\ }\href {\doibase
  10.1111/j.1365-2966.2006.11040.x} {\bibfield  {journal} {\bibinfo  {journal}
  {\mnras}\ }\textbf {\bibinfo {volume} {373}},\ \bibinfo {pages} {369--381}
  (\bibinfo {year} {2006})},\ \Eprint
  {http://arxiv.org/abs/arXiv:astro-ph/0606505} {arXiv:astro-ph/0606505}
  \BibitemShut {NoStop}%
\bibitem [{\citenamefont {{Crocce}}\ \emph {et~al.}(2010)\citenamefont
  {{Crocce}}, \citenamefont {{Fosalba}}, \citenamefont {{Castander}},\ and\
  \citenamefont {{Gazta{\~n}aga}}}]{2010MNRAS.403.1353C}%
  \BibitemOpen
  \bibfield  {author} {\bibinfo {author} {\bibfnamefont {M.}~\bibnamefont
  {{Crocce}}}, \bibinfo {author} {\bibfnamefont {P.}~\bibnamefont {{Fosalba}}},
  \bibinfo {author} {\bibfnamefont {F.~J.}\ \bibnamefont {{Castander}}}, \ and\
  \bibinfo {author} {\bibfnamefont {E.}~\bibnamefont {{Gazta{\~n}aga}}},\
  }\bibfield  {title} {\enquote {\bibinfo {title} {{Simulating the Universe
  with MICE: the abundance of massive clusters}},}\ }\href {\doibase
  10.1111/j.1365-2966.2009.16194.x} {\bibfield  {journal} {\bibinfo  {journal}
  {\mnras}\ }\textbf {\bibinfo {volume} {403}},\ \bibinfo {pages} {1353--1367}
  (\bibinfo {year} {2010})},\ \Eprint {http://arxiv.org/abs/0907.0019}
  {arXiv:0907.0019 [astro-ph.CO]} \BibitemShut {NoStop}%
\bibitem [{\citenamefont {{Jenkins}}(2010)}]{2010MNRAS.403.1859J}%
  \BibitemOpen
  \bibfield  {author} {\bibinfo {author} {\bibfnamefont {A.}~\bibnamefont
  {{Jenkins}}},\ }\bibfield  {title} {\enquote {\bibinfo {title} {{Second-order
  Lagrangian perturbation theory initial conditions for resimulations}},}\
  }\href {\doibase 10.1111/j.1365-2966.2010.16259.x} {\bibfield  {journal}
  {\bibinfo  {journal} {\mnras}\ }\textbf {\bibinfo {volume} {403}},\ \bibinfo
  {pages} {1859--1872} (\bibinfo {year} {2010})},\ \Eprint
  {http://arxiv.org/abs/0910.0258} {arXiv:0910.0258 [astro-ph.CO]} \BibitemShut
  {NoStop}%
\bibitem [{\citenamefont {{Hahn}}\ and\ \citenamefont
  {{Abel}}(2011)}]{2011arXiv1103.6031H}%
  \BibitemOpen
  \bibfield  {author} {\bibinfo {author} {\bibfnamefont {O.}~\bibnamefont
  {{Hahn}}}\ and\ \bibinfo {author} {\bibfnamefont {T.}~\bibnamefont
  {{Abel}}},\ }\bibfield  {title} {\enquote {\bibinfo {title} {{Multi-scale
  initial conditions for cosmological simulations}},}\ }\href@noop {}
  {\bibfield  {journal} {\bibinfo  {journal} {ArXiv e-prints}\ } (\bibinfo
  {year} {2011})},\ \Eprint {http://arxiv.org/abs/1103.6031} {arXiv:1103.6031
  [astro-ph.CO]} \BibitemShut {NoStop}%
\bibitem [{\citenamefont {et~al}(2011)}]{LasDamasMF}%
  \BibitemOpen
  \bibfield  {author} {\bibinfo {author} {\bibfnamefont {{C. McBride}}\
  \bibnamefont {et~al}},\ }\href@noop {} {\bibfield  {journal} {\bibinfo
  {journal} {In preparation}\ } (\bibinfo {year} {2011})}\BibitemShut {NoStop}%
\bibitem [{\citenamefont {{Springel}}(2005)}]{2005MNRAS.364.1105S}%
  \BibitemOpen
  \bibfield  {author} {\bibinfo {author} {\bibfnamefont {V.}~\bibnamefont
  {{Springel}}},\ }\bibfield  {title} {\enquote {\bibinfo {title} {{The
  cosmological simulation code GADGET-2}},}\ }\href {\doibase
  10.1111/j.1365-2966.2005.09655.x} {\bibfield  {journal} {\bibinfo  {journal}
  {\mnras}\ }\textbf {\bibinfo {volume} {364}},\ \bibinfo {pages} {1105--1134}
  (\bibinfo {year} {2005})},\ \Eprint
  {http://arxiv.org/abs/arXiv:astro-ph/0505010} {arXiv:astro-ph/0505010}
  \BibitemShut {NoStop}%
\bibitem [{\citenamefont {{Q. Mao}}\ \emph {et~al.}(2011)\citenamefont {{Q.
  Mao}}, \citenamefont {{A.A. Berlind}}, \citenamefont {{C. McBride}},
  \citenamefont {{R. Scherrer}}, \citenamefont {{R. Scoccimarro}},\ and\
  \citenamefont {{M. Manera}}}]{LasDamasS3S4}%
  \BibitemOpen
  \bibfield  {author} {\bibinfo {author} {\bibnamefont {{Q. Mao}}}, \bibinfo
  {author} {\bibnamefont {{A.A. Berlind}}}, \bibinfo {author} {\bibnamefont
  {{C. McBride}}}, \bibinfo {author} {\bibnamefont {{R. Scherrer}}}, \bibinfo
  {author} {\bibnamefont {{R. Scoccimarro}}}, \ and\ \bibinfo {author}
  {\bibnamefont {{M. Manera}}},\ }\bibfield  {title} {\enquote {\bibinfo
  {title} {Constraining primordial non-gaussianity with skewness and kurtosis
  measurements of large-scale structure},}\ }\href@noop {} {\bibfield
  {journal} {\bibinfo  {journal} {In preparation}\ } (\bibinfo {year}
  {2011})}\BibitemShut {NoStop}%
\bibitem [{\citenamefont {{Bardeen}}\ \emph {et~al.}(1986)\citenamefont
  {{Bardeen}}, \citenamefont {{Bond}}, \citenamefont {{Kaiser}},\ and\
  \citenamefont {{Szalay}}}]{1986ApJ...304...15B}%
  \BibitemOpen
  \bibfield  {author} {\bibinfo {author} {\bibfnamefont {J.~M.}\ \bibnamefont
  {{Bardeen}}}, \bibinfo {author} {\bibfnamefont {J.~R.}\ \bibnamefont
  {{Bond}}}, \bibinfo {author} {\bibfnamefont {N.}~\bibnamefont {{Kaiser}}}, \
  and\ \bibinfo {author} {\bibfnamefont {A.~S.}\ \bibnamefont {{Szalay}}},\
  }\bibfield  {title} {\enquote {\bibinfo {title} {{The statistics of peaks of
  Gaussian random fields}},}\ }\href {\doibase 10.1086/164143} {\bibfield
  {journal} {\bibinfo  {journal} {\apj}\ }\textbf {\bibinfo {volume} {304}},\
  \bibinfo {pages} {15--61} (\bibinfo {year} {1986})}\BibitemShut {NoStop}%
\bibitem [{\citenamefont {{Cole}}\ and\ \citenamefont
  {{Kaiser}}(1989)}]{1989MNRAS.237.1127C}%
  \BibitemOpen
  \bibfield  {author} {\bibinfo {author} {\bibfnamefont {S.}~\bibnamefont
  {{Cole}}}\ and\ \bibinfo {author} {\bibfnamefont {N.}~\bibnamefont
  {{Kaiser}}},\ }\bibfield  {title} {\enquote {\bibinfo {title} {{Biased
  clustering in the cold dark matter cosmogony}},}\ }\href@noop {} {\bibfield
  {journal} {\bibinfo  {journal} {\mnras}\ }\textbf {\bibinfo {volume} {237}},\
  \bibinfo {pages} {1127--1146} (\bibinfo {year} {1989})}\BibitemShut {NoStop}%
\bibitem [{\citenamefont {{Bond}}\ \emph {et~al.}(1991)\citenamefont {{Bond}},
  \citenamefont {{Cole}}, \citenamefont {{Efstathiou}},\ and\ \citenamefont
  {{Kaiser}}}]{1991ApJ...379..440B}%
  \BibitemOpen
  \bibfield  {author} {\bibinfo {author} {\bibfnamefont {J.~R.}\ \bibnamefont
  {{Bond}}}, \bibinfo {author} {\bibfnamefont {S.}~\bibnamefont {{Cole}}},
  \bibinfo {author} {\bibfnamefont {G.}~\bibnamefont {{Efstathiou}}}, \ and\
  \bibinfo {author} {\bibfnamefont {N.}~\bibnamefont {{Kaiser}}},\ }\bibfield
  {title} {\enquote {\bibinfo {title} {{Excursion set mass functions for
  hierarchical Gaussian fluctuations}},}\ }\href {\doibase 10.1086/170520}
  {\bibfield  {journal} {\bibinfo  {journal} {\apj}\ }\textbf {\bibinfo
  {volume} {379}},\ \bibinfo {pages} {440--460} (\bibinfo {year}
  {1991})}\BibitemShut {NoStop}%
\bibitem [{\citenamefont {{Paranjape}}\ \emph {et~al.}(2011)\citenamefont
  {{Paranjape}}, \citenamefont {{Lam}},\ and\ \citenamefont
  {{Sheth}}}]{2011arXiv1105.1990P}%
  \BibitemOpen
  \bibfield  {author} {\bibinfo {author} {\bibfnamefont {A.}~\bibnamefont
  {{Paranjape}}}, \bibinfo {author} {\bibfnamefont {T.~Y.}\ \bibnamefont
  {{Lam}}}, \ and\ \bibinfo {author} {\bibfnamefont {R.~K.}\ \bibnamefont
  {{Sheth}}},\ }\bibfield  {title} {\enquote {\bibinfo {title} {{Halo
  abundances and counts-in-cells: The excursion set approach with correlated
  steps}},}\ }\href@noop {} {\bibfield  {journal} {\bibinfo  {journal} {ArXiv
  e-prints}\ } (\bibinfo {year} {2011})},\ \Eprint
  {http://arxiv.org/abs/1105.1990} {arXiv:1105.1990 [astro-ph.CO]} \BibitemShut
  {NoStop}%
\bibitem [{\citenamefont {{Sheth}}\ and\ \citenamefont
  {{Lemson}}(1999)}]{1999MNRAS.304..767S}%
  \BibitemOpen
  \bibfield  {author} {\bibinfo {author} {\bibfnamefont {R.{}K.}\ \bibnamefont
  {{Sheth}}}\ and\ \bibinfo {author} {\bibfnamefont {G.}~\bibnamefont
  {{Lemson}}},\ }\bibfield  {title} {\enquote {\bibinfo {title} {{Biasing and
  the distribution of dark matter haloes}},}\ }\href@noop {} {\bibfield
  {journal} {\bibinfo  {journal} {\mnras}\ }\textbf {\bibinfo {volume} {304}},\
  \bibinfo {pages} {767--792} (\bibinfo {year} {1999})}\BibitemShut {NoStop}%
\bibitem [{\citenamefont {{Mo}}\ \emph {et~al.}(1997)\citenamefont {{Mo}},
  \citenamefont {{Jing}},\ and\ \citenamefont {{White}}}]{MoJinWhi97}%
  \BibitemOpen
  \bibfield  {author} {\bibinfo {author} {\bibfnamefont {H.~J.}\ \bibnamefont
  {{Mo}}}, \bibinfo {author} {\bibfnamefont {Y.~P.}\ \bibnamefont {{Jing}}}, \
  and\ \bibinfo {author} {\bibfnamefont {S.~D.~M.}\ \bibnamefont {{White}}},\
  }\bibfield  {title} {\enquote {\bibinfo {title} {{High-order correlations of
  peaks and haloes: a step towards understanding galaxy biasing}},}\
  }\href@noop {} {\bibfield  {journal} {\bibinfo  {journal} {\mnras}\ }\textbf
  {\bibinfo {volume} {284}},\ \bibinfo {pages} {189--201} (\bibinfo {year}
  {1997})}\BibitemShut {NoStop}%
\bibitem [{\citenamefont {{Scoccimarro}}\ \emph {et~al.}(2001)\citenamefont
  {{Scoccimarro}}, \citenamefont {{Sheth}}, \citenamefont {{Hui}},\ and\
  \citenamefont {{Jain}}}]{ScoSheHui01}%
  \BibitemOpen
  \bibfield  {author} {\bibinfo {author} {\bibfnamefont {R.}~\bibnamefont
  {{Scoccimarro}}}, \bibinfo {author} {\bibfnamefont {R.{}K.}\ \bibnamefont
  {{Sheth}}}, \bibinfo {author} {\bibfnamefont {L.}~\bibnamefont {{Hui}}}, \
  and\ \bibinfo {author} {\bibfnamefont {B.}~\bibnamefont {{Jain}}},\
  }\bibfield  {title} {\enquote {\bibinfo {title} {{How Many Galaxies Fit in a
  Halo? Constraints on Galaxy Formation Efficiency from Spatial Clustering}},}\
  }\href@noop {} {\bibfield  {journal} {\bibinfo  {journal} {\apj}\ }\textbf
  {\bibinfo {volume} {546}},\ \bibinfo {pages} {20--34} (\bibinfo {year}
  {2001})}\BibitemShut {NoStop}%
\bibitem [{\citenamefont {{Smith}}\ \emph {et~al.}(2007)\citenamefont
  {{Smith}}, \citenamefont {{Scoccimarro}},\ and\ \citenamefont
  {{Sheth}}}]{2007PhRvD..75f3512S}%
  \BibitemOpen
  \bibfield  {author} {\bibinfo {author} {\bibfnamefont {R.~E.}\ \bibnamefont
  {{Smith}}}, \bibinfo {author} {\bibfnamefont {R.}~\bibnamefont
  {{Scoccimarro}}}, \ and\ \bibinfo {author} {\bibfnamefont {R.~K.}\
  \bibnamefont {{Sheth}}},\ }\bibfield  {title} {\enquote {\bibinfo {title}
  {{Scale dependence of halo and galaxy bias: Effects in real space}},}\ }\href
  {\doibase 10.1103/PhysRevD.75.063512} {\bibfield  {journal} {\bibinfo
  {journal} {\prd}\ }\textbf {\bibinfo {volume} {75}},\ \bibinfo {pages}
  {063512--+} (\bibinfo {year} {2007})},\ \Eprint
  {http://arxiv.org/abs/arXiv:astro-ph/0609547} {arXiv:astro-ph/0609547}
  \BibitemShut {NoStop}%
\bibitem [{\citenamefont {{Manera}}\ \emph {et~al.}(2010)\citenamefont
  {{Manera}}, \citenamefont {{Sheth}},\ and\ \citenamefont
  {{Scoccimarro}}}]{2010MNRAS.402..589M}%
  \BibitemOpen
  \bibfield  {author} {\bibinfo {author} {\bibfnamefont {M.}~\bibnamefont
  {{Manera}}}, \bibinfo {author} {\bibfnamefont {R.~K.}\ \bibnamefont
  {{Sheth}}}, \ and\ \bibinfo {author} {\bibfnamefont {R.}~\bibnamefont
  {{Scoccimarro}}},\ }\bibfield  {title} {\enquote {\bibinfo {title}
  {{Large-scale bias and the inaccuracy of the peak-background split}},}\
  }\href {\doibase 10.1111/j.1365-2966.2009.15921.x} {\bibfield  {journal}
  {\bibinfo  {journal} {\mnras}\ }\textbf {\bibinfo {volume} {402}},\ \bibinfo
  {pages} {589--602} (\bibinfo {year} {2010})},\ \Eprint
  {http://arxiv.org/abs/0906.1314} {arXiv:0906.1314 [astro-ph.CO]} \BibitemShut
  {NoStop}%
\bibitem [{\citenamefont {{Manera}}\ and\ \citenamefont
  {{Gaztanaga}}(2011)}]{2009arXiv0912.0446M}%
  \BibitemOpen
  \bibfield  {author} {\bibinfo {author} {\bibfnamefont {M.}~\bibnamefont
  {{Manera}}}\ and\ \bibinfo {author} {\bibfnamefont {E.}~\bibnamefont
  {{Gaztanaga}}},\ }\bibfield  {title} {\enquote {\bibinfo {title} {{The Local
  Bias Model in the Large Scale Halo Distribution}},}\ }\href@noop {}
  {\bibfield  {journal} {\bibinfo  {journal} {\mnras}\ }\textbf {\bibinfo
  {volume} {415}},\ \bibinfo {pages} {383--398} (\bibinfo {year} {2011})},\
  \Eprint {http://arxiv.org/abs/0912.0446} {arXiv:0912.0446 [astro-ph.CO]}
  \BibitemShut {NoStop}%
\bibitem [{\citenamefont {{Desjacques}}\ \emph {et~al.}(2010)\citenamefont
  {{Desjacques}}, \citenamefont {{Crocce}}, \citenamefont {{Scoccimarro}},\
  and\ \citenamefont {{Sheth}}}]{DesCroSco1011}%
  \BibitemOpen
  \bibfield  {author} {\bibinfo {author} {\bibfnamefont {V.}~\bibnamefont
  {{Desjacques}}}, \bibinfo {author} {\bibfnamefont {M.}~\bibnamefont
  {{Crocce}}}, \bibinfo {author} {\bibfnamefont {R.}~\bibnamefont
  {{Scoccimarro}}}, \ and\ \bibinfo {author} {\bibfnamefont {R.~K.}\
  \bibnamefont {{Sheth}}},\ }\bibfield  {title} {\enquote {\bibinfo {title}
  {{Modeling scale-dependent bias on the baryonic acoustic scale with the
  statistics of peaks of Gaussian random fields}},}\ }\href {\doibase
  10.1103/PhysRevD.82.103529} {\bibfield  {journal} {\bibinfo  {journal}
  {\prd}\ }\textbf {\bibinfo {volume} {82}},\ \bibinfo {pages} {103529--+}
  (\bibinfo {year} {2010})},\ \Eprint {http://arxiv.org/abs/1009.3449}
  {arXiv:1009.3449 [astro-ph.CO]} \BibitemShut {NoStop}%
\bibitem [{\citenamefont {{Ma}}\ \emph {et~al.}(2010)\citenamefont {{Ma}},
  \citenamefont {{Maggiore}}, \citenamefont {{Riotto}},\ and\ \citenamefont
  {{Zhang}}}]{2010arXiv1007.4201M}%
  \BibitemOpen
  \bibfield  {author} {\bibinfo {author} {\bibfnamefont {{C.-P.}}\ \bibnamefont
  {{Ma}}}, \bibinfo {author} {\bibfnamefont {M.}~\bibnamefont {{Maggiore}}},
  \bibinfo {author} {\bibfnamefont {A.}~\bibnamefont {{Riotto}}}, \ and\
  \bibinfo {author} {\bibfnamefont {J.}~\bibnamefont {{Zhang}}},\ }\bibfield
  {title} {\enquote {\bibinfo {title} {{The Bias and Mass Function of Dark
  Matter Halos in Non-Markovian Extension of the Excursion Set Theory}},}\
  }\href@noop {} {\bibfield  {journal} {\bibinfo  {journal} {ArXiv e-prints}\ }
  (\bibinfo {year} {2010})},\ \Eprint {http://arxiv.org/abs/1007.4201}
  {arXiv:1007.4201 [astro-ph.CO]} \BibitemShut {NoStop}%
\bibitem [{\citenamefont {{Paranjape}}\ and\ \citenamefont
  {{Sheth}}(2011)}]{2011arXiv1105.2261P}%
  \BibitemOpen
  \bibfield  {author} {\bibinfo {author} {\bibfnamefont {A.}~\bibnamefont
  {{Paranjape}}}\ and\ \bibinfo {author} {\bibfnamefont {R.~K.}\ \bibnamefont
  {{Sheth}}},\ }\bibfield  {title} {\enquote {\bibinfo {title} {{Halo bias in
  the excursion set approach with correlated steps}},}\ }\href@noop {}
  {\bibfield  {journal} {\bibinfo  {journal} {ArXiv e-prints}\ } (\bibinfo
  {year} {2011})},\ \Eprint {http://arxiv.org/abs/1105.2261} {arXiv:1105.2261
  [astro-ph.CO]} \BibitemShut {NoStop}%
\bibitem [{\citenamefont {{Peacock}}\ and\ \citenamefont
  {{Heavens}}(1990)}]{1990MNRAS.243..133P}%
  \BibitemOpen
  \bibfield  {author} {\bibinfo {author} {\bibfnamefont {J.~A.}\ \bibnamefont
  {{Peacock}}}\ and\ \bibinfo {author} {\bibfnamefont {A.~F.}\ \bibnamefont
  {{Heavens}}},\ }\bibfield  {title} {\enquote {\bibinfo {title} {{Alternatives
  to the Press-Schechter cosmological mass function}},}\ }\href@noop {}
  {\bibfield  {journal} {\bibinfo  {journal} {\mnras}\ }\textbf {\bibinfo
  {volume} {243}},\ \bibinfo {pages} {133--143} (\bibinfo {year}
  {1990})}\BibitemShut {NoStop}%
\bibitem [{\citenamefont {{Maggiore}}\ and\ \citenamefont
  {{Riotto}}(2010{\natexlab{a}})}]{2010ApJ...711..907M}%
  \BibitemOpen
  \bibfield  {author} {\bibinfo {author} {\bibfnamefont {M.}~\bibnamefont
  {{Maggiore}}}\ and\ \bibinfo {author} {\bibfnamefont {A.}~\bibnamefont
  {{Riotto}}},\ }\bibfield  {title} {\enquote {\bibinfo {title} {{The Halo Mass
  Function from Excursion Set Theory. I. Gaussian Fluctuations with
  Non-Markovian Dependence on the Smoothing Scale}},}\ }\href {\doibase
  10.1088/0004-637X/711/2/907} {\bibfield  {journal} {\bibinfo  {journal}
  {\apj}\ }\textbf {\bibinfo {volume} {711}},\ \bibinfo {pages} {907--927}
  (\bibinfo {year} {2010}{\natexlab{a}})},\ \Eprint
  {http://arxiv.org/abs/0903.1249} {arXiv:0903.1249 [astro-ph.CO]} \BibitemShut
  {NoStop}%
\bibitem [{\citenamefont {{Maggiore}}\ and\ \citenamefont
  {{Riotto}}(2010{\natexlab{b}})}]{2010ApJ...717..515M}%
  \BibitemOpen
  \bibfield  {author} {\bibinfo {author} {\bibfnamefont {M.}~\bibnamefont
  {{Maggiore}}}\ and\ \bibinfo {author} {\bibfnamefont {A.}~\bibnamefont
  {{Riotto}}},\ }\bibfield  {title} {\enquote {\bibinfo {title} {{The Halo mass
  function from Excursion Set Theory. II. The Diffusing Barrier}},}\ }\href
  {\doibase 10.1088/0004-637X/717/1/515} {\bibfield  {journal} {\bibinfo
  {journal} {\apj}\ }\textbf {\bibinfo {volume} {717}},\ \bibinfo {pages}
  {515--525} (\bibinfo {year} {2010}{\natexlab{b}})},\ \Eprint
  {http://arxiv.org/abs/0903.1250} {arXiv:0903.1250 [astro-ph.CO]} \BibitemShut
  {NoStop}%
\bibitem [{\citenamefont {{Maggiore}}\ and\ \citenamefont
  {{Riotto}}(2010{\natexlab{c}})}]{2010ApJ...717..526M}%
  \BibitemOpen
  \bibfield  {author} {\bibinfo {author} {\bibfnamefont {M.}~\bibnamefont
  {{Maggiore}}}\ and\ \bibinfo {author} {\bibfnamefont {A.}~\bibnamefont
  {{Riotto}}},\ }\bibfield  {title} {\enquote {\bibinfo {title} {{The Halo Mass
  Function from Excursion Set Theory. III. Non-Gaussian Fluctuations}},}\
  }\href {\doibase 10.1088/0004-637X/717/1/526} {\bibfield  {journal} {\bibinfo
   {journal} {\apj}\ }\textbf {\bibinfo {volume} {717}},\ \bibinfo {pages}
  {526--541} (\bibinfo {year} {2010}{\natexlab{c}})},\ \Eprint
  {http://arxiv.org/abs/0903.1251} {arXiv:0903.1251 [astro-ph.CO]} \BibitemShut
  {NoStop}%
\bibitem [{\citenamefont {{Desjacques}}\ \emph
  {et~al.}(2011{\natexlab{a}})\citenamefont {{Desjacques}}, \citenamefont
  {{Jeong}},\ and\ \citenamefont {{Schmidt}}}]{2011arXiv1105.3628D}%
  \BibitemOpen
  \bibfield  {author} {\bibinfo {author} {\bibfnamefont {V.}~\bibnamefont
  {{Desjacques}}}, \bibinfo {author} {\bibfnamefont {D.}~\bibnamefont
  {{Jeong}}}, \ and\ \bibinfo {author} {\bibfnamefont {F.}~\bibnamefont
  {{Schmidt}}},\ }\bibfield  {title} {\enquote {\bibinfo {title} {{Non-Gaussian
  Halo Bias Re-examined: Mass-dependent Amplitude from the Peak-Background
  Split and Thresholding}},}\ }\href@noop {} {\bibfield  {journal} {\bibinfo
  {journal} {ArXiv e-prints}\ } (\bibinfo {year} {2011}{\natexlab{a}})},\
  \Eprint {http://arxiv.org/abs/1105.3628} {arXiv:1105.3628 [astro-ph.CO]}
  \BibitemShut {NoStop}%
\bibitem [{\citenamefont {{Desjacques}}\ \emph
  {et~al.}(2011{\natexlab{b}})\citenamefont {{Desjacques}}, \citenamefont
  {{Jeong}},\ and\ \citenamefont {{Schmidt}}}]{2011arXiv1105.3476D}%
  \BibitemOpen
  \bibfield  {author} {\bibinfo {author} {\bibfnamefont {V.}~\bibnamefont
  {{Desjacques}}}, \bibinfo {author} {\bibfnamefont {D.}~\bibnamefont
  {{Jeong}}}, \ and\ \bibinfo {author} {\bibfnamefont {F.}~\bibnamefont
  {{Schmidt}}},\ }\bibfield  {title} {\enquote {\bibinfo {title} {{Accurate
  Predictions for the Scale-Dependent Galaxy Bias from Primordial
  Non-Gaussianity}},}\ }\href@noop {} {\bibfield  {journal} {\bibinfo
  {journal} {ArXiv e-prints}\ } (\bibinfo {year} {2011}{\natexlab{b}})},\
  \Eprint {http://arxiv.org/abs/1105.3476} {arXiv:1105.3476 [astro-ph.CO]}
  \BibitemShut {NoStop}%
\bibitem [{\citenamefont {{Smith}}\ \emph
  {et~al.}(2011{\natexlab{a}})\citenamefont {{Smith}}, \citenamefont
  {{Ferraro}},\ and\ \citenamefont {{LoVerde}}}]{2011arXiv1106.0503S}%
  \BibitemOpen
  \bibfield  {author} {\bibinfo {author} {\bibfnamefont {K.~M.}\ \bibnamefont
  {{Smith}}}, \bibinfo {author} {\bibfnamefont {S.}~\bibnamefont {{Ferraro}}},
  \ and\ \bibinfo {author} {\bibfnamefont {M.}~\bibnamefont {{LoVerde}}},\
  }\bibfield  {title} {\enquote {\bibinfo {title} {{Halo clustering and
  $\gnl$-type primordial non-Gaussianity}},}\ }\href@noop {} {\bibfield
  {journal} {\bibinfo  {journal} {ArXiv e-prints}\ } (\bibinfo {year}
  {2011}{\natexlab{a}})},\ \Eprint {http://arxiv.org/abs/1106.0503}
  {arXiv:1106.0503 [astro-ph.CO]} \BibitemShut {NoStop}%
\bibitem [{\citenamefont {{De Simone}}\ \emph {et~al.}(2010)\citenamefont {{De
  Simone}}, \citenamefont {{Maggiore}},\ and\ \citenamefont
  {{Riotto}}}]{2010arXiv1007.1903D}%
  \BibitemOpen
  \bibfield  {author} {\bibinfo {author} {\bibfnamefont {A.}~\bibnamefont {{De
  Simone}}}, \bibinfo {author} {\bibfnamefont {M.}~\bibnamefont {{Maggiore}}},
  \ and\ \bibinfo {author} {\bibfnamefont {A.}~\bibnamefont {{Riotto}}},\
  }\bibfield  {title} {\enquote {\bibinfo {title} {{Excursion Set Theory for
  generic moving barriers and non-Gaussian initial conditions}},}\ }\href@noop
  {} {\bibfield  {journal} {\bibinfo  {journal} {ArXiv e-prints}\ } (\bibinfo
  {year} {2010})},\ \Eprint {http://arxiv.org/abs/1007.1903} {arXiv:1007.1903
  [astro-ph.CO]} \BibitemShut {NoStop}%
\bibitem [{\citenamefont {{Gong}}\ and\ \citenamefont
  {{Yokoyama}}(2011)}]{2011arXiv1106.4404G}%
  \BibitemOpen
  \bibfield  {author} {\bibinfo {author} {\bibfnamefont {J.-O.}\ \bibnamefont
  {{Gong}}}\ and\ \bibinfo {author} {\bibfnamefont {S.}~\bibnamefont
  {{Yokoyama}}},\ }\bibfield  {title} {\enquote {\bibinfo {title} {{Scale
  dependent bias from primordial non-Gaussianity with trispectrum}},}\
  }\href@noop {} {\bibfield  {journal} {\bibinfo  {journal} {ArXiv e-prints}\ }
  (\bibinfo {year} {2011})},\ \Eprint {http://arxiv.org/abs/1106.4404}
  {arXiv:1106.4404 [astro-ph.CO]} \BibitemShut {NoStop}%
\bibitem [{\citenamefont {{Bernardeau}}\ \emph {et~al.}(2010)\citenamefont
  {{Bernardeau}}, \citenamefont {{Crocce}},\ and\ \citenamefont
  {{Sefusatti}}}]{2010PhRvD..82h3507B}%
  \BibitemOpen
  \bibfield  {author} {\bibinfo {author} {\bibfnamefont {F.}~\bibnamefont
  {{Bernardeau}}}, \bibinfo {author} {\bibfnamefont {M.}~\bibnamefont
  {{Crocce}}}, \ and\ \bibinfo {author} {\bibfnamefont {E.}~\bibnamefont
  {{Sefusatti}}},\ }\bibfield  {title} {\enquote {\bibinfo {title} {{Multipoint
  propagators for non-Gaussian initial conditions}},}\ }\href {\doibase
  10.1103/PhysRevD.82.083507} {\bibfield  {journal} {\bibinfo  {journal}
  {\prd}\ }\textbf {\bibinfo {volume} {82}},\ \bibinfo {pages} {083507--+}
  (\bibinfo {year} {2010})},\ \Eprint {http://arxiv.org/abs/1006.4656}
  {arXiv:1006.4656 [astro-ph.CO]} \BibitemShut {NoStop}%
\bibitem [{\citenamefont {{Giannantonio}}\ and\ \citenamefont
  {{Porciani}}(2010)}]{2010PhRvD..81f3530G}%
  \BibitemOpen
  \bibfield  {author} {\bibinfo {author} {\bibfnamefont {T.}~\bibnamefont
  {{Giannantonio}}}\ and\ \bibinfo {author} {\bibfnamefont {C.}~\bibnamefont
  {{Porciani}}},\ }\bibfield  {title} {\enquote {\bibinfo {title} {{Structure
  formation from non-Gaussian initial conditions: Multivariate biasing,
  statistics, and comparison with N-body simulations}},}\ }\href {\doibase
  10.1103/PhysRevD.81.063530} {\bibfield  {journal} {\bibinfo  {journal}
  {\prd}\ }\textbf {\bibinfo {volume} {81}},\ \bibinfo {pages} {063530--+}
  (\bibinfo {year} {2010})},\ \Eprint {http://arxiv.org/abs/0911.0017}
  {arXiv:0911.0017 [astro-ph.CO]} \BibitemShut {NoStop}%
\bibitem [{\citenamefont {{Desjacques}}\ \emph {et~al.}(2009)\citenamefont
  {{Desjacques}}, \citenamefont {{Seljak}},\ and\ \citenamefont
  {{Iliev}}}]{2009MNRAS.tmp..631D}%
  \BibitemOpen
  \bibfield  {author} {\bibinfo {author} {\bibfnamefont {V.}~\bibnamefont
  {{Desjacques}}}, \bibinfo {author} {\bibfnamefont {U.}~\bibnamefont
  {{Seljak}}}, \ and\ \bibinfo {author} {\bibfnamefont {I.~T.}\ \bibnamefont
  {{Iliev}}},\ }\bibfield  {title} {\enquote {\bibinfo {title}
  {{Scale-dependent bias induced by local non-Gaussianity: a comparison to
  N-body simulations}},}\ }\href {\doibase 10.1111/j.1365-2966.2009.14721.x}
  {\bibfield  {journal} {\bibinfo  {journal} {\mnras}\ ,\ \bibinfo {pages}
  {631--+}} (\bibinfo {year} {2009})},\ \Eprint
  {http://arxiv.org/abs/0811.2748} {arXiv:0811.2748} \BibitemShut {NoStop}%
\bibitem [{\citenamefont {{Pillepich}}\ \emph {et~al.}(2010)\citenamefont
  {{Pillepich}}, \citenamefont {{Porciani}},\ and\ \citenamefont
  {{Hahn}}}]{2010MNRAS.402..191P}%
  \BibitemOpen
  \bibfield  {author} {\bibinfo {author} {\bibfnamefont {A.}~\bibnamefont
  {{Pillepich}}}, \bibinfo {author} {\bibfnamefont {C.}~\bibnamefont
  {{Porciani}}}, \ and\ \bibinfo {author} {\bibfnamefont {O.}~\bibnamefont
  {{Hahn}}},\ }\bibfield  {title} {\enquote {\bibinfo {title} {{Halo mass
  function and scale-dependent bias from N-body simulations with non-Gaussian
  initial conditions}},}\ }\href {\doibase 10.1111/j.1365-2966.2009.15914.x}
  {\bibfield  {journal} {\bibinfo  {journal} {\mnras}\ }\textbf {\bibinfo
  {volume} {402}},\ \bibinfo {pages} {191--206} (\bibinfo {year} {2010})},\
  \Eprint {http://arxiv.org/abs/0811.4176} {arXiv:0811.4176} \BibitemShut
  {NoStop}%
\bibitem [{\citenamefont {{Desjacques}}\ and\ \citenamefont
  {{Seljak}}(2010{\natexlab{a}})}]{2010CQGra..27l4011D}%
  \BibitemOpen
  \bibfield  {author} {\bibinfo {author} {\bibfnamefont {V.}~\bibnamefont
  {{Desjacques}}}\ and\ \bibinfo {author} {\bibfnamefont {U.}~\bibnamefont
  {{Seljak}}},\ }\bibfield  {title} {\enquote {\bibinfo {title} {{Primordial
  non-Gaussianity from the large-scale structure}},}\ }\href {\doibase
  10.1088/0264-9381/27/12/124011} {\bibfield  {journal} {\bibinfo  {journal}
  {Classical and Quantum Gravity}\ }\textbf {\bibinfo {volume} {27}},\ \bibinfo
  {pages} {124011--+} (\bibinfo {year} {2010}{\natexlab{a}})},\ \Eprint
  {http://arxiv.org/abs/1003.5020} {arXiv:1003.5020 [astro-ph.CO]} \BibitemShut
  {NoStop}%
\bibitem [{\citenamefont {{Valageas}}(2010)}]{2010A&A...514A..46V}%
  \BibitemOpen
  \bibfield  {author} {\bibinfo {author} {\bibfnamefont {P.}~\bibnamefont
  {{Valageas}}},\ }\bibfield  {title} {\enquote {\bibinfo {title} {{Mass
  function and bias of dark matter halos for non-Gaussian initial
  conditions}},}\ }\href {\doibase 10.1051/0004-6361/200912636} {\bibfield
  {journal} {\bibinfo  {journal} {\aap}\ }\textbf {\bibinfo {volume} {514}},\
  \bibinfo {pages} {A46+} (\bibinfo {year} {2010})},\ \Eprint
  {http://arxiv.org/abs/0906.1042} {arXiv:0906.1042 [astro-ph.CO]} \BibitemShut
  {NoStop}%
\bibitem [{\citenamefont {{White}}(2002)}]{2002ApJS..143..241W}%
  \BibitemOpen
  \bibfield  {author} {\bibinfo {author} {\bibfnamefont {M.}~\bibnamefont
  {{White}}},\ }\bibfield  {title} {\enquote {\bibinfo {title} {{The Mass
  Function}},}\ }\href {\doibase 10.1086/342752} {\bibfield  {journal}
  {\bibinfo  {journal} {\apjs}\ }\textbf {\bibinfo {volume} {143}},\ \bibinfo
  {pages} {241--255} (\bibinfo {year} {2002})},\ \Eprint
  {http://arxiv.org/abs/arXiv:astro-ph/0207185} {arXiv:astro-ph/0207185}
  \BibitemShut {NoStop}%
\bibitem [{\citenamefont {{Reed}}\ \emph {et~al.}(2007)\citenamefont {{Reed}},
  \citenamefont {{Bower}}, \citenamefont {{Frenk}}, \citenamefont {{Jenkins}},\
  and\ \citenamefont {{Theuns}}}]{2007MNRAS.374....2R}%
  \BibitemOpen
  \bibfield  {author} {\bibinfo {author} {\bibfnamefont {D.~S.}\ \bibnamefont
  {{Reed}}}, \bibinfo {author} {\bibfnamefont {R.}~\bibnamefont {{Bower}}},
  \bibinfo {author} {\bibfnamefont {C.~S.}\ \bibnamefont {{Frenk}}}, \bibinfo
  {author} {\bibfnamefont {A.}~\bibnamefont {{Jenkins}}}, \ and\ \bibinfo
  {author} {\bibfnamefont {T.}~\bibnamefont {{Theuns}}},\ }\bibfield  {title}
  {\enquote {\bibinfo {title} {{The halo mass function from the dark ages
  through the present day}},}\ }\href {\doibase
  10.1111/j.1365-2966.2006.11204.x} {\bibfield  {journal} {\bibinfo  {journal}
  {\mnras}\ }\textbf {\bibinfo {volume} {374}},\ \bibinfo {pages} {2--15}
  (\bibinfo {year} {2007})},\ \Eprint
  {http://arxiv.org/abs/arXiv:astro-ph/0607150} {arXiv:astro-ph/0607150}
  \BibitemShut {NoStop}%
\bibitem [{\citenamefont {{Tinker}}\ \emph {et~al.}(2008)\citenamefont
  {{Tinker}}, \citenamefont {{Kravtsov}}, \citenamefont {{Klypin}},
  \citenamefont {{Abazajian}}, \citenamefont {{Warren}}, \citenamefont
  {{Yepes}}, \citenamefont {{Gottl{\"o}ber}},\ and\ \citenamefont
  {{Holz}}}]{2008ApJ...688..709T}%
  \BibitemOpen
  \bibfield  {author} {\bibinfo {author} {\bibfnamefont {J.}~\bibnamefont
  {{Tinker}}}, \bibinfo {author} {\bibfnamefont {A.~V.}\ \bibnamefont
  {{Kravtsov}}}, \bibinfo {author} {\bibfnamefont {A.}~\bibnamefont
  {{Klypin}}}, \bibinfo {author} {\bibfnamefont {K.}~\bibnamefont
  {{Abazajian}}}, \bibinfo {author} {\bibfnamefont {M.}~\bibnamefont
  {{Warren}}}, \bibinfo {author} {\bibfnamefont {G.}~\bibnamefont {{Yepes}}},
  \bibinfo {author} {\bibfnamefont {S.}~\bibnamefont {{Gottl{\"o}ber}}}, \ and\
  \bibinfo {author} {\bibfnamefont {D.~E.}\ \bibnamefont {{Holz}}},\ }\bibfield
   {title} {\enquote {\bibinfo {title} {{Toward a Halo Mass Function for
  Precision Cosmology: The Limits of Universality}},}\ }\href {\doibase
  10.1086/591439} {\bibfield  {journal} {\bibinfo  {journal} {\apj}\ }\textbf
  {\bibinfo {volume} {688}},\ \bibinfo {pages} {709--728} (\bibinfo {year}
  {2008})},\ \Eprint {http://arxiv.org/abs/0803.2706} {arXiv:0803.2706}
  \BibitemShut {NoStop}%
\bibitem [{\citenamefont {{Tinker}}\ \emph {et~al.}(2010)\citenamefont
  {{Tinker}}, \citenamefont {{Robertson}}, \citenamefont {{Kravtsov}},
  \citenamefont {{Klypin}}, \citenamefont {{Warren}}, \citenamefont {{Yepes}},\
  and\ \citenamefont {{Gottl{\"o}ber}}}]{2010ApJ...724..878T}%
  \BibitemOpen
  \bibfield  {author} {\bibinfo {author} {\bibfnamefont {J.~L.}\ \bibnamefont
  {{Tinker}}}, \bibinfo {author} {\bibfnamefont {B.~E.}\ \bibnamefont
  {{Robertson}}}, \bibinfo {author} {\bibfnamefont {A.~V.}\ \bibnamefont
  {{Kravtsov}}}, \bibinfo {author} {\bibfnamefont {A.}~\bibnamefont
  {{Klypin}}}, \bibinfo {author} {\bibfnamefont {M.~S.}\ \bibnamefont
  {{Warren}}}, \bibinfo {author} {\bibfnamefont {G.}~\bibnamefont {{Yepes}}}, \
  and\ \bibinfo {author} {\bibfnamefont {S.}~\bibnamefont {{Gottl{\"o}ber}}},\
  }\bibfield  {title} {\enquote {\bibinfo {title} {{The Large-scale Bias of
  Dark Matter Halos: Numerical Calibration and Model Tests}},}\ }\href
  {\doibase 10.1088/0004-637X/724/2/878} {\bibfield  {journal} {\bibinfo
  {journal} {\apj}\ }\textbf {\bibinfo {volume} {724}},\ \bibinfo {pages}
  {878--886} (\bibinfo {year} {2010})},\ \Eprint
  {http://arxiv.org/abs/1001.3162} {arXiv:1001.3162 [astro-ph.CO]} \BibitemShut
  {NoStop}%
\bibitem [{\citenamefont {{Heavens}}\ \emph {et~al.}(1998)\citenamefont
  {{Heavens}}, \citenamefont {{Matarrese}},\ and\ \citenamefont
  {{Verde}}}]{1998MNRAS.301..797H}%
  \BibitemOpen
  \bibfield  {author} {\bibinfo {author} {\bibfnamefont {A.~F.}\ \bibnamefont
  {{Heavens}}}, \bibinfo {author} {\bibfnamefont {S.}~\bibnamefont
  {{Matarrese}}}, \ and\ \bibinfo {author} {\bibfnamefont {L.}~\bibnamefont
  {{Verde}}},\ }\bibfield  {title} {\enquote {\bibinfo {title} {{The non-linear
  redshift-space power spectrum of galaxies}},}\ }\href@noop {} {\bibfield
  {journal} {\bibinfo  {journal} {\mnras}\ }\textbf {\bibinfo {volume} {301}},\
  \bibinfo {pages} {797--808} (\bibinfo {year} {1998})},\ \Eprint
  {http://arxiv.org/abs/arXiv:astro-ph/9808016} {arXiv:astro-ph/9808016}
  \BibitemShut {NoStop}%
\bibitem [{\citenamefont {{McDonald}}(2006)}]{2006PhRvD..74j3512M}%
  \BibitemOpen
  \bibfield  {author} {\bibinfo {author} {\bibfnamefont {P.}~\bibnamefont
  {{McDonald}}},\ }\bibfield  {title} {\enquote {\bibinfo {title} {{Clustering
  of dark matter tracers: Renormalizing the bias parameters}},}\ }\href
  {\doibase 10.1103/PhysRevD.74.103512} {\bibfield  {journal} {\bibinfo
  {journal} {\prd}\ }\textbf {\bibinfo {volume} {74}},\ \bibinfo {pages}
  {103512--+} (\bibinfo {year} {2006})},\ \Eprint
  {http://arxiv.org/abs/arXiv:astro-ph/0609413} {arXiv:astro-ph/0609413}
  \BibitemShut {NoStop}%
\bibitem [{\citenamefont {{Sefusatti}}(2009)}]{2009PhRvD..80l3002S}%
  \BibitemOpen
  \bibfield  {author} {\bibinfo {author} {\bibfnamefont {E.}~\bibnamefont
  {{Sefusatti}}},\ }\bibfield  {title} {\enquote {\bibinfo {title} {{One-loop
  perturbative corrections to the matter and galaxy bispectrum with
  non-Gaussian initial conditions}},}\ }\href {\doibase
  10.1103/PhysRevD.80.123002} {\bibfield  {journal} {\bibinfo  {journal}
  {\prd}\ }\textbf {\bibinfo {volume} {80}},\ \bibinfo {pages} {123002--+}
  (\bibinfo {year} {2009})},\ \Eprint {http://arxiv.org/abs/0905.0717}
  {arXiv:0905.0717 [astro-ph.CO]} \BibitemShut {NoStop}%
\bibitem [{\citenamefont {{Sefusatti}}\ \emph {et~al.}(2010)\citenamefont
  {{Sefusatti}}, \citenamefont {{Crocce}},\ and\ \citenamefont
  {{Desjacques}}}]{2010MNRAS.406.1014S}%
  \BibitemOpen
  \bibfield  {author} {\bibinfo {author} {\bibfnamefont {E.}~\bibnamefont
  {{Sefusatti}}}, \bibinfo {author} {\bibfnamefont {M.}~\bibnamefont
  {{Crocce}}}, \ and\ \bibinfo {author} {\bibfnamefont {V.}~\bibnamefont
  {{Desjacques}}},\ }\bibfield  {title} {\enquote {\bibinfo {title} {{The
  matter bispectrum in N-body simulations with non-Gaussian initial
  conditions}},}\ }\href {\doibase 10.1111/j.1365-2966.2010.16723.x} {\bibfield
   {journal} {\bibinfo  {journal} {\mnras}\ }\textbf {\bibinfo {volume}
  {406}},\ \bibinfo {pages} {1014--1028} (\bibinfo {year} {2010})},\ \Eprint
  {http://arxiv.org/abs/1003.0007} {arXiv:1003.0007 [astro-ph.CO]} \BibitemShut
  {NoStop}%
\bibitem [{\citenamefont {{Smith}}\ \emph
  {et~al.}(2011{\natexlab{b}})\citenamefont {{Smith}}, \citenamefont
  {{Desjacques}},\ and\ \citenamefont {{Marian}}}]{2011PhRvD..83d3526S}%
  \BibitemOpen
  \bibfield  {author} {\bibinfo {author} {\bibfnamefont {R.~E.}\ \bibnamefont
  {{Smith}}}, \bibinfo {author} {\bibfnamefont {V.}~\bibnamefont
  {{Desjacques}}}, \ and\ \bibinfo {author} {\bibfnamefont {L.}~\bibnamefont
  {{Marian}}},\ }\bibfield  {title} {\enquote {\bibinfo {title} {{Nonlinear
  clustering in models with primordial non-Gaussianity: The halo model
  approach}},}\ }\href {\doibase 10.1103/PhysRevD.83.043526} {\bibfield
  {journal} {\bibinfo  {journal} {\prd}\ }\textbf {\bibinfo {volume} {83}},\
  \bibinfo {pages} {043526--+} (\bibinfo {year} {2011}{\natexlab{b}})},\
  \Eprint {http://arxiv.org/abs/1009.5085} {arXiv:1009.5085 [astro-ph.CO]}
  \BibitemShut {NoStop}%
\bibitem [{\citenamefont {{Grossi}}\ \emph {et~al.}(2009)\citenamefont
  {{Grossi}}, \citenamefont {{Verde}}, \citenamefont {{Carbone}}, \citenamefont
  {{Dolag}}, \citenamefont {{Branchini}}, \citenamefont {{Iannuzzi}},
  \citenamefont {{Matarrese}},\ and\ \citenamefont
  {{Moscardini}}}]{2009MNRAS.398..321G}%
  \BibitemOpen
  \bibfield  {author} {\bibinfo {author} {\bibfnamefont {M.}~\bibnamefont
  {{Grossi}}}, \bibinfo {author} {\bibfnamefont {L.}~\bibnamefont {{Verde}}},
  \bibinfo {author} {\bibfnamefont {C.}~\bibnamefont {{Carbone}}}, \bibinfo
  {author} {\bibfnamefont {K.}~\bibnamefont {{Dolag}}}, \bibinfo {author}
  {\bibfnamefont {E.}~\bibnamefont {{Branchini}}}, \bibinfo {author}
  {\bibfnamefont {F.}~\bibnamefont {{Iannuzzi}}}, \bibinfo {author}
  {\bibfnamefont {S.}~\bibnamefont {{Matarrese}}}, \ and\ \bibinfo {author}
  {\bibfnamefont {L.}~\bibnamefont {{Moscardini}}},\ }\bibfield  {title}
  {\enquote {\bibinfo {title} {{Large-scale non-Gaussian mass function and halo
  bias: tests on N-body simulations}},}\ }\href {\doibase
  10.1111/j.1365-2966.2009.15150.x} {\bibfield  {journal} {\bibinfo  {journal}
  {\mnras}\ }\textbf {\bibinfo {volume} {398}},\ \bibinfo {pages} {321--332}
  (\bibinfo {year} {2009})},\ \Eprint {http://arxiv.org/abs/0902.2013}
  {arXiv:0902.2013 [astro-ph.CO]} \BibitemShut {NoStop}%
\bibitem [{\citenamefont {{Fry}}\ and\ \citenamefont
  {{Gaztanaga}}(1993)}]{FryGaz9308}%
  \BibitemOpen
  \bibfield  {author} {\bibinfo {author} {\bibfnamefont {J.~N.}\ \bibnamefont
  {{Fry}}}\ and\ \bibinfo {author} {\bibfnamefont {E.}~\bibnamefont
  {{Gaztanaga}}},\ }\bibfield  {title} {\enquote {\bibinfo {title} {{Biasing
  and hierarchical statistics in large-scale structure}},}\ }\href {\doibase
  10.1086/173015} {\bibfield  {journal} {\bibinfo  {journal} {\apj}\ }\textbf
  {\bibinfo {volume} {413}},\ \bibinfo {pages} {447--452} (\bibinfo {year}
  {1993})},\ \Eprint {http://arxiv.org/abs/arXiv:astro-ph/9302009}
  {arXiv:astro-ph/9302009} \BibitemShut {NoStop}%
\bibitem [{\citenamefont {{Matarrese}}\ \emph {et~al.}(1986)\citenamefont
  {{Matarrese}}, \citenamefont {{Lucchin}},\ and\ \citenamefont
  {{Bonometto}}}]{1986ApJ...310L..21M}%
  \BibitemOpen
  \bibfield  {author} {\bibinfo {author} {\bibfnamefont {S.}~\bibnamefont
  {{Matarrese}}}, \bibinfo {author} {\bibfnamefont {F.}~\bibnamefont
  {{Lucchin}}}, \ and\ \bibinfo {author} {\bibfnamefont {S.~A.}\ \bibnamefont
  {{Bonometto}}},\ }\bibfield  {title} {\enquote {\bibinfo {title} {{A
  path-integral approach to large-scale matter distribution originated by
  non-Gaussian fluctuations}},}\ }\href {\doibase 10.1086/184774} {\bibfield
  {journal} {\bibinfo  {journal} {\apjl}\ }\textbf {\bibinfo {volume} {310}},\
  \bibinfo {pages} {L21--L26} (\bibinfo {year} {1986})}\BibitemShut {NoStop}%
\bibitem [{\citenamefont {{McDonald}}(2008)}]{2008PhRvD..78l3519M}%
  \BibitemOpen
  \bibfield  {author} {\bibinfo {author} {\bibfnamefont {P.}~\bibnamefont
  {{McDonald}}},\ }\bibfield  {title} {\enquote {\bibinfo {title} {{Primordial
  non-Gaussianity: Large-scale structure signature in the perturbative bias
  model}},}\ }\href {\doibase 10.1103/PhysRevD.78.123519} {\bibfield  {journal}
  {\bibinfo  {journal} {\prd}\ }\textbf {\bibinfo {volume} {78}},\ \bibinfo
  {pages} {123519--+} (\bibinfo {year} {2008})},\ \Eprint
  {http://arxiv.org/abs/0806.1061} {arXiv:0806.1061} \BibitemShut {NoStop}%
\bibitem [{\citenamefont {{Scoccimarro}}(2000)}]{2000ApJ...542....1S}%
  \BibitemOpen
  \bibfield  {author} {\bibinfo {author} {\bibfnamefont {R.}~\bibnamefont
  {{Scoccimarro}}},\ }\bibfield  {title} {\enquote {\bibinfo {title}
  {{Gravitational Clustering from $\chi^2$ Initial Conditions}},}\ }\href@noop
  {} {\bibfield  {journal} {\bibinfo  {journal} {\apj}\ }\textbf {\bibinfo
  {volume} {542}},\ \bibinfo {pages} {1--8} (\bibinfo {year}
  {2000})}\BibitemShut {NoStop}%
\bibitem [{\citenamefont {{Lacey}}\ and\ \citenamefont
  {{Cole}}(1993)}]{1993MNRAS.262..627L}%
  \BibitemOpen
  \bibfield  {author} {\bibinfo {author} {\bibfnamefont {C.}~\bibnamefont
  {{Lacey}}}\ and\ \bibinfo {author} {\bibfnamefont {S.}~\bibnamefont
  {{Cole}}},\ }\bibfield  {title} {\enquote {\bibinfo {title} {{Merger rates in
  hierarchical models of galaxy formation}},}\ }\href@noop {} {\bibfield
  {journal} {\bibinfo  {journal} {\mnras}\ }\textbf {\bibinfo {volume} {262}},\
  \bibinfo {pages} {627--649} (\bibinfo {year} {1993})}\BibitemShut {NoStop}%
\bibitem [{\citenamefont {{Regan}}\ \emph {et~al.}(2011)\citenamefont
  {{Regan}}, \citenamefont {{Schmittfull}}, \citenamefont {{Shellard}},\ and\
  \citenamefont {{Fergusson}}}]{RegSchShe1108}%
  \BibitemOpen
  \bibfield  {author} {\bibinfo {author} {\bibfnamefont {D.~M.}\ \bibnamefont
  {{Regan}}}, \bibinfo {author} {\bibfnamefont {M.~M.}\ \bibnamefont
  {{Schmittfull}}}, \bibinfo {author} {\bibfnamefont {E.~P.~S.}\ \bibnamefont
  {{Shellard}}}, \ and\ \bibinfo {author} {\bibfnamefont {J.~R.}\ \bibnamefont
  {{Fergusson}}},\ }\bibfield  {title} {\enquote {\bibinfo {title} {{Universal
  Non-Gaussian Initial Conditions for N-body Simulations}},}\ }\href@noop {}
  {\bibfield  {journal} {\bibinfo  {journal} {ArXiv e-prints}\ } (\bibinfo
  {year} {2011})},\ \Eprint {http://arxiv.org/abs/1108.3813} {arXiv:1108.3813
  [astro-ph.CO]} \BibitemShut {NoStop}%
\bibitem [{\citenamefont {{Desjacques}}\ and\ \citenamefont
  {{Seljak}}(2010{\natexlab{b}})}]{2010PhRvD..81b3006D}%
  \BibitemOpen
  \bibfield  {author} {\bibinfo {author} {\bibfnamefont {V.}~\bibnamefont
  {{Desjacques}}}\ and\ \bibinfo {author} {\bibfnamefont {U.}~\bibnamefont
  {{Seljak}}},\ }\bibfield  {title} {\enquote {\bibinfo {title} {{Signature of
  primordial non-Gaussianity of ${\phi}^{3}$ type in the mass function and bias
  of dark matter haloes}},}\ }\href {\doibase 10.1103/PhysRevD.81.023006}
  {\bibfield  {journal} {\bibinfo  {journal} {\prd}\ }\textbf {\bibinfo
  {volume} {81}},\ \bibinfo {pages} {023006--+} (\bibinfo {year}
  {2010}{\natexlab{b}})},\ \Eprint {http://arxiv.org/abs/0907.2257}
  {arXiv:0907.2257 [astro-ph.CO]} \BibitemShut {NoStop}%
\bibitem [{\citenamefont {{LoVerde}}\ and\ \citenamefont
  {{Smith}}(2011)}]{2011arXiv1102.1439L}%
  \BibitemOpen
  \bibfield  {author} {\bibinfo {author} {\bibfnamefont {M.}~\bibnamefont
  {{LoVerde}}}\ and\ \bibinfo {author} {\bibfnamefont {K.~M.}\ \bibnamefont
  {{Smith}}},\ }\bibfield  {title} {\enquote {\bibinfo {title} {{The
  Non-Gaussian Halo Mass Function with $\fnl$, $\gnl$ and $\tnl$}},}\
  }\href@noop {} {\bibfield  {journal} {\bibinfo  {journal} {ArXiv e-prints}\ }
  (\bibinfo {year} {2011})},\ \Eprint {http://arxiv.org/abs/1102.1439}
  {arXiv:1102.1439 [astro-ph.CO]} \BibitemShut {NoStop}%
\end{thebibliography}%
\vskip 2pc

\appendix

\section{Generating Non-Local Primordial non-Gaussianity for non-scale invariant models}
\label{NSI}

When the power spectrum is not scale-invariant, the non-local operators used in the main text, Eqs.~(\ref{sqrtbox}-\ref{nabalinv}), can be easily generalized by defining a generic operator ${\cal P}_m$,

\beq
[{\cal P}_m\, A](\x)\equiv  \int {\rm e}^{-i \k \cdot \x}\ [P(k)]^m \, A(\k)\, d^3k
\label{genP}
\eeq
where $m=\pm1,\pm2/3,\pm 1/3$ for the templates we have used in this paper. The kernels in real space then read,

\begin{widetext}

\beqa
K_{\rm eq}[\phi,\phi]&=& -3(1-u)\, \phi^2 -2 \, {\cal P}_{2/3}([{\cal P}_{-1/3}\phi]^2)
+(4-3u)\, {\cal P}_{1/3}(\phi [{\cal P}_{-1/3}\phi]) + 2\, {\cal P}_{2/3}(\phi [{\cal P}_{-2/3}\phi])\nonumber \\ & & 
-3u\,  {\cal P}_{1}(\phi [{\cal P}_{-1}\phi] -  [{\cal P}_{-2/3}\phi] \ [{\cal P}_{-1/3}\phi] )
\label{KeqNSI}
\eeqa
and 
\beqa
K_{\rm ort}[\phi,\phi] &= & -9(1-u)\, \phi^2 -8 \, {\cal P}_{2/3}([{\cal P}_{-1/3}\phi]^2)
+(10-9u)\, {\cal P}_{1/3}(\phi [{\cal P}_{-1/3}\phi]) + 8\, {\cal P}_{2/3}(\phi [{\cal P}_{-2/3}\phi])\nonumber \\ & & 
-9u\,  {\cal P}_{1}(\phi [{\cal P}_{-1}\phi] -  [{\cal P}_{-2/3}\phi] \ [{\cal P}_{-1/3}\phi] )
\label{KortNSI}
\eeqa
For scale-invariant power spectra, ${\cal P}_{\pm2/3} = -\nabla^{\mp2}$ and ${\cal P}_{\pm1/3} = \partial^{\mp1}$ (Eqs.~\ref{sqrtbox}-\ref{nabalinv}), which yields the results quoted in the text (Eqs.~\ref{KEQ}-\ref{KORT}). As discussed in the text, we set the free parameter $u$ to zero for simplicity. To minimize sub-leading one-loop corrections in the power spectrum that go as $k^{-2}$ one may instead choose $u$ as given in Eq.~(\ref{minimize}). Other possible choices are discussed in Appendix~\ref{sqz}.
\end{widetext}

\section{More General Initial Conditions, Including Scale-Dependent $\fnl$ and  $\gnl$}
\label{ICsSDfnl}

In this Appendix, we give a short, self-contained recipe for
generating initial conditions for fairly general, but factorizable, primordial
non-gaussianity. 
Let's divide the discussion into two parts.
One is how to simulate efficiently a kernel of the factorizable type.
The other is how to determine the right kernel to use for a bispectrum
of the factorizable type.

First, how to simulate. Suppose one wants to simulate:
\begin{eqnarray}
&& \Phi(x) = \phi(x) + \int d^3 k_1 d^3 k_2\, K_{12}\, \tilde \phi({\bf k_1})
\tilde \phi({\bf k_2}) \nonumber \\ && \quad \quad \quad \quad \quad
\quad \quad \quad e^{- i ({\bf k}_1 + {\bf k_2}) \cdot {\bf x}}
\label{Phifnl}
\end{eqnarray}
with a kernel $K_{12}$ of the form:
\begin{eqnarray}
\label{kernel}
K_{12} = \sum D\,  f_1(k_1)\,  f_2 (k_2)\,  f_3 (|{\bf k}_1 + {\bf k}_2|)
\end{eqnarray}
where $f_1$, $f_2$ and $f_3$ are arbitrary functions, and
the summation $\sum$ denotes  the fact that the kernel
could be the sum of many terms of such factorizable form with
different functions $f_1$, $f_2$, $f_3$ and constants $D$. 
Our algorithm would allow us to simulate each of these, and one
can simply add to obtain the desired non-Gaussian $\Phi$.
Note that we have subsumed $f_{\rm NL}$ into the definition $K_{12}$
to allow for the fact that even $f_{\rm NL}$ could be scale dependent.

The simulation algorithm goes as follows. First, generate a Gaussian
random $\tilde \phi$ in Fourier space. Then, Fourier transform
$\tilde \phi({\bf k_1}) f_1(k_1)$ and $\tilde \phi({\bf k_2}) f_2(k_2)$ separately back into real space
and multiply the results: let's denote this product 
{\it in real space} by $[f_1 \phi]({\bf x}) \times [f_2 \phi]({\bf x})$. 
Then, Fourier transform this back to Fourier space and 
multiply by $f_3$, and finally Fourier transform back:
\begin{eqnarray}
&& D \int d^3 k_3 e^{-i {\bf k_3} \cdot {\bf x}} f_3(k_3) 
\nonumber \\ &&
\int d^3 y e^{i
  {\bf k_3} \cdot {\bf y}} 
[f_1 \phi]({\bf y}) \times  [f_2 \phi]({\bf y}) 
\end{eqnarray}
This real space quantity successfully realizes $\int d^3 k_1 d^3 k_2 K_{12} \tilde \phi({\bf k_1}) \tilde \phi({\bf k_2}) e^{- i ({\bf k}_1 + {\bf k_2}) \cdot {\bf x}}$
with the kernel given by $K_{12} = D f_1(k_1) f_2(k_2) f_3(|{\bf k}_1 + {\bf k}_2|)$.
Note that the algorithm requires only Fourier transforms and multiplications (in Fourier
or real space). Nowhere do we need to perform expensive convolutions.

Now, for a given bispectrum of the factorizable type, how do we decide what kernel to use?
Suppose one wants to simulate a bispectrum of the form
\begin{eqnarray}
\label{B123}
B_{123} &=& \sum 2C [g_1(k_1) g_2(k_2) g_3(k_3) + {\rm 5 \, \, perm.}] \nonumber \\ & &
\end{eqnarray}
where $g_1$, $g_2$ and $g_3$ are arbitrary functions,
and the summation $\sum$ denotes the fact that one might have a bispectrum that
is a sum over terms of the above form, but with different functions
$g_1$, $g_2$, $g_3$ and
constants $C$~\footnote{Even non-separable bispectra can actually be well approximated in this way by using an appropriate basis~\cite{2010arXiv1008.1730F,2010PhRvD..82b3502F}.
}. Once again, if we know how to determine the kernel for one of these, we can simply
add to obtain the total. A kernel that produces the bispectrum $2C
[g_1(k_1) g_2(k_2) g_3(k_3) + {\rm 5 \, \, perm.}]$ is:
\begin{eqnarray}
K_{12} &=& C {g_3(k_3) \over P_1 P_2}\left[g_1(k_1) g_2(k_2) + g_2(k_1) g_1(k_2) \right] \nonumber \\ & & 
\end{eqnarray}
where $P_1$ and $P_2$ are the $\phi$ power spectrum at $k_1$ and $k_2$. 
The kernel is thus of
the form in Eq. (\ref{kernel}), and can therefore be simulated using
the FFT-based algorithm above.

This is not the only choice of kernel that will produce the
desired bispectrum. There are in fact two other kernels possible,
which involve permutations of $1$, $2$ and $3$, and the most general
kernel is a linear combination of the three. 
However, for simplicity, one could adopt the kernel described
above, but take care to check that the resulting one-loop $\Phi$ power
spectrum will not have an IR limit that is more divergent than the
tree level one (see Section~\ref{1Lconst} and Appendix~\ref{sqz} for examples). This corresponds to choosing a $g_3$ (since there are
3 functions involved, one has the freedom to choose which to call
$g_3$) such that $g_3(k)^2 / P_\phi (k)$ is not IR divergent.

Finally, we note that following the same algorithm it is straightforward to construct non-local PNG initial conditions for $\gnl$ models. That is, a term in the Bardeen potential at cubic order in Eq.~(\ref{Phifnl}) of the form
\beq
\int d^3k_1 d^3k_2\, L_{123}\,  \tilde \phi({\bf k_1})  \tilde \phi({\bf k_2})  \tilde \phi({\bf k_3})
e^{- i ({\bf k}_1 + {\bf k_2}+ {\bf k_3}) \cdot {\bf x}}
\label{Lgnl}
\eeq
where $L$ is the cubic kernel (proportional to $\gnl$) and assumed to be a sum of factorizable terms
\beq
L_{123} = \sum E\,  f_1(k_1) f_2 (k_2) f_3 (k_3) f_4 (|{\bf k}_1 + {\bf k}_2+ {\bf k_3}|)
\eeq
can also be easily implemented following the procedure outlined above for non-local $\fnl$-PNG. For work on {\em local} $\gnl$ N-body simulations see~\cite{2010PhRvD..81b3006D,2011arXiv1102.1439L}.

\begin{figure}[!t]
\centering
\includegraphics[width=0.95\linewidth]{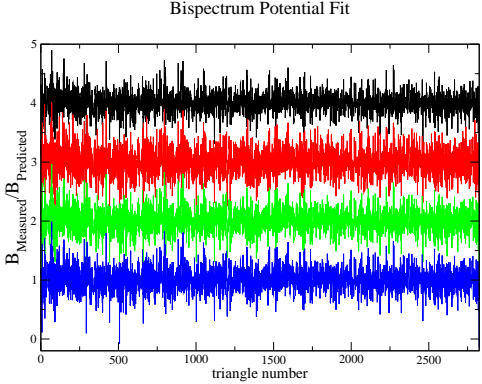}
 \caption{The ratio between the Bardeen potential Bispectrum measured in the initial conditions and the tree-level prediction for all triangles with sides smaller than 60 times the fundamental mode for the four different operators discussed in the text (with amplitudes set by $\fnl=-400$), from top to bottom:  A (black), B (red), C (green) and D (blue), respectively. All ratios are close to one but are shifted by integer units for clarity.}
 \label{figbispphi}
 \end{figure}

\section{Further Tests on Initial Conditions}

\label{BispPhi}

We have done a number of tests on our initial conditions generator. The most direct test is to make sure that the bispectrum of the Bardeen potential is indeed the one desired for a given template. Here we do this for each of the operators that enter into the local, equilateral and orthogonal model. Let's consider the local model, for which we used
\beq
\label{termA}
\Phi_A = \phi +\fnl \phi^2 
\eeq
and the corresponding bispectrum
\beq
\label{BispA}
B_A = 2 \fnl (P_1 P_2 + P_2 P_3 + P_3 P_1 ) 
\eeq
the second generator discussed in Section~\ref{secGen}
\beq
\label{termB}
\Phi_B = \phi + \fnl \nabla^{-2} (\partial \phi)^2 
\eeq
for which the bispectrum reads
\beq
\label{BispB}
B_B= - 6 \fnl (P_1 P_2 P_3 )^{2/3} 
\eeq
We also consider the two first operators corresponding to the third generator in Section~\ref{3rdterm} (see Eq.~\ref{3rdfnl}), i.e. 
\beq
\label{termC}
\Phi_C = \phi + \fnl \partial^{-1} (\phi \partial \phi)
\eeq
with bispectrum
\beq
\label{BispC}
B_C = \fnl (P_1^{1/3} P_2^{2/3}  P_3 + {\rm cyc.} ) 
\eeq
and 
\beq
\label{termD}
\Phi_D = \phi + \fnl \nabla^{-2} (\phi \nabla^2 \phi)
\eeq
with bispectrm
\beq
\label{BispD}
B_D = \fnl (P_1^{1/3} P_2^{2/3}  P_3 + {\rm cyc.} ) 
\eeq
The last operator in Eq.~(\ref{3rdfnl}) is not included in these tests as we choose $u=2s=0$, and thus its amplitude vanishes (i.e. it does not appear in Eqs.~\ref{KEQ} or~\ref{KORT}).

To beat down cosmic variance we generated 100 realizations each of initial conditions for these four operators with $\fnl =-50, -400,-3000$ and measured the bispectrum for each of them. A simple $\chi^2$ fit of the ratio of the numerically generated $B_\Phi$ to its tree-level predictions  (see Fig.~\ref{figbispphi}, corresponding to the $\fnl=-400$ case), gave the  values shown in Table~\ref{fnlfit} for $\fnl$, when using {\em all} triangles with sides up to $k=0.16 \kvecMpc$ corresponding to sixty times the fundamental mode of the box. The best fit values are close to the input values even for large values of $\fnl$ where loop corrections may start to become important.

\begin{table}[!t]
\caption{Best fit values of $\fnl$ obtained by fitting the generated $B_\Phi$ to tree-level predictions for each of the four operators.}
\begin{ruledtabular}
\begin{tabular}{c c c c c}
Input $\fnl$ &   A &  B &  C &  D \\
\hline
-50 & -48.5  &  -48.0 & -48.4 & -48.4 \\
-400 & -390  &  -397  & -392  & -392 \\ 
-3000 & -3224 & -3042 & -3078  & -3087 \\
\label{fnlfit}
\end{tabular} 
\end{ruledtabular} 
\end{table}

\begin{widetext}

\section{New scale-independent corrections for local PNG}
\label{IBlocApp}

While it is often assumed that for local PNG of $\fnl$ type the change in the small-scale density perturbation $\delta_s$ corresponds to a local rescaling of the Gaussian small-scale matter fluctuations, $\delta_s^{\rm NG} = (1+2\fnl\phi_\ell)\,  \delta_s^{\rm G}$, this is only true asymptotically at large-scales. In reality there is no local rescaling neither in real nor Fourier space, because of three different effects encoded by $I_{21}$, which give quadratic corrections in $(k_\ell/k_s)$ due to,

\begin{enumerate}
\item contributions from the second term in $\nabla^2 (\Phi-\phi) = 2\fnl (\phi\nabla^2\phi + |\nabla\phi|^2)$, which in previous literature has been neglected on the (incorrect) assumption that $\nabla\phi=0$ for halos,

\item contributions from the transfer function $T(|\k-\q|)$,

\item contributions from the smoothing kernel $W_{\rm TH}(|\k-\q|)$.

\end{enumerate}

All of these are included in the main text exactly, by computing $I_{12}$ numerically. Here  we separate each of these contributions analytically by doing a Taylor expansion. The transfer function expansion in powers of $(k/q)$ reads,
\beq
T(|\k-\q|) = T(q) \times \Bigg\{ 1 -\Big( {{\k \cdot \q} \over 2 \, q^2}\Big) \delta n+ \Big( {k\over 2q} \Big)^2 \Big[
\delta n \, [1-2 (\hat{\k}\cdot \hat{\q})^2] + (\hat{\k}\cdot \hat{\q})^2( \dot{\delta n}+\delta n^2/2)
\Big]
\Bigg\}
\eeq
where $\delta n$ and $\dot{\delta n}$ are the contributions to the effective spectral index and its running coming from the transfer function,
%\end{widetext}
\beq
\delta n(q) \equiv n_{\rm eff}(q) - n_s , \ \ \ \ \  \ \ \ \ \
\dot{\delta n}(q) \equiv {d n_{\rm eff} \over d \ln q}  \label{neff}
\eeq
where for simplicity we have assumed a constant $n_s$ (no primordial running), and as usual $n_{\rm eff} \equiv d \ln P/d \ln k$, with $P$ the linear density power spectrum. We can use this to calculate, for example, $I_{21}$ as a large-scale expansion in powers of $k$.  For local PNG we have that the angular integral over $\hat{\k}\cdot \hat{\q}$ is given by

%\begin{widetext}
\beqa
\int \Big({d\Omega\over 4 \pi}\Big) W_{\rm TH}(|\k-\q|R)\, (\k-\q)^2\, T(|\k-\q|) &=& q^2\, W_{\rm TH}(qR)W_{\rm TH}(kR)\,T(q) \nonumber \\ && \times
\Big(1+{1\over 3} w_k + {k^2 \over q^2} \Big[ 1 + {1\over 3} w_q (1+\delta n/2) + {5 \delta n\over 12} + {(\delta n)^2\over 24}  + {\dot{\delta n}\over 12}  \Big] \Big)
\nonumber \\ & & 
\label{res}
\eeqa
% \end{widetext}
where $w_k=d \ln W_{\rm TH}(kR)/ d\ln k$ and we have used the standard summation theorem of Bessel functions to integrate over the top-hat window function (see Appendix~C in~\cite{BerColGaz02}). The cross-term $\k \cdot \q$ inside $(\k-\q)^2$ takes into account the $|\nabla\phi|^2$ contributions mentioned above. Since we are interested in small-$k$ limit we can use that $w_k \approx -k^2R^2/5$ and $W_{\rm TH}(kR) \approx 1 + w_k/2$ and thus, 

%\begin{widetext}

\beq
I_{21}(k,m) \approx 4\fnl \times \Bigg\{ \sigma_m^2 \Big(1+{5\over 6} w_k \Big) + k^2
\int {d^3q \over q^2}  P_\phi(q)[M_m(q)]^2 \Big[ 1 + {1\over 3} w_q (1+\delta n/2) + {5 \delta n\over 12} + {(\delta n)^2\over 24}  + {\dot{\delta n}\over 12}  \Big] \Bigg\}
\label{res2}
\eeq
We see from this that to zero-th order in $k$, we recover Eq.~(\ref{IBls}). Since the scale-dependent bias goes as $I_{21}/M(k)\propto I_{21}/k^2$ (see Eq.~\ref{DeltaB}), the next-to-leading order terms contribute to scale-independent bias, with a coefficient determined by integrating the {\em velocity} power spectrum against contributions from filtering and transfer function effects.

\end{widetext}

\section{Squeezed Limit of Kernels and Scale-Dependent Bias}
\label{sqz}

Recall that  in the main text we raised the issue of whether when implementing the PBS the split variable should be  the Gaussian field $\phi$ (as it has been done in the past), or $\Phi$ (as we discuss in Section~\ref{ConstPhi}),  which may give rise to different scalings even in the squeezed limit (agreeing with local bias models).  These differences may arise because the former depends on the cross-bispectrum ($B_{\Phi\Phi\phi}$) while the latter on the full bispectrum  ($B_{\Phi\Phi\Phi}$), but how different these are from each other depends on the details of the long-mode kernel in the squeezed limit. In this appendix we want to explore these subtleties. 

Let's look at the squeezed limit of the kernels (which determine the low-$k$ limit) for equilateral and orthogonal PNG, Eqs.~(\ref{KeqGen}-\ref{KortGen}), before and after imposing one-loop correction constraints. First, equilateral.
There are actually {\em two} kernels that enter, depending on whether the coupling between $\k_1$ and $\k_2$ is to a large-scale ($k_\ell$) or small-scale mode ($k_s\gg k_\ell$),
\beqa
K_{\rm eq}^{(s)}(\k_s,\k_\ell) &\approx& {3\over 2} (u-2s) +(1-3t)\Big( {k_\ell\over k_s} \Big) \nonumber \\ & & + 3\, (s+t) \Big({k_\ell\over k_s}\Big)^2 + \ldots
\label{KEQs}
\eeqa
and
\beqa
K_{\rm eq}^{(\ell)}(\k_s,\k_s') &\approx& {-3} (u-2s) \Big({k_s\over k_\ell}\Big)^3 -2(1-3t)\Big( {k_s\over k_\ell} \Big)^2 \nonumber \\ & & -6\, (s+t-1) \Big({k_s\over k_\ell}\Big) + \ldots
\nonumber \\ & = & -2 K_{\rm eq}^{(s)}(\k_s,\k_\ell) {P_\phi(k_\ell) \over P_\phi(k_s)} + 6 \Big({k_s\over k_\ell}\Big) 
\nonumber \\ & & + \ldots
\label{KEQl}
\eeqa
where $|\k_s+\k_s'|=k_\ell$. In the first kernel, to simplify the expressions we assumed that the third side is $k_s$ (i.e. Eq.~\ref{KEQs} is for a fixed angle between $\k_s$ and $\k_\ell$, the full expression as a function of angle is given below after one-loop correction constraints are imposed). Equation~(\ref{KEQl}), on the other hand, is for $|\k_s'|=k_s$ and third side equal to $k_\ell$, in this case the angle between $\k_s$ and $\k_s'$ is restricted to be nearly $\pi$ to make a low-$k$ mode. Thus we are considering precisely {\em the same triangle} for both kernels, to highlight that they can differ significantly, even in the squeezed limit.

We see from these two expressions that these kernels can have very different behavior, and that  assuming that kernels are totally symmetric in $k_1,k_2,k_3$  is very restrictive, since it forces equality of these two expressions. Even after imposing one-loop constraints the behavior of these kernels will generically differ from each other in the squeezed limit. 

Before we get into one-loop corrections, note that requiring that $\langle \Phi(\k) \rangle =0$  leads to
\beq
\dD(\k) \lim_{k_\ell \rightarrow 0} \int K_{\rm eq}^{(\ell)}(\k_s,\k_s') P_\phi(k_s) d^3k_s
\label{meanConstr}
\eeq
which from Eq.~(\ref{KEQl}) will typically diverge in the infrared (IR). However, this IR divergence is no different that in the local case, in which Eq.~(\ref{meanConstr}) leads to the variance of the $\phi$ field, which is also IR divergent, and can be cured by putting the field in a box, i.e. imposing that $P_\phi(k)$ vanish for $k<\epsilon=2\pi/L_{\rm box}$. The same is true here, because the factors of $(k_s/k_\ell)^n$ in Eq.~(\ref{KEQl}) actually arise from $P_\phi(k_\ell)/P_\phi(k_s)$ ratios when solving for the kernels and thus cutting the $P_\phi(k_\ell)$ power spectrum below some IR cutoff $\epsilon$ regulates $\langle \Phi(\k) \rangle $ in the same way as for local models. Therefore, we won't impose any constraint on the free parameters of the kernel from these considerations.

From Eqs.~(\ref{KEQs}-\ref{KEQl}) we can compute the squeezed limit of the relevant equilateral PNG bispectra,
\begin{widetext}
\beqa
B_{\Phi\Phi\phi}^{\rm eq}(k_s,k_s,k_\ell) &\approx& \fnl \Big[ 2 K_{\rm eq}^{(s)}(\k_s,\k_\ell) P_\phi(k_\ell) P_\phi(k_s) %\nonumber \\ & & 
+2 K_{\rm eq}^{(s)}(\k'_s,\k_\ell) P_\phi(k_\ell) P_\phi(k'_s)\Big], %\nonumber \\ & & 
\label{BcrossSQZ}
\eeqa
which depends on the parameters $s,t,u$ and
\beqa
B_{\Phi\Phi\Phi}^{\rm eq}(k_s,k_s,k_\ell) &\approx& \fnl \Big[ 2 K_{\rm eq}^{(s)}(\k_s,\k_\ell) P_\phi(k_\ell) P_\phi(k_s) %\nonumber \\ & & 
+2 K_{\rm eq}^{(s)}(\k'_s,\k_\ell) P_\phi(k_\ell) P_\phi(k'_s)  %\nonumber \\ & &
+ 2 K_{\rm eq}^{(\ell)}(\k_s,\k_s') [P_\phi(k_s)]^2 \Big] \nonumber \\ & =&
12\fnl\, \Big({k_\ell\over k_s}\Big)^2 P_\phi(k_\ell) P_\phi(k_s) + \ldots % \nonumber \\ & &
\label{BSQZ}
\eeqa
\end{widetext}
which, of course doesn't depend on $s, t, u$, as varying these parameters leaves the tree-level bispectrum of $\Phi$ unchanged. Note that the relevant bispectrum for the calculation of the bias when doing the PBS in $\phi$ is $B_{\widehat\delta \widehat\delta \phi}= [M_m(k_s)]^2 B_{\Phi\Phi\phi}$, thus we are interested in the squeezed limit of $B_{\Phi\Phi\phi}$ and how it relates to that in $B_{\Phi\Phi\Phi}$. We see from Eq.~(\ref{KEQs}) that there is a priori no reason for Eqs.~(\ref{BSQZ}) and~(\ref{BcrossSQZ}) to agree, unless the first two terms in Eq.~(\ref{KEQs}) vanish and the last one has a suitable coefficient. 

Let us now impose the constraints from one-loop corrections, which are easy to see from Eq.~(\ref{KEQl}) since that's the kernel whose square enters into the calculation of the one-loop power spectrum. We see, in agreement with the detailed calculations of the power spectrum presented in Section~\ref{1Lconst}, that to avoid IR corrections more important than $k_\ell^{-3}$ we must require $u=2s$ (to avoid $k_\ell^{-6}$) and $t=1/3$  (to avoid $k_\ell^{-4}$), which then says that 
\beqa
K_{\rm eq}^{(s)}(\k_s,\k_\ell) &\approx& {3\over 2}  \Big(u+{2\over 3}\Big) \Big({k_\ell\over k_s}\Big)^2 + \ldots
\label{KEQs2}
\eeqa
and
\beqa
K_{\rm eq}^{(\ell)}(\k_s,\k_s') &\approx& (4-3u) \Big({k_s\over k_\ell}\Big) + \ldots
\label{KEQl2}
\eeqa
Let us now restore the full angular dependence in Eq.~(\ref{KEQs2}), and include the cosine of the angle between short and long modes $x\equiv \hat{k}_s \cdot \hat{k}_\ell$,
\begin{widetext}
\beqa
K_{\rm eq}^{(s)}(\k_s,\k_\ell) &\approx& 2 (-2+3u)\, x \Big({k_\ell\over k_s}\Big) + 
\Big[(7x^2+2x-1) % \nonumber \\ & & 
-{3\over 2}u\,(9x^2+2x-3)\Big] \Big({k_\ell\over k_s}\Big)^2
 + \ldots % \nonumber \\ &\equiv  &  A(u,x)\, \Big({k_\ell\over k_s}\Big) + B(u,x)  \, \Big({k_\ell\over k_s}\Big)^2
\label{KEQs3}
\eeqa
from which Eq.~(\ref{KEQs2}) can be obtained by using $x = -k_\ell/2k_s$ for the triangle considered above. We see from Eq.~(\ref{KEQs3}) that a term of ${\cal O}(k_\ell/k_s)$ is now present, it is however proportional to the cosine variable $x$, which as long the triangle is not isosceles is of order unity. This means that  the squeezed limit in Eq.~(\ref{BcrossSQZ}) is given by
\beqa
B_{\Phi\Phi\phi}^{\rm eq} &\approx&\fnl [4-12\,x^2+6u\,(1+x^2)]  \, \Big({k_\ell\over k_s}\Big)^2 %\nonumber \\ & &  
\times \,
 P_\phi(k_\ell) P_\phi(k_s) + {\cal O}\Big({k_\ell\over k_s}\Big)^3, 
\label{BcrossSQZ2}
\eeqa
\end{widetext}
where the linear term in $(k_\ell/k_s)$ proportional to $x$ cancels when summing over the two terms in Eq.~(\ref{BcrossSQZ}) since $\k_s$ and $\k'_s$ are nearly antiparallel. On the other hand,  restoring the full angular dependence for the Bardeen potential bispectrum, Eq.~(\ref{BSQZ}), we have
\beqa
B_{\Phi\Phi\Phi}^{\rm eq} &\approx& 
12\fnl\, (1-x^2)\, \Big({k_\ell\over k_s}\Big)^2   \, P_\phi(k_\ell) P_\phi(k_s) \nonumber \\ &&
\label{BSQZ2}
\eeqa
We see that Eqs.~(\ref{BcrossSQZ2}) and~(\ref{BSQZ2}) disagree in amplitude for all values of $u$, although they agree on the scaling.   Does this mean that PBS predicts scale-independent bias corrections for the equilateral template? It's not so trivial, because Eq.~(\ref{BcrossSQZ2}) depends on $u$, and in fact for our choice here ($u=0$) the angular average of the  $(k_\ell/k_s)^2$ amplitude vanishes! Thus, there is no correction to the bias at this order, and the leading order will come from terms of order  $(k_\ell/k_s)^3$. The cross bispectrum to this order for $u=0$ reads,\begin{widetext}
\beqa
B_{\Phi\Phi\phi}^{\rm eq}|_{u=0} &\approx&\fnl \Big[(4-12\,x^2)  \, \Big({k_\ell\over k_s}\Big)^2 
 %\nonumber \\ & &  
 + 2x\, (21x^2+2x-11) \, \Big({k_\ell\over k_s}\Big)^3 \Big]  %\nonumber \\ & &
 \times \,
 P_\phi(k_\ell) P_\phi(k_s) + \ldots
\label{BcrossSQZ3}
\eeqa
\end{widetext}
The full calculation to this order will have to include the cross term between the $(k_\ell/k_s)^2$ and the next-to-leading corrections ${\cal O}(k_\ell/k_s)$ coming from $M_m(|\k-\q|)$ discussed in Appendix~\ref{IBlocApp}. However, ignoring such cross terms and using only the angular average of the second term in Eq.~(\ref{BcrossSQZ3}) gives a very good approximation to the asymptotic value of $I_{21}$ in the low-$k$ limit as shown by the bottom panel in Fig.~\ref{IBnonloc}.

Therefore we have an interesting and nontrivial example for the equilateral template where the scaling of the kernel, Eq.~(\ref{KEQs3}), would suggest a $k^{-1}$ {\em scale-dependent} bias, this term however cancels in the cross bispectrum, Eq.~(\ref{BcrossSQZ2}), which suggests a {\em scale-independent} bias, but this amplitude has for our choice $u=0$ a vanishing angular average, therefore one has to go to the next order and finds a {\em scale-dependent} bias contribution going as $k^{+1}$ with a small amplitude. Therefore, in this particular example, the scaling of the possibly scale-dependent bias when doing the PBS in $\phi$ ($k^{+1}$) disagrees from when doing the PBS in $\Phi$ ($k^0$), which agrees with local bias models. This is made possible because $K^{(\ell)}$ is sufficiently singular compared to $K^{(s)}$ in the squeezed limit, i.e.  $K^{(\ell)}\sim K^{(s)} (k_s/k_\ell)^3$, to compensate for the ratio of power spectra between short and long modes. These differences could be in principle be checked in simulations by running initial conditions with different values of $u$. However, because for equilateral PNG these corrections are small, it may be difficult to do in practice  (see Fig.~\ref{BiasEq}). 

Let us now turn to the orthogonal template. We have for the kernels,
\beqa
K_{\rm ort}^{(s)}(\k_s,\k_\ell) &\approx& {9\over 2} (u-2s) +(1-9t)\Big( {k_\ell\over k_s} \Big) + \ldots \nonumber \\ & & 
\label{KORTs}
\eeqa
and
\beqa
K_{\rm ort}^{(\ell)}(\k_s,\k_s') &\approx& {-9} (u-2s) \Big({k_s\over k_\ell}\Big)^3 -2 (4-9t)\Big( {k_s\over k_\ell} \Big)^2 \nonumber \\ & & -18\, (s+t-1) \Big({k_s\over k_\ell}\Big) + \ldots
\label{KORTl}
\eeqa
which after imposing one-loop power spectrum constraints ($u=2s$ and $t=4/9$) give, restoring the angular dependence
\beqa
K_{\rm ort}^{(s)}(\k_s,\k_\ell) &\approx& [-3 +(18u-13)x] \Big( {k_\ell\over k_s} \Big) + \ldots \nonumber \\ & & 
\label{KORTs2}
\eeqa
and
\beqa
K_{\rm ort}^{(\ell)}(\k_s,\k_s') &\approx& [10-9u(1+x^2)] \Big({k_s\over k_\ell}\Big) + \ldots
\label{KORTl2}
\eeqa
again, showing very different scaling of the kernels. However, because the $K^{(\ell)}$ kernel is only enhanced by $({k_s/ k_\ell})^2$ compared to $K^{(s)}$ it cannot compensate for the suppression factor  $P_\phi(k_s)/P_\phi(k_\ell)$ in the $\Phi$ bispectrum and thus now both squeezed-limit bispectra agree to leading order,
\beqa
B_{\Phi\Phi\phi}^{\rm ort}\approx B_{\Phi\Phi\Phi}^{\rm ort} &\approx& -12\fnl \Big({k_\ell\over k_s}\Big) P_\phi(k_\ell) P_\phi(k_s), \nonumber \\ & & 
\label{BORTSQZ}
\eeqa
where again the linear terms in $x$ in Eq.~(\ref{KORTs2}) cancel in the cross bispectrum, leading to a $u$-independent result. 
In this case, therefore, the  scaling of the scale-dependence bias with $k_\ell$ would agree in the PBS with the split performed in $\phi$ or $\Phi$.

\end{document}